\documentclass[a4paper,11pt,oneside]{article}

\title{Derivation and thermodynamically consistent coupling of a  Debye-H\"uckel type energy to a steric electrolyte model  		and application to the apparent molar volume and phase boundaries}
\author{Wolfgang Dreyer \and R\"udiger M\"uller\\
{rue.m.sci@gmail.com} } %%%%%%%%%% % Body of Paper Begins

\usepackage{graphicx,xcolor}
\usepackage{amsmath,amsfonts,amssymb}

\usepackage{microtype}
\usepackage{mhchem}		%
\usepackage{bm}			%
\usepackage[normalem]{ulem}

\def\BD{{}}		% for debuging, i.e. check consistent usage of macros
\def\PreprintClearpage{}
\def\PrePrintEmptyLine{
}
\def\BK{{}}

\newcommand{\Ukg}{\,\text{kg}}
\newcommand{\Ug}{\,\text{g}}
\newcommand{\UK}{\,\text{K}}

\newcommand{\UJ}{\,\text{J}}
\newcommand{\Us}{\,\text{s}}

\newcommand{\Um}{\,\text{m}}

\newcommand{\Unm}{\,\text{nm}}
\newcommand{\Upm}{\,\text{pm}}
\newcommand{\Ul}{\,\text{L}}

\newcommand{\UV}{\,\text{V}}
\newcommand{\UC}{\,\text{C}}

\newcommand{\Umol}{\,\text{mol}}
\newcommand{\Uev}{\,\text{eV}}

\def\tilde{\widetilde}
\def\eps{\varepsilon}

\def\cF{\mathcal{F}}
\def\cI{\mathcal{I}}
\def\cL{\mathcal{L}}
\def\cN{\mathcal{N}}
\def\cO{\mathcal{O}}

\def\div{\operatorname{div}}

\def\e{\textrm{e}}
\def\F{\textrm{F}}

\def\jump#1{[ \! [ #1 ] \! ]}

\def\vau{{\BD\upsilon\BK{}}}
\def\kB{{\BD{}k_B\BK{}}}

\def\Avo{{\BD{}\cN_A\BK{}}}

\def\Mol{\BD{}m\BK{}}
\def\Ratio{\BD\sigma\BK{}}
\def\Diss{\BD\delta\BK{}}
\def\LMicro{\ell}
\def\nInv{{\BD{}\nu\BK{}}}
\def\Thermal#1{{\BD{}\beta_{#1}^{T^\Ref}\!\BK{}}}
\def\Thermal#1{{\BD{}\beta_{#1}\BK{}}}

\def\Ref{{\BD{}\mathrm{ref}\BK{}}}
\def\Deb{{\BD{}\mathrm{D}\BK{}}}
\def\SM{{\BD{}\mathrm{sim}\BK{}}}

\def\Visc{\mathrm{visc}}

\def\Pure{\mathrm{pure}}
\def\Ideal{{\BD{}\mathrm{ideal}\BK{}}}
\def\El{\mathrm{el}}
\def\Visc{\mathrm{visc}}

\begin{document}
\maketitle

\begin{abstract}
% We propose to determine the size parameters of a steric electrolyte model from the experimental data of apparent molar volume obtained from mass density measurements.
% Thereby, we avoid the difficulties associated with modeling the complex structure of the double layer.
% To this end,
We derive a Debye-Hückel-like model of the electric ion-ion interaction for non-constant dielectric susceptibility, which does not depend on any kind of charging process due to its foundation in the general framework of non-equilibrium electro-thermodynamics.
The derivation, however, leads to a novel thermodynamic consistency condition for the temperature dependence of the susceptibility.
Due to its contributions to the total pressure, the consistent coupling of this new contribution to the free energy requires subtle modifications in the derivation of the simple mixture model for electrolytes.
The coupled model is then applied to the apparent molar volume for various related monovalent salts over a wider range of salt concentrations and temperatures, and classical tests of the electrolyte theory at phase boundaries are investigated.
\end{abstract}

\section{Introduction}\label{sect:intro}
Mixing salt in water and observing the change in volume and mass density seems to be one of the simplest experiments imaginable, cf.~\cite{RedlichMeyer:1964}.
But even this experiment can reveal shortcomings and limitations of standard continuum models of electrolytes.
In particular, frequently used steric Poisson-Boltzmann models, cf.\ e.g.\ \cite{KrIgIg:1996,BoAnOr97,KiBaAj07I,DGM13},
can not be expected to yield meaningful results in predicting mass density or apparent molar volume of electrolytes,
despite the fact that they are developed with the goal to handle volume exclusion effects.
In principle, the experiment can be considered as sufficiently explained by the Debye-H\"uckel theory \cite{DH23},
at least in the limit of strong dilution.
Nevertheless, the Debye-H\"uckel theory is often considered too simplified and, for example,
the role of non-constant susceptibility is still actively discussed \cite{ValiskoBoda:2014,ShLy:2015,KoMaMoTh:2018,SiLiKo:2023}.
While classical Debye-H\"uckel theory is usually considered applicable only for dilute solutions,
steric Poisson-Boltzmann like models are designed for concentrated electrolytes.
Thus it appears worthwhile to couple both approaches.
With the aim of comprehensive and fundamental revision of the modeling
within the context of non-equilibrium electro-thermodynamics,
the main goals here are:
\begin{enumerate}
\item	Derivation of a Debye-H\"uckel-like electric interaction energy that consistently incorporates temperature and concentration dependent susceptibility from the outset.
\item	Establishing the non-obvious consistent coupling of the steric simple mixture model with the obtained interaction energy.
\item	Incorporating in general incomplete dissociation in order to apply linear material laws for mixtures
		while still retaining the nonlinear observed behavior.
\item	Testing the developed theory over a wider concentration and temperature range
		using the phenomena of apparent molar volume and phase equilibrium at phase boundaries.
\end{enumerate}

\subsection{Some historic background of models}\label{sect:hist}

\paragraph*{Ideal electrolyte theory.}
The microscopic concept of dissociation of a salt into ions was introduced by Arrhenius \cite{Arrhenius:1887}
and greatly supported by van't Hoff's observation of the analogy
between diluted electrolyte solutions and osmosis at semi-permeable membranes.
Thus, in the early electrolyte theory ideality was expressed
by the logarithmic chemical potential inherited from the ideal gases and Ostwald's mass action law.
The theory relied on incomplete dissociation to account
for the Kohlrausch square root law observed in conductivity measurements.
However, the classical ideal electrolyte model was not able to explain different experimental data consistently.
In particular, when assuming the dissociation degree determined by conductivity measurement,
then the Ostwald mass action law was not able to yield a reaction constant, cf.\cite{Redlich:1946}.
We remark that in light of the non-equilibrium thermodynamic theory, which was only developed decades later,
equilibrium quantities like the free energy or chemical potentials
should not be determined by non-equilibrium properties like the electrolyte conductivity, cf.\ e.g.~\cite{DGM19}.
The classical ideal theory leads on a macroscopic scale to a Boltzmann distribution of ions and
where ions do not necessarily have to be point charges,
but volume exclusion effects can be neglected due to the strong dilution assumption
and thus there is no mechanism in the model to limit charge accumulation at strongly charged surfaces.

\paragraph*{Steric electrolyte models.}
In the electrochemical double layer, electrolytes are in general concentrated
and there is voltage drop over a typically very short distance, giving rise to locally extremely strong electric fields.
The standard Poisson-Boltzmann model of ideal electrolytes can adequately account for the action of this electric field on ions
but the missing of volume exclusion effects prevent the model to reasonably limit the local boundary layer charge.
Therefore, in steric modified Poisson-Boltzmann models a size parameter is added
to change the mixing entropy of ideal gases into a lattice gas approach \cite{KrIgIg:1996,BoAnOr97,KiBaAj07I}.
Extensions of lattice gas models to ions of different size have also been developed, cf.\ e.g.\ \cite{Chu:2007,Li:2009,LiEi14} %
and the review in \cite{ZhangHuang:2018}.
Instead of the lattice gas approach,
in the very general simple mixture model \cite{DGM13,LGD16,DGM18},
the mixing entropy accounts for solvation of ions although formally looking identic to the ideal gas case,
and  adds an elastic energy contribution to the free energy
in order to account for elastic interaction of constituents of finite size.
In an incompressible limit of the elastic law,
the non-ideality is expressed in a dependence of the chemical potentials on the local pressure
the model can be reduced to a lattice gas.
An axiomatic derivation of the simple mixture is provided in \cite{BoDrDr23}
and it is shown that the incompressible limit requires linear dependence of the volume on the mixtures composition.
Steric modified Poisson-Boltzmann models which account for the
concentration dependence with linear dielectric decrement
are proposed in \cite{BeYaAnPo11,HavRLu:2012,NaAn15,FiChBuMo:2016}
and extended to different ion sizes developed in \cite{%
GuSt:2018,LM22}.
For steric models taking into account the dependency of the susceptibility
on the strong electric field in the double layer, we refer to \cite{LM22} and references therein.

\paragraph*{Debye-H\"uckel theory.}
Because the double layer screens outer electric fields from the electrolyte bulk,
it leaves in equilibrium the bulk electrolyte free of any electric field
and the above models of electrolytes reduce to mixtures of uncharged particles,
although the electrolytes consists besides of the solvent of negatively charged anions  and positive cations.
As a way to take electric ion-ion interactions into account
within a homogeneous, macroscopically electroneutral electrolyte,
Debye and H\"uckel introduced in 1923 a mean field model \cite{DH23}.
They determined the electric energy gained by an arbitrary selected ion
under the assumptions that:
(i.) the surrounding environment can be described in a continuum way
	and is solely characterized by its constant susceptibility
(ii.) the electrolytic environment can be described by Boltzmann distribution
	and linearization is admissible.
Then, the free energy density is deduced analytically
by integrating the inner energy density  with respect to temperature.
By this it was possible to explain non-ideal behavior at phase boundaries
without having to rely on incomplete dissociation.
Most important, it was possible to formulate limiting laws for strong dilution,
that did not depend on the specific salt, but only on the valency of the ions.
The Debye-H\"uckel theory was rapidly adopted by others and tested in different experimental applications.
Deviations from the limit law were found even at moderate salt concentrations,
especially for multivalent ions, and an ion-specific behavior of the heat of dilution was observed
\cite{GrHa:1925,LangeMeixner:1929}.
For the apparent molar volume, the limit law even predicted the wrong sign of the slope
\cite{RedlichMeyer:1964} %
and the correct sign was obtained only when the susceptibility depends on the pressure \cite{RedlichRosenfeld:1931}.
However, a susceptibility depending on temperature is not compatible with the derivation of the free energy in \cite{DH23}.
Debye therefore introduced the thought experiment of a hypothetical charging process \cite{Debye:1924},
which allows non-constant susceptibility, and arrived at the same result as before.
Although the justification of the charging process remained somewhat controversial,
cf. e.g.\ \cite{GrHa:1925,KoMaMoTh:2018} and the references therein,
it is now generally accepted.
Already in 1925 H\"uckel proposed a linear dependence of the susceptibility on the salt concentration
and a splitting of energy in ionic and solvation contributions \cite{Hueckel:1925}.
In \cite{ShLy:2015}, the charging process is carried out
with non-linear concentration-dependent susceptibility based on experimental data.
The splitting of potentials and concentration dependent susceptibility are applied in
\cite{VinczeValiskoBoda:2010,ValiskoBoda:2014,ValiskoBoda:2023}
together with a non-primitive model based on molecular simulations.
An extended Debye-H\"uckel theory based on a Poisson–Fermi approach %
was  proposed in \cite{LiuEisenberg:2015,LiuEisenberg:2018}.
Various modifications and extensions related to the susceptibility and the charging process are reviewed in \cite{KoMaMoTh:2018,SiLiKo:2023}.

The electric ion-ion interaction was also employed in a non-equilibrium context to derive
non-ideal ionic conductivity without having to relay on incomplete dissociation.
By adding Brownian motion in \cite{OnsagerII:1927},
the square root law of the conductivity by Debye-H\"uckel \cite{DH23II} was improved
and dependence of conductivity on ionic radius was removed.
We will not consider the non-equilibrium case further here.

\subsection{Aims and scope}\label{sect:scope} %
In the following, we aim at developing a macroscopic continuum model within the framework of non-equilibrium electro-thermodynamics.
The model must consistently couple steric effects of finite volumes with the electrical interaction between ions.
Effects originating from the microscopic scale,
which are explicitly resolved within the so-called non-primitive models cf.~e.g.\ \cite{ValiskoBoda:2014},
must be taken into account implicitly in the susceptibility.
The charging process is avoided in the derivation.

\paragraph*{Solvated ions.}
The first step in modeling is to define the constituents of the mixture,
in particular whether solvated ions or just the bare unsolvated ions should be considered as constituents.
The application of a solvated radius in the model is rejected in \cite{ValiskoBoda:2014,ValiskoBoda:2023}
because it might hinder the contact of ions to structural groups or the formation of ion pairs.
To the contrary, \cite{Zavitsas:2001} emphasises the importance
of modeling the reduction of free solvent molecules due to solvation
since this way already large amounts of the non-ideality of the electrolytes can be removed.
Similar ideas can also be found in \cite{RoSt2002}
and it is also comparable with a simple mixture model \cite{LGD16,DGM18} but with non-integer solvation numbers applied.
In a the simple mixture that we use, size parameters for each of the constituents are needed.
While in previous work \cite{LGD16,LM22},
solvation numbers and specific volumes were adjusted to data from differential capacitance of the double layer
we here want in contrast to deduce specific volumes of possibly solvated ions
from much more directly related experiments on the mass density or the apparent molar volume.
In particular, adding a solute to the solvent,
the solution volume may increase by less than the volume of added solute, or it may even decrease.
For example, the volume decreases in aqueous $\ce{LiF}$ and $\ce{NaF}$ solutions
such that a negative volume would have to be assigned to the bare $\ce{F^-}$ ion.
In contrast, the solvated ion still has a positive volume,
although it is less than the volume of the individual solvent molecules in the solvation shell combined.

\paragraph*{Incomplete dissociation.}
At the time of \cite{DH23},
incomplete dissociation on the one hand and
electrostatic ion-ion interaction on the other hand,
were considered exclusive alternatives to explain non-ideal behavior, c.f.\ \cite{Redlich:1946}.
Later, the concept of ion pairing was (re)introduced by Bjerrum \cite{Bjerrum:1926}.
Nowadays, the dominant opinion in the literature, with the exception of \cite{Heyrovska:1996},
is such that while there is ion pairing in weaker electrolytes and in some strong electrolytes at higher concentrations,
there is no ion pairing of alkali halides in water, cf.~\cite{MarcusHefter:2006,MayRowland:2017}. %
We include incomplete dissociation in the general model,
with the possibility that a-posteriori its relevance in certain applications may prove to be low.

\subsection{Outline}\label{sect:outline} %
We start in Sect.~\ref{sect:gen-thermodyn} by characterizing the assumptions
and stating the electro-thermodynamic framework.
In Sect.~\ref{sect:constitutive},
the axiomatic approach of \cite{BoDrDr23} is slightly modified
to derive in Sect.~\ref{sect:reps} and \ref{sect:simple-mix} a simple mixture model
that can be coupled to the
Debye-H\"uckel type free energy contribution which is derived in Sect.~\ref{sect:phi} and \ref{sect:DH}.
The full coupled model is then applied to the apparent molar volume related to mass density measurement of aqueous electrolytes  in Sect.~\ref{sect:appl}.
Concentration and temperature dependence are investigated numerically
and a set of ion-specific parameters is deduced for several related salts.
In Sect.~\ref{sect:appl-phase} the model is tested at phase boundaries
through experiments on the vapour pressure reduction  and the freezing point depression.
We conclude in Sect.~\ref{sect:discussion} with a discussion of the model, its application and remaining open issues.

\PreprintClearpage
\section{General thermodynamic model}	\label{sect:gen-thermodyn}

We apply the general electro-thermodynamic model framework based on \cite{dGM84,Mu85}
and particularly adapted for electrochemical bulk-surface systems in \cite{DGM18,ML23}.
It relies on the exploitation of an entropy principle for the restriction of constitutive quantities
like fluxes, stress or reaction rates.
A particular ingredient in the entropy principle is
the specification of a constitutive function for the entropy density, i.e.
the definition of its set of independent variables.

We denote the spatial domain occupied by an electrolyte by $\Omega\subseteq\mathbb{R}^3$
and use index sets $\cI$ to refer to the different constituents of the mixture in $\Omega$.
Each constituent in $\Omega$ has the (atomic) mass $m_\alpha$, for $\alpha\in\cI$,
and the net charge $z_\alpha e_0$, where $z_\alpha$ is the charge number of the constituent
and $e_0$ is the elementary charge.

\subsection{Assumptions and thermodynamic consistency}\label{sect:thermodyn}
We consider the \emph{entropy density} $\rho s$ to be a function of
the \emph{inner energy density} $\rho u$,
the (local) particle number densities per $\Um^3$ denoted as $n_\alpha$ for $\alpha\in \cI$ and
the \emph{vector of polarization} $\bm{P}$
as the independent variables that determine the Thermodynamic state in $\Omega$.

\paragraph*{Temperature.}
The constitutive modeling is greatly eased by the introduction of
the \emph{temperature} $T$ and the \emph{free energy density} $\rho\psi$
by means the Legendre transformation of the entropy density $\rho s$
with respect to the inner energy density $\rho u$, viz.\
\begin{align}\label{Mod01}
	\rho\psi	&= \rho u -T\, \rho s~, &
	&\text{ with } &
	\frac{1}{T}	&= \frac{\partial }{\partial \rho u} \rho s~. &
\end{align}
Then we obtain the consistency conditions
\begin{align}\label{Mod03}
	\rho s	&= -	\frac{\partial }{\partial T} \rho\psi~,&
	\rho u	&= -T^2	\frac{\partial }{\partial T} \bigg( \frac{\rho\psi}{T} \bigg)~. &
\end{align}

\paragraph*{Assumptions on the electric field.}
\begin{itemize}
\item	in the considered applications,
		all relevant features of electrochemical system can be well described in the
		\emph{quasi-electrostatic} limit of Maxwell equations.
		Thus, the electromotoric force equals the electric field $\bm{E}$
		and there exists an \emph{electrostatic potential} $\varphi$, such that
\PrePrintEmptyLine
		\begin{align} \label{Mod11}
			\bm{E}			&= -\nabla \varphi~, &
			\jump{\varphi}	&= \operatorname{const. } \text{ on intersecting surfaces,} &
		\end{align}
\PrePrintEmptyLine
		where the constant is frequently chosen as zero.
\item	The timescale of polarization relaxation in liquid electrolytes is typically
		in the range of $10^{-8} \Us$ (cf.~\cite{BuHeMa99}),
		and therefore orders of magnitude smaller than the relevant
		observed time scales of the considered application.
		Thus, we assume quasi-equilibrium of polarization relaxation.
\item	We assume that the dielectric properties of the liquid electrolyte
		are reasonably well represented by a scalar susceptibility $\chi$
		which may be a function of temperature and the electrolyte composition.
		Since the focus here is not on boundary layers where the electric field strength can be the extremely high,
		we assume in order to reduce complexity that $\chi$ is independent of the electric field $\bm{E}$.
\end{itemize}
As discussed in \cite{LM22,ML23} these assumptions allow to use the electric field $\bm{E}$ instead of $\bm{P}$
as an independent variable for the constitutive functions and to relate $\bm{E}$ to the
the vector of polarization by
		\begin{align} \label{Mod13}
			\bm{P}			&=-\frac{\partial \rho\psi}{\partial \bm{E}} ~, &
			\bm{P}			&= \chi \cdot \eps_0 \bm{E}~. &
		\end{align}

\paragraph*{Thermodynamic consistency.}
The simple relations \eqref{Mod13} suggest to choose for the free energy a constitutive function
depending on temperature, the composition of the mixture and the electric field as
\PrePrintEmptyLine
\begin{align}\label{TC123}
	\rho\psi&%
			=	\rho  \hat{\psi}( T, (n_\alpha)_{\alpha\in\cI} ) -\chi\, \tfrac{\eps_0}{2}|\bm{E}|^2 ~. &
\end{align}
\PrePrintEmptyLine
From \eqref{Mod01} and \eqref{Mod03} we obtain an analogous splitting of the entropy density and the inner energy density as
\PrePrintEmptyLine
\begin{subequations}
\label{TC125}
\begin{align}
	\label{TC125a}
	\rho s	&= -	\tfrac{\partial }{\partial T} \rho\hat{\psi}
				+	\tfrac{\partial }{\partial T} \chi\, \tfrac{\eps_0}{2}|\bm{E}|^2	&
			&= \rho \hat{s}	+				\tfrac{\partial }{\partial T}\chi		\, \tfrac{\eps_0}{2}|\bm{E}|^2 ~, &
\\
	\rho u	&= -T^2	\tfrac{\partial }{\partial T} \bigg( \frac{\rho\hat{\psi}}{T} \bigg)
				-T^2\tfrac{\partial }{\partial T} \bigg( \frac{-\chi\, \tfrac{\eps_0}{2}|\bm{E}|^2}{T} \bigg)	&
			&= \rho \hat{u}	-\big( \chi -T\,\tfrac{\partial }{\partial T}\chi \big)	\, \tfrac{\eps_0}{2}|\bm{E}|^2 ~, &
	\label{TC125b}
\end{align}
\end{subequations}
\PrePrintEmptyLine
such that $ \rho\hat{\psi} = \rho\hat{u} -T\,\rho\hat{s}$.
Moreover, an analogous splitting holds for the chemical potentials, i.e.\
\begin{align}\label{TC127}
	\mu_\alpha	&= \frac{\partial \rho\psi}{\partial n_\alpha}
	\quad		= \hat{\mu}_\alpha -\tfrac{\partial }{\partial n_\alpha} \chi \, \tfrac{\eps_0}{2}|\bm{E}|^2 ~. &
\end{align}
\PrePrintEmptyLine
By means of a Gibbs-Duhem relation, we introduce a quantity that we refer to as pressure, viz.
\begin{align}
	p	&:= -\rho\psi	+\sum_{\alpha\in\cI} n_\alpha \mu
	\notag
	\\*
	\label{TC129}
	&= -\rho\hat{\psi}	+\sum_{\alpha\in\cI} n_\alpha \hat{\mu}_\alpha
		\quad
		+\tfrac{1}{2}{\eps_0}|\bm{E}|^2 \Big(
			\chi	-\sum_{\alpha\in\cI} n_\alpha\,\tfrac{\partial }{\partial n_\alpha} \chi
			\Big) &
	&= \hat{p} + p^\El ~. &
\end{align}
In order to get a correct interpretation of the pressure, $p$ has to be related to the momentum balance.

\subsection{Balance equations of charged mixtures}	\label{sect:mix}
We summarise here the model equations adapted to the
above simplifying assumptions, i.e. fast polarisation relaxation.
The \emph{total mass density}, the \emph{barycentric velocity} and the \emph{free charge density}
are defined as
\begin{align}
	\label{PCM12}
	\rho		&=		\sum_{\alpha\in\cI} m_{\alpha} n_{\alpha}~, &
	\bm{\vau}	&=		\sum_{\alpha\in\cI} m_{\alpha} n_{\alpha} \bm{\vau}_\alpha ~, &
	n^\F		&= e_0	\sum_{\alpha\in\cI} z_{\alpha} n_{\alpha}~, &
\end{align}
where $\bm{\vau_\alpha}$ denotes the velocity of constituent $\alpha\in\cI$.
Within the volume domain $\Omega$  the balance equations are for time $t>0$
\PrePrintEmptyLine
\begin{subequations}
\label{EC111}
\begin{align}
\label{EC111A}
	-\div( (1+\chi)\eps_0 \nabla\varphi) &= n^\F~,
\\[1ex]
\label{EC111a}
	\partial_t \rho_\alpha	+\div( \rho_\alpha\bm{\vau} +\bm{J}_\alpha)
	&= m_\alpha \sum_{k}	\nu_\alpha^k R^k ~,
\\
\label{EC111c}
	\partial_t(	\rho \bm{\vau} )
	+\div\big(	\rho \bm{\vau} \otimes \bm{\vau} -\bm{\sigma}^\Visc \big)	+\nabla p
	&= \rho \bm{f}
		-n^\F\nabla\varphi		+	(D^2\varphi)\,\chi\eps_0\nabla\varphi~,
\\[1ex]
	\partial_t(	\rho u)
	+\div\left(	\rho u\, \bm{\vau} +\bm{Q}	\right)
	&=
		-\bm{J}^\F\cdot\nabla\varphi\	+\bm{\sigma}^\Visc :\nabla\bm{\vau}\	-p\,\div(\bm{\vau})
\notag \\*
\label{EC111d}
	& \rule{3ex}{0ex}
	+\left( \partial_t (\chi\eps_0\nabla\varphi)	+(\bm{\vau} \cdot \nabla)(\chi\eps_0\nabla\varphi) \right)\cdot \nabla\varphi
\\*\notag
	& \rule{3ex}{0ex}
		+\chi\,\eps_0|\nabla\varphi|^2 \, \div(\bm{\vau})~,
\end{align}
\end{subequations}
\PrePrintEmptyLine
Each one of the reactions with the reaction rates $R^k$ in \eqref{EC111a} has to conserve mass and charge
and not all of the diffusive mass fluxes $\bm{J}_\alpha$ are linearly independent. %
We thus have for the stoichiometric coefficients $\nu_\alpha^k$ and for the fluxes the constraints
\PrePrintEmptyLine
\begin{align}
	\sum_{\alpha\in\cI} m_\alpha \nu_\alpha^k  &= 0 ~, &
	\sum_{\alpha\in\cI} z_\alpha \nu_\alpha^k  &= 0 ~,  &
	\sum_{\alpha\in\cI} \bm{J}_\alpha &= 0  ~, &
\end{align}
\PrePrintEmptyLine
such that one of the partial mass balances \eqref{EC111a} can be replaced by the total mass balance.
The right hand side of the momentum balance \eqref{EC111c} contains external body forces $\rho \bm{f}$ like gravitational force
and in addition Lorentz force and the Kelvin force term $(\nabla\bm{E})\,\bm{P}$.
Taking the $p^\El$ contribution out of the pressure term $\nabla p$ on the left hand side
and combining it with the Kelvin force term on the right hand side,
we get the Korteweg-Helmholtz force density, cf.~\cite[Sect.~3.2]{LM22} and references therein.
The pressure term and all forces except of the external body force
can be moved into the divergence to form the total stress tensor, viz.\
\begin{align}
	\label{EC151}
	\partial_t (\rho\bm{\vau})
		+\div\left(	\rho\bm{\vau}\otimes\bm{\vau}
			-\left( \bm{\sigma}^\Visc	-\hat{p}\,\bm{1}
					-( \tfrac{\eps_0}{2}|\bm{E}|^2 +p^\El )\,\bm{1} +(\eps_0\bm{E} +\bm{P})\otimes\bm{E}	\right) \right)
	&= \rho\bm{f} ~,
\end{align}
where $\bm{1}$ denotes the unit tensor.
Then, the field independent part of the total stress tensor
consists of $-\hat{p}$ on the diagonal and the viscous stress tensor.
The free current in \eqref{EC111d} is $\bm{J}^\F = \e_0 \sum_{\alpha\in\cI} z_\alpha \bm{J}_\alpha$.

Finally the remaining constitutive quantities are determined by the entropy principle.
Due to the linear dependence of the fluxes $\bm{J}_\alpha$,
one constituent is selected as a reference.
Typically the solvent is chosen and we refer here to the chosen constituent by the index $0\in\cI$.
The constitutive equations are %
for each counting index $k$ and  $\alpha\in\cI\setminus\{0\}$
\PrePrintEmptyLine
\begin{subequations} \label{EC123}
\PrePrintEmptyLine
\begin{align}
\label{EC123d}
	R^k	&= R_{0}^k \cdot \bigg(1-\exp\Big(\frac{A^k}{\kB T} \sum_{\alpha\in\cI^\pm} \nu_\alpha^k\, \mu_\alpha \Big)\bigg) ~,
\\
\label{EC123bb}
	\bm J_\alpha &=	%
			-\!\!\sum_{\gamma\in\cI\setminus\{0\}} M_{\alpha\gamma}\cdot\bigg(
				\nabla			\left( \frac{\mu_\gamma}{T}	-\frac{\mu_0}{T}	\frac{m_\gamma}{m_0}	\right)
				+\frac{e_0}{T}	\left(			z_\gamma		-z_{0}			\frac{m_\gamma}{m_0}	\right) \nabla\varphi
				\bigg)
	~,
\\
\label{EC123c}
	\bm{\sigma}^\Visc	&=	\eta_b \, \div( \bm{\vau} )\,\bm{1}\
						+	\eta_s\cdot\left( \nabla\bm{\vau} +(\nabla\bm{\vau})^T \right)~,
\\
\label{EC123b}
	\bm{Q}		&=\	-\frac{\kappa}{T^2}\, \nabla T ~.
\end{align}
\end{subequations}
\PrePrintEmptyLine
The phenomenological coefficients satisfy
$A^k>0$, $\kappa>0$,
$( \eta_b + \tfrac{3}{2} \eta_s ) > 0$, $\eta_s > 0$
and $M_{\alpha\gamma}$ form a positive definite matrix.
Most relevant is here that the equilibrium condition following from \eqref{EC123d} is the mass action law.
Boundary conditions for the system \eqref{EC111} follow from surface balances
and the application of an entropy principle on the surface, cf.~\cite{DGM18,ML23}.

\PreprintClearpage
\section{Derivation of the constitutive model}	\label{sect:constitutive}

\subsection{Outline of our strategy}
There are typically two distinct regimes of electrolytes in equilibrium
where we propose to apply different constitutive equations:
\begin{itemize}
\item[(BL)]
	\emph{Boundary layers} at domain boundaries or intersecting surfaces.
	In these layers there is locally a macroscopic charge imbalance, i.e.\ $n^\F \neq 0$.
	The according strong electric field, that reaches the order of $10^{9}\UV\Um^{-1}$ %
	by far dominates the electric interaction of individual ions.
	Thus, we can apply above splitting of the constitutive equations
	\eqref{Mod13}$_\text{(right)}$, \eqref{TC123}-\eqref{TC129},
	with a free energy function $\rho\hat{\psi}(T, (n_\alpha)_{\alpha\in\cI} )$
	like it is used for uncharged particles.
	In particular, we apply a simple mixture model.
\item[(HB)]
	\emph{Homogeneous bulk} domain are shielded from external electric influence by the boundary layers.
	On a macroscopic scale, the electrolyte thus appears locally electro-neutral, i.e. $n^\F=0$
	and the electric field vanishes, viz.\ $\bm{E}=0$.
	This reduces the free energy function \eqref{Mod13}$_\text{(right)}$ to
	$\rho\psi = \rho\hat{\psi}(T, (n_\alpha)_{\alpha\in\cI} )$.
	The missing of dominant macroscopic electric forces
	now allows some otherwise neglected higher order effects like the electric interaction of ions to become relevant.
	We thus combine the simple mixture model with a Debye-H\"uckel like model
	that takes into account a microscopic electric field in the vicinity of ions and
	transfers the electric field energy contribution after spatial averaging into a field independent macroscopic one,
	i.e., we have
	\begin{align}\label{CM11}
			\rho\hat{\psi}(	T, (n_\alpha)_{\alpha\in\cI} )
		=	\rho\psi^\SM(	T, (n_\alpha)_{\alpha\in\cI} )
		+	\rho\psi^\Deb(	T, (n_\alpha)_{\alpha\in\cI} )	~,
	\end{align}
	where $\rho\hat{\psi}^\SM$ and $\rho\hat{\psi}^\Deb$
	denote the contributions of the simple mixture model and the Debye-H\"uckel like model, respectively.
\end{itemize}

A simple mixture for electrolytes, including entropy of mixing, elastic interactions between the constituents
and solvation of ions,
has been introduced and discussed in \cite{DGM13,LGD16} for the isothermal case
and more general in \cite{BoDrDr23}.
In the spirit of \cite{BoDrDr23},
we will not directly postulate a free energy function,
but rather make assumptions on specific heat, pressure, specific energy and specific entropy.
Then, we are able to deduce a free energy function compatible with these assumptions.
But since we have to be expect both $\rho\psi^\SM$ and $\rho\psi^\Deb$  in \eqref{CM11}
to contribute to the pressure,
some subtile differences compared to \cite{BoDrDr23} are necessary here.

The derivation of Debye-H\"uckel like model consists of two parts:
The first part starts the same way as in \cite{DH23}
with selecting an arbitrary individual ion of the mixture and
calculating its electric field interaction with the ions in surrounding neighborhood.
In the second part, our approach differs from \cite{DH23} in the way
that we base the calculation of the inner energy and entropy %
on the splitting \eqref{TC125}$_\text{(right)}$,
which we assume it also applies to the electric field in the vicinity of the ions  on the more microscopic scale.
This finally allows a direct identification of the macroscopic energy contributions after spatial averaging.

\begin{table}
\centering
\caption{\label{tab:const}
	Physical constants.}
\begin{tabular}{ll}
\hline
\hline
\\[-2ex]
Boltzmann constant	& $\kB = 1.380649\times 10^{-23} \UJ \UK^{-1}$ 		%
\\
elementary charge	& $e_0 = 1.602176634\times 10^{-19} \UC$ 				%
\\
Avogadro number		& $\Avo = 6.02214076\times 10^{23} \Umol^{-1}$ 			%
\\
dielectric constant	& $\varepsilon_0 = 8.8541878188\times 10^{-12} \UC\UV^{-1}\Um^{-1}$ %
\\
\hline
\end{tabular}
\end{table}
\subsection{Change of variables and representation equations.} \label{sect:reps}
The derivation of a free energy is greatly eased by a change of variables.
We denote the total number density and its inverse by
\begin{align}\label{CM31}
	n		&= \sum_{\alpha\in\cI} n_\alpha ~, &
	\nInv 	&=\frac{1}{n} ~, &
\end{align}
respectively, where $\nInv$ has the dimensions of a volume.
Then, we introduce as new variables the mole fractions
\begin{align}\label{CM32}
	y_\alpha &=\frac{1}{n}\, n_\alpha	=   \nInv \, n_\alpha &
	&\text{ implying the constraint } \quad \sum_{\alpha\in\cI}	y_\alpha = 1 ~. &
\end{align}
For abbreviation we set $\bm{y} = (y_\alpha)_{\alpha\in\cI}$
and introduce $M(\bm{y})	= \sum_{\alpha\in\cI} m_\alpha y_\alpha$.
We consider three different sets of basic variables
to cover the state space, viz.
\PrePrintEmptyLine
\begin{align}\label{CM33}
	(T, (n_\alpha)_{\alpha\in\cI} )			\quad\longleftrightarrow\quad
	(T, \nInv, \bm{y} )						\quad\longleftrightarrow\quad
	(T, p, \bm{y} ) ~.
\end{align}
\PrePrintEmptyLine
While the first one of the variable transformations in \eqref{CM33} is most simple,
the second one needs a constitutive equation which has not been specified up to now.
This is the so-called \emph{thermal equation of state} relating the pressure to the chosen variables.
Depending on the chosen variables, this constitutive law is given in the abstract form
\begin{align}\label{CM37}
	p	= \bar{p}(T, \nInv, \bm{y}) 		\quad\longleftrightarrow\quad
	\nInv= \tilde{\nInv}(T, p, \bm{y})~.
\end{align}
\PrePrintEmptyLine
For a generic quantity $f$,
we denote the function acting on the first set of variables in \eqref{CM33} as
$f = \hat{f}(T, (n_\alpha)_{\alpha\in\cI} )$.
Then, we define two new functions,
\PrePrintEmptyLine
\begin{align}\label{CM35}
	\bar{f}(T, \nInv, \bm{y})	&:= \hat{f}(T, \tfrac{1}{\nInv} \bm{y}) &
	& \text{ and }	&
	\tilde{f}(T, p, \bm{y}) 	&:= \bar{f}(T, \tilde{\nInv}(T,p,\bm{y}), \bm{y}) . &
\end{align}
\PrePrintEmptyLine
Obviously the thermal equation of state must satisfy the Gibbs-Duhem equation \eqref{TC129} for $\bm{E} = 0$.
Thus, by using \eqref{CM35} and applying chain rule, we easily obtain from \eqref{TC125} and \eqref{TC129}
\PrePrintEmptyLine
\begin{align}\label{CM41}
	M(\bm{y}) \tfrac{\partial}{\partial \nInv} \bar{\psi}	&= -\bar{p} ~, &
	M(\bm{y}) \tfrac{\partial}{\partial \nInv} \bar{u}		&= T\tfrac{\partial}{\partial T} \bar{p} -\bar{p} ~, &
	M(\bm{y}) \tfrac{\partial}{\partial \nInv} \bar{s}		&=	\tfrac{\partial}{\partial T} \bar{p} ~. &
\end{align}
\PrePrintEmptyLine
We conclude that the volume derivatives of
$\bar{\psi}(T, \nInv, \bm{y})$, $\bar{u}(T, \nInv, \bm{y})$ and $\bar{s}(T, \nInv, \bm{y})$
are fully determined by the pressure $\bar{p}(T, \nInv, \bm{y})$.

\paragraph*{Definition of the specific heats.}
The specific enthalpy is defined by
$M(\bm{y})\,\tilde{h}(T, p, \bm{y}) = M(\bm{y})\,\tilde{u}(T, p, \bm{y}) +p\cdot\tilde{\nInv}(T, p, \bm{y})$.
The derivatives of specific internal energy and the specific enthalpy with respect to temperature $T$ represent
the specific heat at constant volume and
the specific heat at constant pressure, respectively, i.e.
\PrePrintEmptyLine
\begin{align}\label{CM53}
	c_v(T, \nInv, \bm{y})	&:=	\frac{\partial \bar{u}(T, \nInv, \bm{y})}{	\partial T}~, &
	c_p( T, p, \bm{y})			&:=	\frac{\partial \tilde{h}(T, p, \bm{y})}{	\partial T}~. &
\end{align}
\PrePrintEmptyLine
While $c_p$ can often be measured easily in experiments,
in many cases $c_v$ is more easy to treat in the constitutive theory.
According to \cite{BoDrDr23} %
the important difference of the specific heats satisfies
\PrePrintEmptyLine
\begin{align}\label{CM55}
	c_p - c_v 	= \frac{T}{M(\bm{y})} \frac{\partial \tilde{\nInv }}{\partial T} \frac{\partial \bar{p} }{\partial T}
					\quad \geq 0 ~.
\end{align}
\PrePrintEmptyLine
If in particular the pressure is a linear function of temperature,
then we conclude from \eqref{CM41} that $\tfrac{\partial}{\partial \nInv} c_v (T, \nInv, \bm{y}) = 0$.
Using \eqref{TC125}, we infer from the definition \eqref{CM53} that the specific entropy
$\bar{s}(T, \nInv, x) := \hat{s}(T, \nInv^{-1} \bm{y})$ satisfies
\PrePrintEmptyLine
\begin{align}\label{SM231}
				\tfrac{\partial}{\partial T}		\bar{s} &= \frac{c_v}{T} ~. &
\end{align}
\PrePrintEmptyLine

\paragraph*{Representations of specific internal energy and specific entropy.}
Integrating \eqref{CM41} and \eqref{CM53} with respect to $\nInv$ and $T$, respectively,
we deduce a representation for the specific internal energy,
\PrePrintEmptyLine
\begin{align}
		M(\bm{y})\bar{u}(T,		\nInv,		\bm{y})
	=&	M(\bm{y})\bar{u}(T^\Ref,\nInv^\Ref,	\bm{y})
	+	M(\bm{y})	\int_{T^\Ref}^T 		c_v(\theta, \nInv^\Ref, \bm{y}) \,d\theta
\notag\\*
	&+				\int_{\nInv^\Ref}^\nInv	( T \tfrac{\partial}{\partial T} \bar{p} -\bar{p})(T, \xi, \bm{y} ) \,d\xi
	~.
\label{SM213}
\end{align}
\PrePrintEmptyLine
Integrating the specific heat at constant volume
and the thermal equation of state with respect to $T$ and $\nInv$, respectively,
we derive for specific entropy the representation
\PrePrintEmptyLine
\begin{align}
		M(\bm{y})\bar{s}(T,		\nInv,		\bm{y})
	=&	M(\bm{y})\bar{s}(T^\Ref,\nInv^\Ref,	\bm{y})
	+	M(\bm{y})	\int_{T^\Ref}^T 		\frac{c_v}{\theta}(\theta, \nInv^\Ref, \bm{y}) \,d\theta
\notag\\*
	&+				\int_{\nInv^\Ref}^\nInv \tfrac{\partial}{\partial T} \bar{p}(T, \xi, \bm{y}) \,d\xi
	~.
\label{SM233}
\end{align}
\PrePrintEmptyLine
The specific free energy may be calculated from \eqref{SM213} and \eqref{SM233} as
$\bar{\psi}(T, \nInv, \bm{y}) = \bar{u}(T, \nInv, \bm{y}) - T \bar{s}(T, \nInv, \bm{y})$.
Thus a constitutive model can be based on assumptions on the functions
$\bar{p}(T, \nInv, \bm{y})$, $c_v(T, \nInv^\Ref, \bm{y})$,
$\bar{u}(T^\Ref , \nInv^\Ref, \bm{y})$ and $\bar{s}(T^\Ref, \nInv^\Ref, \bm{y})$.

\subsection{The simple mixture model} \label{sect:simple-mix}
For the simple mixture model, we make the following assumptions
similar to \cite{BoDrDr23}:

\paragraph*{Thermal equation of state.}
To describe changes of the molar volume due to thermal expansion, elastic compression and changes of the composition,
we assume the following simple constitutive law for the pressure
\begin{subequations}
\label{SM331}
\begin{align}\label{SM333}
	\bar{p}^\SM(T,\nInv,\bm{y})
	&=	p^\Ref
		+K\cdot	\bigg( \dfrac{\nInv^\SM(T,\bm{y})}{\nInv}  	%
				-1 \bigg) ~,
\\
	\nInv^\SM(T,\bm{y})			&=\sum_{\alpha\in\cI} \vau_\alpha^\Ref\Thermal{\alpha}(T)\cdot  y_\alpha ~.
\end{align}
\end{subequations}
Here,  $K > 0$ denotes the isotropic compression modulus,
and $\vau_\alpha^\Ref$ for $\alpha \in\cI$ are the molar volumes
in a reference state with $T=T^\Ref$ and $p^\SM=p^\Ref$.
The functions $\Thermal{\alpha}(T)$ describing the thermal expansion of constituent $\alpha\in\cI$
have to satisfy $\Thermal{\alpha}(T^\Ref) = 1$. %
In the vicinity of the reference temperature, a linearization may be applied as
\begin{align}\label{SM335}
	\Thermal{\alpha}(T) 	\approx		1 +\left(	\frac{T}{T^\Ref} -1	\right) \, T^\Ref \beta_\alpha^\Ref ~.
\end{align}

\paragraph*{Specific heat at constant volume $\nInv = \nInv^\Ref = \nInv^\SM(T^\Ref,\bm{y})$.}
We choose for $c_v(T, \nInv^\SM,\bm{y})$ a simple linear approach with respect to $y_\alpha$,
that does not depend on temperature,
i.e.\
\begin{align}\label{SM351}
 	M(\bm{y})\, c_v(T, \nInv^\SM,\bm{y})		&=\sum_{\alpha\in\cI} m_\alpha c_\alpha^\Ref\, y_\alpha ~,
\end{align}
where the constant $c_\alpha^\Ref = c_\alpha^\Ref(\nInv^\SM)$
are the specific heat constants of the pure constituents.\footnote{\label{SM353}%
	While for pure uncharged substances these constants may be found in tables,
	they will not be available for solvated ions.}
If in \eqref{SM333} the pressure depends only linearly on temperature,
then we conclude from \eqref{CM41} that
$c_v(T, \nInv, \bm{y})$ does not depend on the volume and
$c_\alpha^\Ref$ is independent of $\nInv^\SM$.

\paragraph*{Specific entropy and specific internal energy at $T = T^\Ref$ and $\nInv = \nInv^\Ref = \nInv^\SM(T^\Ref,\bm{y})$.}
The essential ingredient of a simple mixture is %
a Boltzmann-typ logarithmic mixing contribution
in addition to the purely linear approch, viz.
\begin{subequations}
\label{SM371}
\begin{align}
	\label{SM371a}
 	M(\bm{y})\, \bar{u}(T^\Ref,\nInv^\SM,\bm{y})		&=\sum_{\alpha\in\cI} m_\alpha u_\alpha^\Ref\, y_\alpha	~,&
\\
 	M(\bm{y})\, \bar{s}(T^\Ref,\nInv^\SM,\bm{y})		&=\sum_{\alpha\in\cI} m_\alpha s_\alpha^\Ref\, y_\alpha
														- \sum_{\alpha\in\cI} \kB \ln(y_\alpha)\, y_\alpha 		~, &
	\label{SM371b}
\end{align}
\end{subequations}
where the constants
$u_\alpha^\Ref = u_\alpha(T^\Ref, \nInv^\SM)$ and
$s_\alpha^\Ref = s_\alpha(T^\Ref, \nInv^\SM)$, for $\alpha \in\cI$,
are the energy constants
and the entropy constants for the pure constituents, respectively.\footnote{Same as in footnote \ref{SM353} applies here.}

\paragraph*{Free energy density and chemical potentials of a simple mixture.}
Application of the above the simple mixture assumptions \eqref{SM331}-\eqref{SM371}
to the representations \eqref{SM213} and \eqref{SM233} with $\nInv^\Ref = \nInv^\SM(T^\Ref,\bm{y})$
yields the free energy density.
We rewrite the obtained result in terms of the variable ($T, (n_\alpha)_{\alpha\in\cI} )$
because this is the set of variables to calculate the chemical potentials.
We have
\begin{align}
	\rho\hat{\psi}^\SM
	&=\quad	\sum_{\alpha\in\cI} n_\alpha\ 	m_\alpha 	\psi_\alpha^\Ref
	\quad+	\sum_{\alpha\in\cI} n_\alpha\ 	m_\alpha 	c_\alpha^\Ref \cdot \bigg( T -T^\Ref -T\ln\Big( \tfrac{T}{T^\Ref}\Big) \bigg)
\notag
\\*&
\label{SM522}
	+\kB T	\sum_{\alpha\in\cI} n_\alpha	\ln\left( \tfrac{\,n_\alpha}{n} \right)
	\ -	\sum_{\alpha\in\cI} n_\alpha\ 	m_\alpha 	s_\alpha^\Ref \cdot (T-T^\Ref)
\\*&
\notag
	+	(K -p^\Ref) \Big( 1 -	\sum_{\alpha\in\cI} n_\alpha\,	\vau_\alpha^\Ref\Thermal{\alpha}(T)	\Big)
	+	K						\sum_{\alpha\in\cI}	n_\alpha\,	\vau_\alpha^\Ref\Thermal{\alpha}(T) \cdot
		\ln\Big( 				\sum_{\alpha\in\cI} n_\alpha 	\vau_\alpha^\Ref\Thermal{\alpha}(T)	\Big)
\end{align}
where we set the reference energies $\psi_\alpha^\Ref$
\begin{align}
\label{SM524}
	\psi_\alpha^\Ref	&= u_\alpha^\Ref -T^\Ref s_\alpha^\Ref ~. &
\end{align}
With the definition of the reference Gibbs energies $g_\alpha^\Ref	= m_\alpha\, \psi_\alpha^\Ref	+\vau_\alpha^\Ref\,p^\Ref$, the chemical potentials then are
\begin{align}
	\hat{\mu}_\alpha^\SM
	&=
		g_\alpha^\Ref +p^\Ref \vau_\alpha^\Ref \cdot (\Thermal{\alpha}(T) -1)
	+	m_\alpha	c_\alpha^\Ref \cdot \bigg( T -T^\Ref -T\ln\Big( \tfrac{T}{T^\Ref}\Big) \bigg)
\notag
\\*&
\label{SM528}
	+	\kB T \ln\left( \tfrac{n_\alpha}{n} \right)	-	m_\alpha 	s_\alpha^\Ref \cdot (T-T^\Ref)
\\*&
\notag
	+	K\,	\vau_\alpha^\Ref\Thermal{\alpha}(T)	\cdot \ln\Bigg( \sum_{\gamma\in\cI} n_\gamma\,\vau_\gamma^\Ref\Thermal{\alpha}(T) \Bigg)
	~.
\end{align}

\subsection{Electric field contribution of a single ion}	\label{sect:phi}
We aim at characterizing the energetic difference caused by the fact that
the electrolyte is not a neutral medium but rather  a mixture of charged particles.
Therefore, we consider a bulk electrolyte outside of boundary layers,
such that it is locally electroneutral on the macroscopic scale.
As the main material property of the electrolyte we only consider its dielectric susceptibility
which on the macroscopic scale may depend on the local temperature and composition, i.e. the number densities $n_\gamma$.
Then, we turn to a more microscopic scale, where we take the number densities $n_\gamma$ from the more macroscopic scale
to now define the far field $n_\gamma^\infty = n_\gamma$ %
and  assume that the temperature $T$, the susceptibility $\chi$
and its temperature derivative $\tfrac{\partial}{\partial T} \chi$
are homogeneous in space,
depending only on $T$ and the far field number densities $n_\gamma^\infty$. %

We select an arbitrary solvated ion from any of the species $\gamma\in\cI$ and further
refer it by using the subscript $*$. %
For this selected ion carrying the total charge $z_*e_0$,
we choose the reference frame such that this ion is fixed at the origin.
Except while undergoing the dissociation reaction,
no other charged particles are allowed to enter into
a sphere of radius $R_*>0$ containing the selected bare center ion together with its solvation shell.
The radius $R_*$ may depend on the temperature.
As it will turn out, it is not necessary to postulate any assumptions on the
the spatial charge distribution %
inside of the solvated ion, i.e. for $r<R_*$.

The electric potential $\phi$ due to the single selected ion in an unbounded space is in general determined by
the Poisson boundary value problem
\begin{subequations} \label{Deb:05}
\begin{align}
\label{Deb:05a}
	&&
	-(1+\chi) \eps_0 \Delta \phi &= n^\F&
	& \text{for } r>0~, &
\\
\label{Deb:05b}
	\jump{\phi} &= 0 ~,&
	\jump{(1+\chi) \eps_0\,\tfrac{\partial}{\partial r} \phi} &= 0 &
	& \text{at } r = R_*~, &
\\
	&&
	\phi &\to 0 &
	& \text{for } r \to \infty~, &
\end{align}
\end{subequations}
where $n^\F$ denotes the free charge density.
In order to account for the character of electrolytes as mixture of a charged particles,
we have to find the difference of the electric energy between
i) ions in a electrically neutral medium, i.e.\ $n^\F=0$,  and
ii) ions in locally charged medium.

\paragraph*{Ion in neutral medium.}
Assume first that the selected ion is surrounded
by an electrically neutral medium with susceptibility $\chi$,
i.e.\ $n^\F = 0$ for all $R_*<r$.
Then, the net charge $z_* e_0$ inside the inner domain $r<R_*$ causes
in the outer domain the electric potential
\begin{align}\label{Deb:21}
	\phi_0(r) &= \frac{z_*e_0}{4\pi\,\eps_0} \,
			\frac{1}{1+\chi} \, \frac{1}{r} &
			\text{for } R_*<r~.
\end{align}

\paragraph*{Ion in electrolyte.}
In a continuum description of electrolytes,
a net charge $z_*e_0$ inside a sphere of radius $R_*$
causes the formation of an oppositely charged electrolyte layer around this sphere.
It is well known
that electrolyte boundary layers at strongly charged surfaces may become highly concentrated
and steric effects can not be neglected in order to limit charge accumulation in the layer.
In the present problem however,
the total charge of the whole boundary layer is necessarily limited
to just the opposite charge of the single selected ion,
which makes volume exclusion effects insignificant here.
Thus, steric Poisson-Boltzmann models like e.g.\ \cite{DGM13},
will not do not offer a decisive advantage over
the classical Poisson-Boltzmann approach.\footnote{%
	The  negligibly impact of steric effects on the electric field around the selected ion
	is illustrated by numerical solution of the problem \eqref{Deb:05}in the supporting material.}
To derive an analytic expressions for the energy,
we linearize the Poisson equation \eqref{Deb:05a} around the far field state.
We note that before already the susceptibility $\chi$ and $\tfrac{\partial}{\partial T} \chi$
have been frozen to their far field values.
For given reference temperature $T^\Ref$ and reference number density $n^\Ref$,
we introduced the Debye screening length $\LMicro$ as a reference length
and define $\Lambda$ by
\begin{align}\label{Deb:06}
	\LMicro				&= \sqrt{ \frac{\eps_0\,\kB T^\Ref}{e_0^2\,n^\Ref} } ~, &
	(\Lambda\LMicro)^2	&= \frac{T^\Ref}{T}\frac{1}{1+\chi} \sum_{\gamma\in\cI} z_\gamma^2 \frac{n_\gamma^\infty}{n^\Ref}~. &
\end{align}
We remark that while $\LMicro$ %
depends on the chosen reference state, the inverse length $\Lambda$ does not.
Next, we approximate
\begin{align} \label{Deb:07}
	n^\F &=		e_0 \sum z_\gamma n_\gamma 	\exp\big( -z_\gamma \tfrac{e_0}{\kB T} \phi \big)
		\quad
		\approx	-(1+\chi)\eps_0 \, \Lambda^2 \, \phi
	& \text{ for } R_*<r ~.
\end{align}
The solution of the linearized Poisson problem \eqref{Deb:05} and \eqref{Deb:07} is
\begin{align}\label{Deb:11}
	\phi(r) &= \frac{z_*e_0}{4\pi\,\eps_0} \, \frac{1}{1+\chi} \,
			 \frac{1}{1+\Lambda R_*} \, \frac{\exp\left( -\Lambda (r -R_*) \right)}{r} \quad
			\text{ for } R_*<r~.
\end{align}

\paragraph{Interaction energy.} %
Our basic assumption is that
the general thermodynamic relations \eqref{TC123}-\eqref{TC125} from Sect.~\ref{sect:thermodyn}
are applicable on the scale of single ion problem.
Thus, the densities of internal energy and entropy are
$(\hat{\chi} + T	\tfrac{\partial}{\partial T} \hat{\chi})\tfrac{\eps_0}{2}|\nabla\phi|^2$ and
$					\tfrac{\partial}{\partial T} \hat{\chi}	\tfrac{\eps_0}{2}|\nabla\phi|^2$, respectively.
Internal energy and entropy are additive quantities,
%\todo{Check, Ref}
such that the $r$-dependence of the electric field may simply be smoothed out
by taking the volume integral over both densities.
Finally, we define internal energy and entropy due to ion-ion interaction
as the difference between the two above scenarios. %
Hence, the interaction of an individual ion with its surrounding
leads to the following contributions to energy and entropy
\begin{subequations}
\label{Deb:103}
\begin{align}
	U_*^W	&=	\tfrac{\eps_0}{2}
			\int_{\mathbb{R}^3} \big( \chi +T\,\tfrac{\partial}{\partial T}\chi \big) (|\nabla \phi|^2 -|\nabla\phi_0|^2)\,dx~, &
\\
	S_*^W	&=	\tfrac{\eps_0}{2}
			\int_{\mathbb{R}^3} 				\tfrac{\partial}{\partial T}\chi 		(|\nabla \phi|^2 -|\nabla\phi_0|^2)\,dx~. &
\end{align}
\end{subequations}
By means of partial integration this can be reformulated in the way
\begin{align}
\label{Deb:1032}
	U_*^W&=	\underbrace{2\pi \int_{r<R_*} \frac{ \chi +T\,\tfrac{\partial}{\partial T}\chi}{1+\chi} (\phi -\phi_0)\ n^F\,r^2\,dr}_{= {U_*^W}'} \
		+	\underbrace{2\pi \int_{r>R_*} \frac{ \chi +T\,\tfrac{\partial}{\partial T}\chi}{1+\chi} \phi \ n^F\,r^2\,dr}_{= {U_*^W}''} &
\end{align}
By straightforward calculation, cf.\ the supplementary material, we get
\begin{subequations}
\label{Deb:105}
\begin{align}
	{U_*^W}'	&=	-
				z_*^2 \frac{e_0^2}{8\pi\eps_0}	\frac{ \chi +T\,\tfrac{\partial}{\partial T}\chi }{(1+\chi)^2}\	\Lambda\, \frac{1}{1+\Lambda R_*} ~,	 &
\\
	{U_*^W}''	&=-
				z_*^2\frac{e_0^2}{8\pi\eps_0}	\frac{ \chi +T\,\tfrac{\partial}{\partial T}\chi }{(1+\chi)^2}\	\frac{\Lambda}{2}\, 	\frac{1}{(1+\Lambda R_*)^2} ~.
			&
\end{align}
\end{subequations}
We define for abbreviation
\begin{align} %
	a		&:= %
				\frac{e_0^2}{8\pi\eps_0} \frac{n^\Ref}{\LMicro} ~,			&
	F_*(R_*,\Lambda\LMicro)
	&:= \Lambda\LMicro \frac{3 +2\Lambda R_*				}{(1+\Lambda R_*)^2}	~.
	\label{Deb:107}
\end{align}

In total, we get from the interaction of the selected ion with its surrounding the contributions
to internal energy and entropy
\begin{subequations}
\label{Deb:109}
\begin{align}
	U_*^W
		&=	-\frac{a}{2} \frac{z_*^2}{n^\Ref}	\frac{ \chi +T\,\tfrac{\partial}{\partial T}\chi }{(1+\chi)^2}\
			F_*(R_*,\Lambda\LMicro) ~,
\\
	S_*^W
		&=	-\frac{a}{2} \frac{z_*^2}{n^\Ref}	\frac{ \tfrac{\partial}{\partial T}\chi }{(1+\chi)^2}\
			F_*(R_*,\Lambda\LMicro) ~.
\end{align}
\end{subequations}

\subsection{Free energy contribution due to electric interaction}	\label{sect:DH}
The above construction can be repeated for every ion in the mixture.
In the volume $V$ occupied by the electrolyte,
let there be $N_\gamma / V = n_\gamma$ ions of species $\gamma\in\cI$
and set $n_\gamma^\infty = n_\gamma$.
Recalling \eqref{Deb:06}, we introduce the dimensionless quantities
\begin{align} \label{Deb:06a}
	\tau	&=	\frac{T}{T^\Ref} ~, &
	\lambda &=	\Lambda\LMicro 	= \sqrt{ \frac{1}{(1+\chi)\ \tau} \sum_{\gamma\in\cI} z_\gamma^2 \frac{n_\gamma}{n^\Ref}}~. &
\end{align}
and take \eqref{Deb:107} with $R_\gamma = R_\gamma(\tau)$ and $\lambda=\lambda(\tau,(n_\gamma)_{\gamma})$
to define for $\gamma\in\cI$
\begin{align} \label{Deb:112c}
		F_*(R_\gamma,\lambda)
		&= \lambda \, \frac{3 +2\lambda R_\gamma/\LMicro}{(1 +\lambda R_\gamma/\LMicro)^2} ~.
\end{align}
To obtain internal energy and entropy due to the electrical interaction between the ions,
we must sum up the contributions of all constituents $\gamma\in\cI$, viz.\
\begin{align}\label{Deb:111}
	U^\Deb &= \sum {N_\gamma}{} U_\gamma^W ~, &
	S^\Deb &= \sum {N_\gamma}{} S_\gamma^W ~, &
\end{align}
where $U_\gamma^W$ and $S_\gamma^W$ are given by \eqref{Deb:109}.
Dividing by the volume of the electrolyte mixture and setting
$U^W /V = \rho \hat{u}^\Deb$,
$S^W /V = \rho \hat{s}^\Deb$
yields contributions to the densities of internal energy and entropy, viz.
\begin{subequations}
\label{Deb:112}
\begin{align}
\label{Deb:112a}
	\rho \hat{u}^\Deb
		&=	- \frac{a}{2}\,
			\frac{\chi +T\,\tfrac{\partial}{\partial T}\chi}{(1+\chi)^2}\
			\sum_{\gamma\in\cI} z_\gamma^2 \frac{n_\gamma}{n^\Ref}\,		F_*(R_\gamma,\lambda)
		~,
\\
	\rho \hat{s}^\Deb
		&=	-
			\frac{a}{2}\,
			\frac{\tfrac{\partial}{\partial T} \chi}{(1+\chi)^2}\
			\sum_{\gamma\in\cI} z_\gamma^2 \frac{n_\gamma}{n^\Ref}\,		F_*(R_\gamma,\lambda) ~.
\label{Deb:112b}
\end{align}
\end{subequations}
Finally, applying $\rho\hat{\psi}^\Deb = \rho\hat{u}^\Deb -T\rho\hat{s}^\Deb$,
the resulting free energy density is
\begin{align}\label{Deb:113}
	\rho\hat{\psi}^\Deb
		&=	-\frac{a}{2}
			\frac{\chi}{(1+\chi)^2}\
			\sum_{\gamma\in\cI} z_\gamma^2 \frac{n_\gamma}{n^\Ref}		F(R_\gamma,\lambda)
			\quad \text{ with } \ F(R_\gamma,\lambda) = F_*(R_\gamma,\lambda)~.
\end{align}
The main reason to distinguish $F(R_\gamma,\lambda)$ from $F_*(R_\gamma,\lambda)$
is for later comparison with the standard derivation procedure of a Debye-H\"uckel energy in the way of \cite{DH23}.

\paragraph*{Temperature dependence of the susceptibility.}
We note that unlike the simple mixture model above,
neither the internal energy density $\rho\hat{u}^\Deb$
nor the  entropy density $\rho\hat{s}^\Deb$ in \eqref{Deb:112} %
are calculated by means of general representations like \eqref{SM213} and \eqref{SM233}.
Thus, in order to guarantee thermodynamic consistency,
the constitutive functions \eqref{Deb:112} and \eqref{Deb:113} need to satisfy
the thermodynamic conditions \eqref{TC125}
which must now be exploited for vanishing macroscopic electric field $\bm{E} \approx 0$.
Due to the above construction \eqref{Deb:113} for $\rho\hat{\psi}^\Deb$,
the validity of \eqref{TC125a} directly implies \eqref{TC125b}.
Thus it remains to check \eqref{TC125a}.
Applying \eqref{Deb:112b} and \eqref{Deb:113} to
$\rho\hat{s}^\Deb	= -	\tfrac{\partial}{\partial T} \rho\hat{\psi}^\Deb$,
we get
\begin{align}\label{Deb:121}
			\tfrac{\partial}{\partial T} \left( \frac{ 1}{1+\chi} \right)\
			\sum_{\gamma\in\cI} z_\gamma^2 \frac{n_\gamma}{n^\Ref}		F(R_\gamma,\lambda)
	&=
			\tfrac{\partial}{\partial T} \left(
			\left( \frac{1}{1+\chi} -\frac{1}{(1+\chi)^2} \right)\
			\sum_{\gamma\in\cI} z_\gamma^2 \frac{n_\gamma}{n^\Ref}		F(R_\gamma,\lambda)
			\right)
	 ~,
\end{align}
what we can rearrange as
\begin{align}\label{Deb:125}
	0	&=	2\frac{1+\chi}{\chi}\frac{\,\tfrac{\partial}{\partial T} \chi }{(1+\chi)^2} \
			\sum_{\gamma\in\cI} z_\gamma^2 \frac{n_\gamma}{n^\Ref}		F(R_\gamma,\lambda)
	+
			\tfrac{\partial}{\partial T} \left(
			\sum_{\gamma\in\cI} z_\gamma^2 \frac{n_\gamma}{n^\Ref}		F(R_\gamma,\lambda)
			\right)
	~.
\end{align}
Using the identity
$\frac{\,\tfrac{\partial}{\partial T} \chi }{(1+\chi)\,\chi}
=\tfrac{\partial}{\partial T} \ln\left( \frac{\chi}{1+\chi} \right)$,
we infer
\begin{align}\label{Deb:127}
	0	&=	\tfrac{\partial}{\partial T} \ln \left(
				\left( \frac{\chi}{1+\chi} \right)^2 \
				\sum_{\gamma\in\cI} z_\gamma^2 \frac{n_\gamma}{n^\Ref}		F(R_\gamma,\lambda)
				\right) ~.
\end{align}
We conclude that the argument of the logarithm has to be independent of temperature.
Consequently, the temperature dependence of the susceptibility function $\chi(T, (n_\alpha)_\alpha )$
is determined by the temperature dependence of %
the inverse length $\lambda(T, (n_\alpha)_\alpha )$. %

\paragraph*{Contributions to chemical potentials and pressure.}
We abbreviate
\begin{align}
\label{Deb:135}
	\cL_\chi&= \sum_{\alpha\in\cI} n_\alpha \tfrac{\partial}{\partial n_\alpha}\chi~, &
	D 		&= 1+\chi~, &
\\
	\cF	&= \sum_{\gamma\in\cI}	z_\gamma^2 \frac{n_\gamma}{n^\Ref}	F(R_\gamma,\lambda)~,&
	d\cF	&= \sum_{\gamma\in\cI}	z_\gamma^2 \frac{n_\gamma}{n^\Ref}	%
				 \tfrac{\partial}{\partial\lambda}F(R_\gamma,\lambda)~. &
\label{Deb:136}
\end{align}
We note that whenever the above operator $\cL$ acts on
a homogeneous function $h$ of degree one with respect to the number densities,
then $\cL_h = h$.
We check the identities
\begin{subequations}
\label{Deb:133}
\begin{align}
	\tfrac{\partial}{\partial \chi}	\frac{\chi}{D^2}	&= \frac{1-\chi}{D^3} ~, &
	\tfrac{\partial}{\partial n_\alpha} \lambda
	&= \frac{1}{2} \left( \frac{1}{\tau D\lambda}\frac{z_\alpha^2}{n^\Ref} -\frac{\lambda}{D}\tfrac{\partial}{\partial n_\alpha} \chi \right) ~. &
\end{align}
\end{subequations}
Then, the contributions to the free energy the chemical potentials and pressure, respectively, are
\begin{subequations}
\begin{align}
	\rho\hat{\psi}^\Deb &= -\frac{a}{2} \frac{\chi}{D^2}\,	\cF ~,
\\
\label{Deb:137}
	\hat{\mu}_\alpha^\Deb
		&=	-\frac{a}{2} \frac{\chi}{D^2}\ \Bigg[
			\frac{z_\alpha^2}{n^\Ref}	\left(
				F(R_\alpha,\lambda) + \frac{1}{\tau D\lambda^2}\frac{\lambda}{2} d\cF
				\right)
			- \left(
				\frac{\chi-1}{\chi}		\cF
				+\frac{\lambda}{2}	d\cF	\right)	\frac{\tfrac{\partial}{\partial n_\alpha} \chi}{D}
			\Bigg] ~,
\\%
\label{Deb:138}
	\hat{p}^\Deb
		&=
		-\frac{a}{2}\frac{\chi}{D^2}
			\Bigg( \frac{\lambda}{2} d\cF
				-\left(	\frac{\chi-1}{\chi}	\cF +\frac{\lambda}{2} d\cF	\right)	\frac{\cL_\chi}{D}
			\Bigg)
	~.
\end{align}
\end{subequations}

\subsection{Summary and a first assessment of the material model} \label{sect:const-discuss}
Within boundary layers,
we propose a constitutive model based on the simple mixture model and
an explicit energy contribution due to the electric field, viz.
\begin{align}\label{Deb:390}
\text{(BL)}
\left\{
\begin{aligned}
	\rho\psi	&=	\rho\psi^\SM( T, (n_\alpha)_{\alpha\in\cI} ) \rule{1ex}{0ex}
				-\chi \tfrac{\eps_0}{2}|\bm{E}|^2 ~,
\\
	\mu_\alpha	&=	\rule{1ex}{0ex}\mu_\alpha^\SM( T, (n_\alpha)_{\alpha\in\cI} )
				-\tfrac{\partial }{\partial n_\alpha} \chi \, \tfrac{\eps_0}{2}|\bm{E}|^2 ~,
\\
p	&=	p^\SM \quad
		=	p^\Ref +K\cdot	\bigg( \sum_{\alpha\in\cI} \vau_\alpha^\Ref\Thermal{\alpha}(T)\, n_\alpha  -1 \bigg) ~, &
\end{aligned}
\right.
\end{align}
with $\rho\psi^\SM$ according to \eqref{SM522}.
To this model, a (quasi-)incompressible limit was applied in \cite{DGM13,LGD16,DGM18,LM22}
and analyzed in detail in \cite{BoDrDr23}.
For $p^\Ref\ll K$ a  constraint $\sum_{\alpha\in\cI} \vau_\alpha^\Ref\Thermal{\alpha}(T)\, n_\alpha  =1$ arises,
requiring $p$ to take the role of an independent variable.

In homogeneous bulk domains, where on the macroscopic scale the electric field vanishes, i.e.\ $\bm{E}\approx 0$,
we propose
\begin{align}\label{Deb:320}
\text{(HB)}
\left\{
\begin{aligned}
	\rho\psi	&=	\rho\psi^\SM(	T, (n_\alpha)_{\alpha\in\cI} )
				+	\rho\psi^\Deb(	T, (n_\alpha)_{\alpha\in\cI} )	~,
\\
	\mu_\alpha	&=	\rule{1ex}{0ex}\mu_\alpha^\SM(	T, (n_\alpha)_{\alpha\in\cI} ) \
				+	\rule{1ex}{0ex}\mu_\alpha^\Deb(	T, (n_\alpha)_{\alpha\in\cI} )	~,
\\
	p	&%
		=	p^\Ref +K\cdot	\bigg( \sum_{\alpha\in\cI} \vau_\alpha^\Ref\Thermal{\alpha}(T)\, n_\alpha  -1 \bigg)
		+	p^\Deb(	T, (n_\alpha)_{\alpha\in\cI} ) ~, &
\end{aligned}
\right.
\end{align}
with $\rho\psi^\SM$ according to \eqref{SM522}
and $\rho\psi^\Deb$ according to \eqref{Deb:113}.
Since both contributions $p^\SM$ and $p^\Deb$ in general need to balance each other,
the consideration of a (quasi-)incompressible limit is not suitable in this context.

We choose as reference values room temperature a number density $n^\Ref$ corresponding to a solution of $1\Umol \Ul^{-1}$.
This implies for the Debye screening length
\begin{align} \label{Deb:303}
	T^\Ref	&= 298.15\UK~, &
	n^\Ref	&= 1000\, \Avo ~ &
	\Longrightarrow \quad
	\LMicro &\approx	0.04856\Unm ~.  &
\end{align}
{With the specific volume of water at $T=T^\Ref$  we conclude}
\begin{align}
	\label{Deb:307}
	\vau_S^\Ref &= (55.345\,n^\Ref)^{-1}  &
	\Longrightarrow \quad  %
	R_S(T^\Ref) &= \sqrt[3]{\frac{3}{4\pi} \vau_S^\Ref} \approx	0.1928 \Unm~ 		\approx 4 \LMicro ~. &	%
\end{align}

\paragraph*{Pressure contribution at $\tau=1$.}
Let the total pressure $p$ in the electrolyte be fixed at the outer reference pressure, i.e.\
$p\overset{!}{=}p^\Ref$.
For the elastic bulk modulus of water
$K \approx 2.2\times10^{+9}\UJ\Um^{-3}$,
we have with the above reference values $a/K \approx 0.65$.
If we assume $\cL_\chi/\chi \approx 1$, $\chi-1\approx D$ and $D-\cL_\chi \ll D$,
we can deduce from \eqref{Deb:138} and \eqref{Deb:320} the approximate relation
\begin{align}
	\label{Deb:326}
	\frac{p^\Deb}{K}
	&=-\frac{a}{2K}\frac{\chi}{D^2}
			\left(
				\frac{\lambda}{2}\,d\cF\frac{D-\cL_\chi}{D}
				-\frac{\chi-1}{D} \cF \frac{\cL_\chi}{\chi}
				\right) &
		&\approx %
		\frac{1}{D} \frac{a}{2K} \cF	~. &
\end{align}
For $\lambda<0.15$, the function $F$ is in less than, or in the order of $0.1$, depending on $R_\alpha$.
Assuming
$\cF = \tau D \lambda^2 F(R_\alpha,\lambda)$, for any $\alpha\in\cI$,
we can conclude that
$p^\Deb/K
	\approx \frac{\lambda^2}{2} \frac{a}{K} F(R_\alpha,\lambda)
	\approx 0.001$ for $\lambda<0.15$.
Thus, we can expect the pressure contribution $p^\Deb$
to be smaller than the bulk modulus of water $K$ by a factor in the order of $1000$.
This coincides with the estimate of Debye-H\"uckel \cite[before eq.~(7)]{DH23}.
Accordingly, the absolute change of volume due to $p^\Deb$
can be expected small compared to elastic deformations.
Neglecting this pressure contribution, we conclude from \eqref{Deb:320}
\begin{align}\label{Deb:328}
	&\text{\uline{$\tau=1$, $p=p^\Ref$, \ $p^\Deb \ll K$}: } &
	\sum_{\alpha\in\cI} \vau_\alpha^\Ref\, n_\alpha &=1 -\frac{p^\Deb}{K} \approx 1 ~. &
\end{align}
In the following Sect.~\ref{sect:appl} however, it turns out that for the apparent molar volume of the solution,
the pressure contribution $p^\Deb$ can not be neglected.

\paragraph*{Temperature dependence of susceptibility.}
To study the implications of the thermodynamic compatibility condition \eqref{Deb:127}
on the temperature dependence of the susceptibility,
we choose the most simple configuration of a symmetric $-1:1$-electrolyte.
For concentration less than $1\Umol\Ul^{-1}$, we assume complete dissociation of the salt. %
Thus, we consider a mixture consisting only of the neutral solvent and solvated anions and cations,
i.e.\ $\cI=\{A,C,S\}$ with $-z_A=z_c=1$ and $z_S=0$.
We let
\begin{align}\label{Deb:340}
	\text{\uline{for complete dissociation} } &&
	c &= \frac{1}{2} %
					\frac{n_A +n_C}{1000\Avo} &
\end{align}
denote the electrolyte concentration in $\Umol \Ul^{-1}$.
Since equilibrium in an electrolyte bulk domain implies local electro-neutrality, cf.\ e.g.~\cite{DGM18},
we can conclude $n_A=n_C = c\,n^\Ref$.
For dilute solutions, there is often a decrease of the susceptibility with the electrolyte concentration observed,
cf. \cite{HaRiCo:1948,BuHeMa99}.
Thus, we have a linear relation
\begin{align}
	\text{\uline{at $\tau=1$:}}& &
	\chi &= \chi_S^\Ref -d_A \frac{n_A}{1000\Avo} -d_C \frac{n_C}{1000\Avo} \qquad
		=	\chi_S^\Ref - d \cdot c ~, &
\end{align}
where $d = d_A +d_C$ is called the \emph{dielectric decrement}.
Assuming $p^\Deb \ll K$, we use \eqref{Deb:328}
to rewrite the susceptibility as a homogeneous function of the number densities, viz.
\begin{subequations}
\begin{align}
	\text{\uline{at $\tau=1$:}}& &
	\chi
	&=	\sum_{\alpha\in\cI} \chi_\alpha^\Ref \vau_\alpha^\Ref n_\alpha ~,
\label{Deb:342}
\\
& & \text{with } \quad
	\chi_\alpha^\Ref &= \chi_S^\Ref	-\frac{d_\alpha}{\vau_\alpha^\Ref n^\Ref}	\text{ \quad for } \alpha\in\{A,C\} ~.
\end{align}
\end{subequations}
\begin{figure}
	\centering
	\includegraphics[width=.52\textwidth]{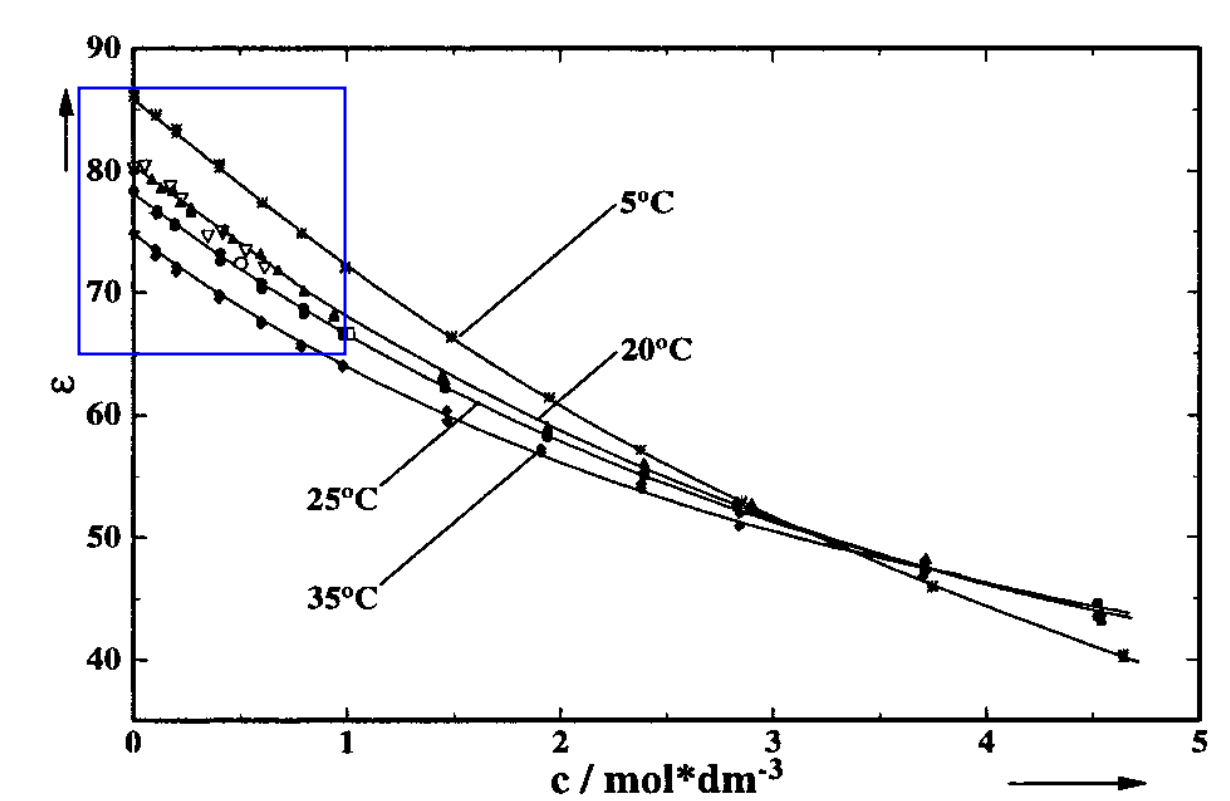}
	\hfill%
	\includegraphics[width=.45\textwidth]{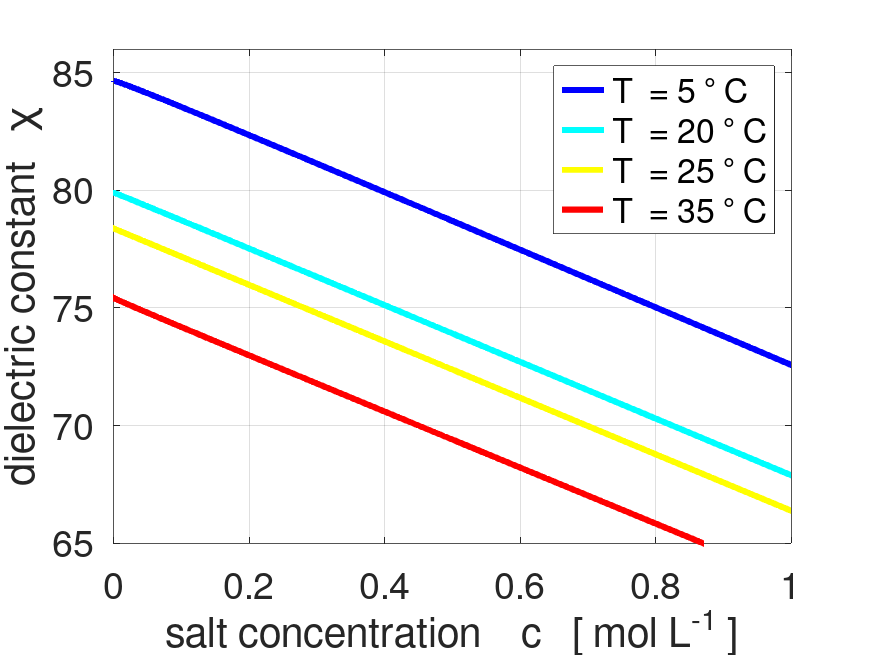}
	\caption{\label{fig:chi_temp}
	Left: Susceptibility $\chi(T,c)$ of aqueous $\ce{NaCl}$ solution,
		Figure from \cite[Fig.~1]{BuHeMa99} reproduced with permission.
		Inside the marked box, the dielectric decrement is close to linear
		for salt concentration less than $1\Umol\Ul^{-1}$.
	Right: computed susceptibility based on \eqref{Deb:3139}, where
		$\chi$ linear at $\tau=1$ %
		with  $d=12\Ul\Umol^{-1}$ and $R_A = R_C = 8\LMicro$.
}
\end{figure}

We abbreviate $D_1=1 +\chi(\tau=1,(n_\alpha)_\alpha)$ and $\lambda_1=\lambda(\tau=1,(n_\alpha)_\alpha)$.
The compatibility condition \eqref{Deb:127} then implies
that $D$ for any $\tau$ and $c$ is determined by
\begin{align}
	\label{Deb:3139}
		\left( \frac{D -1}{D}	\right)^2 \ \sum_{\gamma\in\cI}	z_\gamma^2 \frac{n_\gamma}{n^\Ref}	F(R_\gamma(\tau),\lambda)
	&=	\left( \frac{D_1-1}{D_1}\right)^2 \ \sum_{\gamma\in\cI}	z_\gamma^2 \frac{n_\gamma}{n^\Ref}	F(R_\gamma(1),\lambda_1).
\end{align}
For given $\tau$ and $c$, we set $n_S$, $D_1=1 +\chi(1,c,c,n_S)$ and $\lambda_1=\lambda(1,c,c,n_S)$ according to
\begin{subequations}	\label{Deb:344}
\begin{align}
\label{Deb:344a}
	0	%
	&=\left( (\vau_A^\Ref\Thermal{A}(\tau) +\vau_C^\Ref\Thermal{C}(\tau))\, c + \vau_S^\Ref\Thermal{S}(\tau)\, n_S \right)	%
	-1
	~,
\\*
0 &= D_1 -1 -\chi_S^\Ref\vau_S^\Ref\, n_S -\sum_{\alpha\in\{A,C\}} \chi_\alpha^\Ref \vau_\alpha^\Ref\ c\cdot n^\Ref   ~, &
\\*
0 &= \lambda_1^2 D_1	-2 c ~. &
\intertext{Then, we determine $\lambda=\lambda(\tau,c,c,n_S)$ and $D(\tau,c,c,n_S)$ by solving the system}
0 &= \lambda^2 \tau D	-2 c ~, &
\\*
	0&=	%
	\sum_{\gamma\in\{A,C\}}
	\left[
		\sqrt{\frac{D_1}{\tau D}} %
			\left( \frac{D -1}{D} \right)^2		\frac{F(R_\gamma,\lambda) }{\lambda}
		- 	\left( \frac{D_1 -1}{D_1}\right)^2	\frac{F(R_\gamma(1),\lambda_1)}{\lambda_1}
	\right]
	~.
\label{Deb:344e}
\end{align}
\end{subequations}
% \todo{$T$ dependence of $R_\alpha$ rechnen}
Fig.~\ref{fig:chi_temp} shows the resulting $\chi(\tau,c)$ at four different temperatures,
where the thermal expansion $\Thermal{S}(\tau)$ in \eqref{Deb:344a}
was taken according to the data fit \cite[eq.~(5)]{Apelblat:1999} for
the values of expansion coefficient of pure water measured there.
Literature values of $d$ at $T^\Ref=25^\circ\UC$ for many electrolyte solutions are discussed in \cite{Marcus:2013}
and it is concluded that the accuracy of the given values in the literature is rather low,
i.e. uncertainty is in the order of 1.
We chose $d=12$ for a $\ce{NaCl}$ solutions, compatible with \cite{BuHeMa99,Marcus:2013}.
We observe that for any $\tau\neq1$ the computed susceptibility remains visually indistinguishable
from straight lines. %
This result gives some justification to choose a homogeneous function $\chi(\tau=1,c)$
regardless of whether the reference temperature $T^\Ref$ was chosen more or less arbitrarily.

Although \eqref{Deb:127} and \eqref{Deb:3139} can not directly be used to determine
the thermal variation of the susceptibility $\chi_S(\tau)$ of the pure solvent,
we can safely take the limit for $c\to0$ in \eqref{Deb:344e}.
We infer
\begin{align}
	\frac{\chi_S(\tau)^2}{(\chi_S(\tau) +1)^{5/2}}
	&=
	\sqrt{\tau} \frac{(D_1 -1)^2}{D_1^{3/2}} ~ &
	\text{with } \
	D_1 = 1 + \frac{\chi_S^\Ref}{\vau_S^\Ref \cdot \Thermal{S}(\tau) }
	~,
\label{Deb:347}
\end{align}
and conclude that the thermodynamic compatibility condition \eqref{Deb:127}
does not induce any dependence of the solvent susceptibility  $\chi_S$ on the dissolved salt.
However, \eqref{Deb:127} implies that the temperature dependence of $\chi_S$
is already fully encoded in the thermal expansion of the pure solvent.

\paragraph*{Comparison to standard Debye-H\"uckel model.} %
The density $\rho u^\Deb$ of internal energy is one essential cornerstone of the original Debye-H\"uckel model \cite{DH23},
as well as of the proposed model here.
In both cases, it can be written in the form \eqref{Deb:112a},
where here the quantity $f_\alpha(R_\alpha,\Lambda\LMicro)$ defined by \eqref{Deb:112c}
may also be rewritten as
\begin{equation}
	F_*(R_\alpha,\Lambda\LMicro)
	= \Lambda\LMicro \frac{3 +2 \Lambda R_\alpha}{(1 +\Lambda R_\alpha)^2}
	= \frac{2\Lambda\LMicro}{ 1 +\Lambda R_\alpha}
	+ \frac{ \Lambda\LMicro}{(1 +\Lambda R_\alpha)^2} ~.
\end{equation}
While the first term on the right hand side
also occurs in \cite{DH23}, the second term is missing there.
The absence of the second term might be related to the conception of the selected ion as a point charge
and following evaluation of the electric potential only in the origin.

In order to derive from $\rho u^\Deb$ the free energy density $\rho\psi^\Deb$,
there are two common methods in the literature.
The first method was proposed in \cite{DH23}.
It assumes that the susceptibility is independent of temperature, i.e. $\frac{\partial}{\partial T} \chi = 0$,
such that the formula \eqref{TC125b} for $\bm{E} = 0$ can be integrated with respect to temperature.
This leads to a free energy density of the same structure as in \eqref{Deb:113},
but instead of $F_\alpha(\Lambda\LMicro) =f_\alpha(\Lambda\LMicro)$ one gets
\begin{equation}
	F(R_\alpha,\Lambda\LMicro)	= -T\,\int
	\frac{F_*(R_\alpha,\Lambda\LMicro)}{T^2} \, dT ~.
\end{equation}
Soon after the first derivation,
the importance of a temperature dependent susceptibility was noticed,
an alternative derivation was introduced \cite{Debye:1924}
based on a hypothetical reversible charging process at constant temperature.
Both approaches yield the same result
and they are therefore both generally accepted in the literature,
although some weakness of their justification is usually mentioned.

Instead of \emph{a-priori} assuming $\frac{\partial}{\partial T} \chi = 0$,
we already incorporate the temperature dependence of the susceptibility in
the inner energy density $\rho u^\Deb$ and the density of entropy $\rho s^\Deb$,
based on the general thermodynamic framework of Sect.~\ref{sect:thermodyn} providing \eqref{TC125}.
Then, we can rely on both densities
to directly derive the free energy density by the definition $\rho\psi^\Deb = \rho u^\Deb -T \rho s^\Deb$,
and thereby get \emph{a-posteriori} the thermodynamic compatibility condition \eqref{Deb:127}
restricting the generality of $\frac{\partial}{\partial T} \chi(T,(n_\alpha)_\alpha)$.

\PreprintClearpage
\section{Application to bulk domains of binary electrolytes}\label{sect:appl}

\subsection{Description of electrolytes}\label{sect:electrolyte}

We consider an electrolyte that is prepared by dissolving
some neutral salt $\ce{E}$ in a neutral solvent $\ce{S}$.
The salt dissociates into negatively charged anions and positively charged cations.
Many solvents, in particular water, have a molecular structure that consists of microscopic dipoles.
These dipoles cause a microscopic electrostatic interaction between charged ions and solvent molecules.
This interaction, which is known as solvation,
leads to the formation of larger complexes by clustering of solvent molecules around a center ion.
The solvation has a profound impact on the mixing entropy within the electrolyte model \cite{LGD16}
because solvent molecules that are bounded by a center ion
do not participate in the entropic interaction with the other constituents of the electrolytic mixture.
Therefore, we choose the solvated ions as the constituents of the mixture
and refer to the solvated anions and cations as $\ce{A}$ and $\ce{C}$, respectively.
Neglecting any other species, the index set of the electrolyte
is then given as $\cI = \{\ce{A},\ce{C},\ce{E},\ce{S}\}$.
We use material parameters to represent an aqueous solution of $\ce{Na}\ce{Cl}$ according to Table~\ref{tab:L05}.
\begin{table}
\centering
	\caption{\label{tab:L05}
	Reference values and material parameters for an aqueous $\ce{Na}\ce{Cl}$ electrolyte.}
\begin{tabular}{rlrlr}
\hline\hline
\\[-2ex]
	$T^\Ref$ & $= 25^\circ C$ &
	$n^\Ref$ & $= 1000\,\Avo$ &
\\
\hline
\\[-2ex]
	$m_S$ & $= 18.0152833 \Ug\Umol^{-1}$ &
	$z_S$ & $= 0$ &	 %
\\
	$\vau_S^\Ref$ & $= (55.345\,n^\Ref)^{-1}$ &
	$K$ & $= 2.2\times10^{+9}\UJ\Um^{-3}$ & %
\\
	$\chi_S^\Ref$ & $ = 78.38$ &	 %
\\
\hline
\\[-2ex]
	$m_E$ & $= 58.443 \Ug\Umol^{-1}$ &
	$z_E$ & $= 0$ &	 %
\\
	$z_A$ & $=-z_C = 1$ &	 %
\\
	$d$ &  $= 12 \Ul\Umol^{-1}$ &	 %
\\
\hline
\end{tabular}
\end{table}

\paragraph*{Input ratio of the mixture.}
At any time after the initial state, the actual total volume occupied by the electrolytic mixture is denoted by $V$.
By $N_\alpha$ and $M_\alpha = m_\alpha N_\alpha$,
we denote the total number of molecules and the total mass of species $\alpha\in\cI$
inside the volume $V$, respectively.
The electrolytic solution is prepared from initially $N_S^0$ solvent molecules,
occupying at the temperature $T$ the initial volume
\begin{align}
\label{Dens:202}
	V_S 	&= V_S^\Ref \, \Thermal{S}(T)  ~, &
	V_S^\Ref&= \vau_S^\Ref \, N_S^0  ~, &
\end{align}
with the thermal volume expansion function $\Thermal{S}(T)$
and the specific volume $\vau_S^\Ref$ of the solvent
at the reference temperature $T^\Ref$ as introduced in Sect.~\ref{sect:simple-mix}.
Then, $N_E^0$ molecules of the electrically neutral salt are added.
To characterize the electrolyte mixture,
we introduce the dimensionless \emph{total input ratio} $\Ratio$ and
the \emph{molality} $\Mol$ in $\Umol\Ukg^{-1}$
as
\begin{align}
\label{Dens:204}
	\Ratio	&= 					\frac{N_E^0}{N_S^0} ~, &
	\Mol	&= \frac{1}{\Avo}\,	\frac{N_E^0}{M_S^0} \ =\frac{1}{m_S \Avo}\, \Ratio ~, &
\end{align}
respectively.
A molality $\Mol = 1 \Umol\Ukg^{-1}$ of a salt $E$ in water
corresponds to the input ratio $\Ratio \approx 0.018$. %
Alternatively, it is possible to prepare a mixture with a specified
\emph{molar concentration} $c$ in $\Umol\Ul^{-1}$, with
\begin{align}
\label{Dens:206}
	c	&= \frac{N_E^0}{1000\Avo\ V}~,
\end{align}
by taking $N_E^0$ molecules of the salt
and adding as much solvent as needed until the mixture reaches the actual volume $V$.
In this case, the number $N_S^0$ of solvent molecules is a-priori unknown.

\paragraph*{Volume and mass density.}
To relate in a homogeneous system the actual volume $V$ of the mixture
to the initial volume $V_S$ of the pure solvent,
we introduce dimensionless \emph{volume factor} $f$,
and equivalently the \emph{apparent molar volume} $\Phi_V = 1000\Avo\, \vau_S^\Ref\Thermal{S}(T) \cdot 1000 f$,
with the dimensions $\Ul\Umol^{-1}$, cf.\ e,g,\ \cite{RedlichMeyer:1964,Millero:1971},
such that
\begin{subequations}
\begin{align}
\label{Dens:231}
	V	%
		&=	V_S^\Ref\Thermal{S}(T) \cdot (1 +f\,\Ratio) ~, &
\\
		&= V_S   +\frac{1}{1000}\Phi_V\,\frac{N_E^0}{1000\Avo} ~.&
\end{align}
\end{subequations}
The mass density and the molar concentration of the homogeneous mixture are
\begin{subequations}
\begin{align}
\label{Dens:233}
	\rho
	=	\frac{M_S^0 +M_E^0}{V} \qquad
	&=	\frac{1}{\vau_S^\Ref\Thermal{S}(T)}\,	\frac{m_S+ m_E\,\Ratio}{1 +f\,\Ratio} ~, &
\\
c	%
	&=	\frac{1}{\vau_S^\Ref\Thermal{S}(T)}\, 	\frac{\Ratio}{1 +f\,\Ratio}	\frac{1}{1000 \Avo} ~. &
\label{Dens:233b}
\end{align}
\end{subequations}
Conversely, we can also express the volume factor $f$ in dependence of $\rho$ as
\begin{align}
\label{Dens:235}
	f		=	\frac{m_E}{\vau_S^\Ref\Thermal{S}(T)}\,\frac{1}{\rho}
			+	\left(	\frac{m_S}{\vau_S^\Ref\Thermal{S}(T)}\,\frac{1}{\rho} -1	\right)	\frac{1}{\Ratio}
		~.&
\end{align}
\begin{figure}
	\centering
	\includegraphics[width=.45\textwidth]{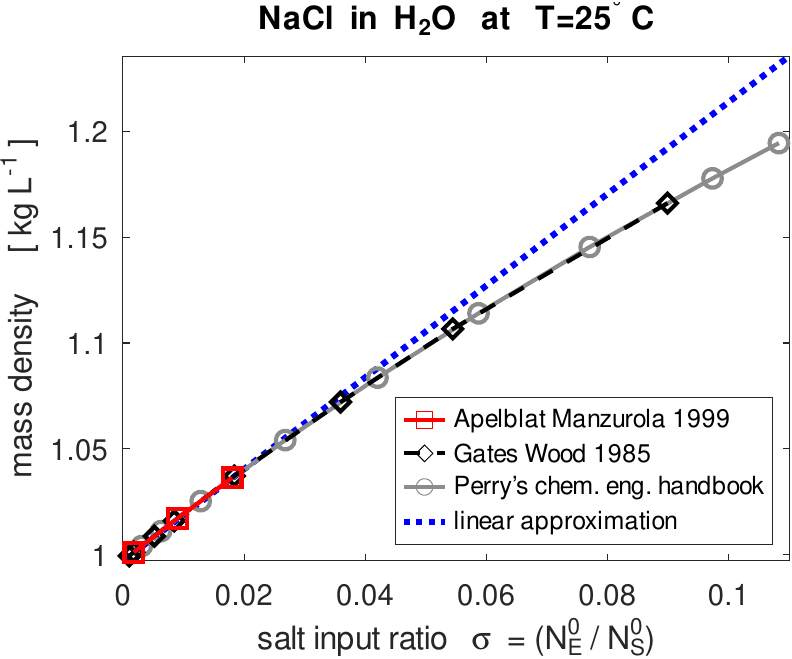}
	\hfill
	\includegraphics[width=.45\textwidth]{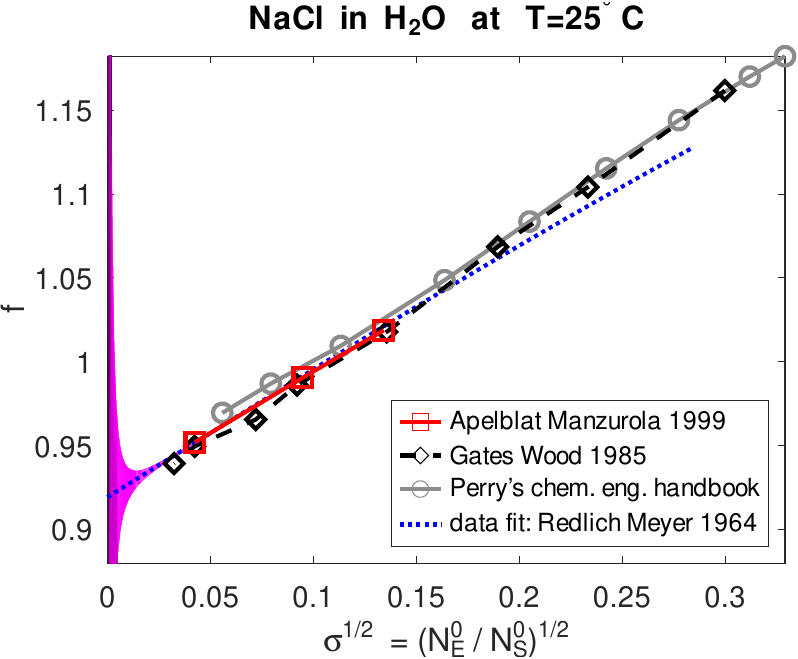}
	\caption{\label{fig:dens-ratio}
 	Left:
		Mass density of aqueous $\ce{NaCl}$ solution at $25^\circ\UC$ over salt concentration.
		from measured density $\rho$ of \cite{ChenChenMillero:1980, AllredWoolley:1981, GatesWood:1985, Apelblat:1999, Perrys}.
	Right:
		Volume change factor $f$ over salt input $\Ratio$
		calculated by \eqref{Dens:235} from measured density $\rho$
		and data fit of \cite{RedlichMeyer:1964}.
	}
\end{figure}
Thus, taking data from density measurements,
it is possible to calculate $f$ as a function of $\Ratio$
without applying any model assumption at all.
In Fig.~\ref{fig:dens-ratio},
we see that the mass density of the solution can already be approximated by a linear relation to the salt input,
at least for concentrations below $2\Umol\Ul^{-1}$. %
The deviation from the linear behavior is caused by variations of the volume factor $f$.
Experimental data shows in Fig.~\ref{fig:dens-ratio} a close to linear behavior of $f$
when plotted as a function of $\sqrt{\Ratio}$.
The values for the linear approximation are taken from \cite{RedlichMeyer:1964}).
The data of \cite{Perrys} show a positive offset
against the linear data fit of $f(\sqrt{\Ratio})$ in \cite{RedlichMeyer:1964}).
For dilute solutions,
the data of \cite{ChenChenMillero:1980, AllredWoolley:1981, GatesWood:1985, Apelblat:1999} agrees well with the linear fit.
Then around salt molality in the range of $1 \Umol\Ukg^{-1}$,
there is a transition and of the experimental data from the linear fit to the data of \cite{Perrys}.

The shaded area in Fig.~\ref{fig:dens-ratio} indicates the possible error in $f$
for input data with an uncertainty of $10^{-6}$ with respect to mass density related to the curve of \cite{RedlichMeyer:1964}).
In particular, we conclude that for strongly diluted solutions with $\Ratio\to 0$,
it is not possible to experimentally validate or falsify predicted behavior of $f$.

\paragraph*{Dissociation reaction.}
The net reaction of dissociation and subsequent solvation reaction is written as
\begin{align}
\label{Dens:212}
	\ce{E} +(\nu_A\kappa_A +\nu_C\kappa_C) \ce{S} & \ce{<=>} \nu_A \ce{A} + \nu_C \ce{C} ~,
\end{align}
where $\kappa_A$ and $\kappa_C$ are the numbers of solvent molecules in the solvation shell
of $\ce{A}$ and $\ce{C}$, respectively.
The stoichiometric coefficients of the reaction are given by
$\nu_A$ and $\nu_C$ on the right hand side of are \eqref{Dens:212} and
\begin{align}
	\nu_E &= -1~, & \nu_S &= -(\nu_A\kappa_A +\nu_C\kappa_C)  &
\end{align}
taken from the left hand side of \eqref{Dens:212}.
We assume that the reaction \eqref{Dens:212} and the electrolyte in total are electroneutral, viz.\
\begin{align}
\label{Dens:214}
	z_A   N_A + z_C   N_C  &= 0~,  &
	z_A \nu_A + z_C \nu_C  &= 0~.  &
\end{align}
The equilibrium condition for the reaction \eqref{Dens:212}
is according to \eqref{EC123d} given as a mass action law
\begin{align}
\label{Dens:218}
	\sum_{\alpha\in\cI} \nu_\alpha ( \mu_{\alpha} -g_\alpha^\Ref ) &= \Delta g &
	\text{with } \quad
	\Delta g &= -\sum_{\alpha\in\cI} \nu_\alpha\, g_{\alpha}^\Ref ~. &
\end{align}
If the Gibbs energy $\Delta g$ is in the order of $+1\Uev$,
then the electrolyte will be effectively completely dissociated
up to a certain concentration, depending on the solvation numbers $\kappa_\alpha$,
where due to a lack of remaining free solvent molecules
the solvation of further center ions is not possible any more.
For homogeneous systems,
the dissociation degree $\Diss$ is formally introduced such that
\begin{align}
\label{Dens:221}
	N_E = (1-\Diss)\,N_E^0~.
\end{align}
In general, the dissociation degree $\Diss$ is a function of $\Ratio$.
By setting %
$N_A^0 = N_C^0 =0$,  we have
\begin{align}
\label{Dens:219}
	N_\alpha	&= N_\alpha^0 +\nu_\alpha \,\Diss\,\Ratio \, N_S^0~
	\quad\text{ for } \alpha\in\cI~.
\end{align}

\subsection{Model based determination of $\Diss$ and $f$}
We define the volume change due to the dissociation reaction as
\begin{align}
	\label{Dens:251}
	\Delta\vau(T)	&:= \sum_{\alpha\in\cI}	\nu_\alpha\	\vau_\alpha^\Ref\Thermal{\alpha}(T)	  ~,	&
\end{align}
and introduce the simple mixture reference volume and the according volume factor as
\begin{align}
	\label{Dens:252}
	V^\SM			&:= \sum_{\alpha\in\cI} N_\alpha \	\vau_\alpha^\Ref\Thermal{\alpha}(T)  ~,		&
	f^\SM 			&:= \tfrac{\vau_E^\Ref\beta_E^T(T)\, +\Delta\vau(T)\, \Diss}{\vau_S^\Ref\Thermal{S}(T)} ~,	&
\end{align}
respectively,
such that by applying \eqref{Dens:219}
\begin{align}
	V^\SM
	&= V_S^\Ref\Thermal{S} \cdot \Big(1 +f^\SM\,\Ratio \Big) ~, &
\label{Dens:253}
\end{align}
The thermal equation of state \eqref{SM331}
can then be rewritten as
\begin{align}\label{Dens:255}
	p^\SM
	&=	p^\Ref
		+K\cdot	\bigg( \dfrac{1 +f^\SM\,\Ratio}{1 +f\ \Ratio} -1 \bigg) ~,
\end{align}

In equilibrium, the dissociation reaction satisfies the mass action law \eqref{Dens:218},
and the pressure in the bulk equals the external pressure, see, e.g.\ \cite{DGM13}.
For a given external pressure $p^\Ref$,
the volume of the mixture is determined from
\begin{subequations}
\label{Dens:257}
\begin{align}
\label{Dens:257a}
	p^\Ref &= p ~,
\\
	\Delta g	&=	\sum_{\alpha\in\cI} \nu_\alpha ( \mu_{\alpha}^\SM -g_\alpha^\Ref )
				+	\sum_{\alpha\in\cI} \nu_\alpha  \mu_{\alpha}^\Deb  &
\label{Dens:257b}
\end{align}
\end{subequations}

\begin{figure}[tb!]
	\centering
	\includegraphics[width=.45\textwidth]{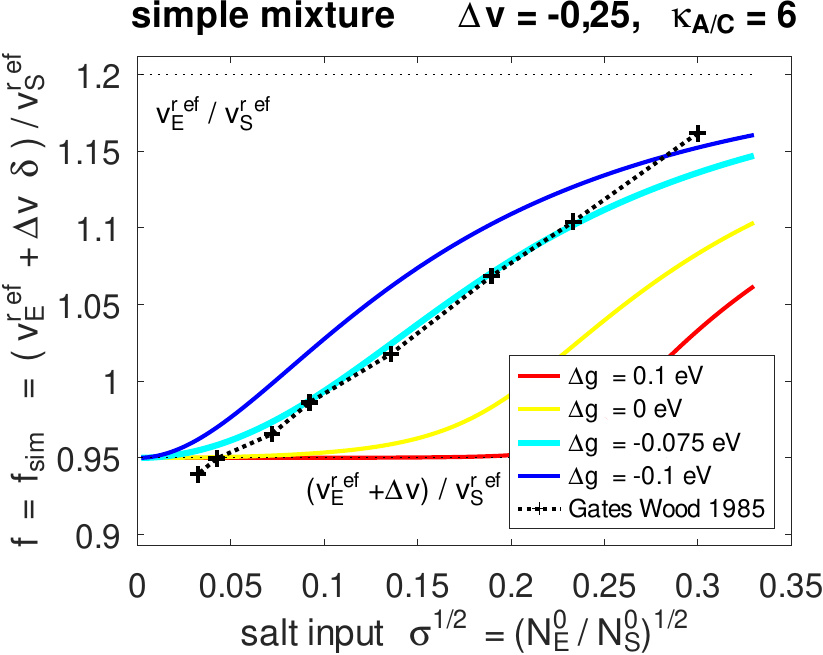}
	\hfill
	\includegraphics[width=.45\textwidth]{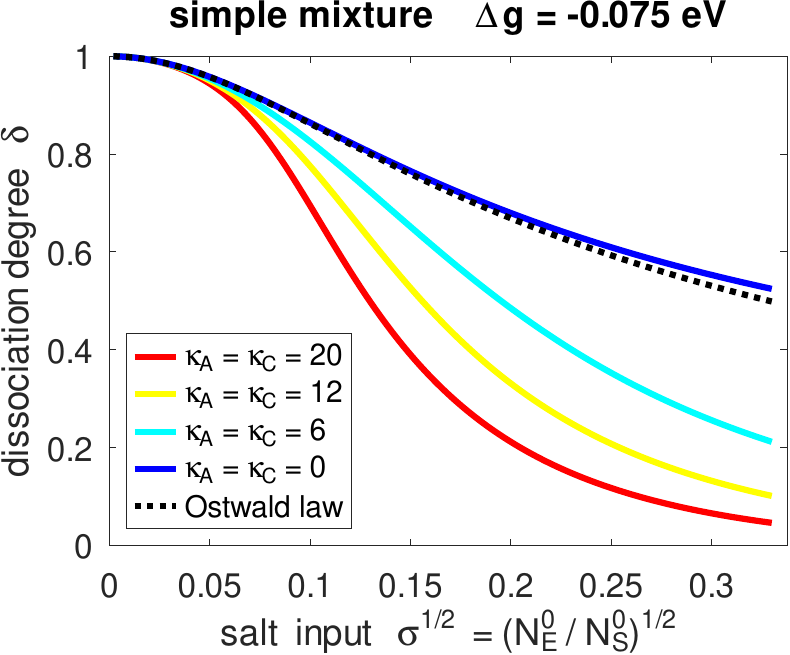}
	\caption{\label{fig:sm}
		Simple mixture model:
		Dissociation degree $\Diss$ decreases more rapid over the salt input $\Ratio$
		for larger %
		$\nu_S =\kappa_A+\kappa_C$ (left).
		Right: Volume change $f$ over the salt input $\Ratio$ for %
		different $\Delta{}g$ compared with experimental data of \cite{GatesWood:1985}.
		}
\end{figure}

\paragraph*{Simple mixture of solvated ions at $T=T^\Ref$.}
In a first step, we neglect the Debye-H\"uckel-type energy contributions and only consider the
isothermal simple mixture model,
i.e.\ the chemical potentials \eqref{SM528} reduce to
\begin{align}
\label{Dens-SM:202}
	\mu_\alpha = \mu_\alpha^\SM
	&=	g_\alpha^\Ref
	+	\kB T^\Ref \ln\left( \tfrac{\,n_\alpha}{n} \right)
	+	\vau_\alpha^\Ref\,K\cdot \ln\Bigg( \sum_{\gamma\in\cI} \vau_\gamma^\Ref n_\gamma \Bigg)
\end{align}
and pressure satisfies $p=p^\SM$.
Then,  condition \eqref{Dens:257a} implies $p^\SM=p^\Ref$, i.e.\ $f=f^\SM$ due to \eqref{Dens:255},
what in turn causes the mechanic last contribution in \eqref{Dens-SM:202}
to vanish independently of $K$. %
Thus, the mechanic contribution to the law of mass action vanishes completely in \eqref{Dens:257b}
and $f$ and $\Diss$ can now be determined from
\begin{subequations}
\begin{align}
\label{Dens-SM:204}
	f	&= %
		\tfrac{\vau_E^\Ref}{\vau_S^\Ref} +\tfrac{\Delta\vau}{\vau_S^\Ref}\ \Diss
\\
 \label{Dens-SM:206}
	\frac{\Delta g}{\kB T^\Ref}
	&=
	\sum_{\alpha\in\cI} \nu_\alpha \ln\bigg( \frac{N_\alpha}{N} \bigg)~. &
\end{align}
\end{subequations}
Remarkably, \eqref{Dens-SM:206} does not depend on the actual volume,
or equivalently on $f$,
but only on the dissociation degree $\Diss$.
However, $\Diss$ depends on $\nu_S$ and thus on the solvation numbers $\kappa_{A/C}$ of the ions.
Moreover, $\kappa_{A/C}$ also have an influence on the specific volume $\vau_{A/C}^\Ref$ of the solvated ions.
If we particularly cveonsider
$\nu_A = \nu_C = 1$, %
then \eqref{Dens-SM:206} yields
\begin{align}
	\frac{\Delta g}{\kB T^\Ref}
	&=	\ln\bigg( \frac{N_{A} N_{C}}{N_{E}\ N} 		\bigg( \frac{N_{S}}{N} \bigg)^{\nu_S} \bigg)
	=	\ln\bigg( \frac{N_{A} N_{C}}{N_{E}\ N_S^0} \cdot \frac{N_S^0}{N} \cdot \bigg( \frac{N_{S}}{N} \bigg)^{\nu_S} \bigg)
\intertext{such that we conclude}
\label{Dens-SM:208}
	\exp\left( \frac{\Delta g}{\kB T^\Ref}  \right)
	&= \frac{ \Diss^2}{ 1-\Diss}\, \Ratio \, \frac{1}{1 +(1 +\sum_{\alpha\in\cI} \nu_\alpha\Diss)\,\Ratio}
	\Big( \frac{1 +(1 +\sum_{\alpha\in\cI} \nu_\alpha\Diss)\,\Ratio}{1+\nu_S\Diss\Ratio} \Big)^{-\nu_S} ~.
\end{align}
Without taking the solvation into account, %
i.e.\ $\nu_S = \kappa_A +\kappa_C =0$,
the last term of \eqref{Dens-SM:208} disappears.
If we also neglect the second last factor in \eqref{Dens-SM:208}, which approaches unity for small $\Ratio$,
we recover the well known Ostwald's law.
The deviation from Ostwald's law and the impact of ion solvation on the dissociation degree are illustrated in Fig.~\ref{fig:sm}.
While we do not have direct experimental data for the dissociation degree,
we note that within the simple mixture model
variations of the volume factor $f = f^\SM$ are exclusively due to changes of $\Diss$.
Thus, in the right plot of Fig.~\ref{fig:sm} we can also directly observe
the impact of the Gibbs energy $\Delta g$ on the dissociation degree
when all particle sizes $\vau_\alpha^\Ref$ and solvation numbers $\kappa_\alpha$ are kept fixed.
We conclude that within this model already for the parameter value $\Delta g = 0.1\Uev$
the electrolyte effectively is completely dissociated.
For strongly diluted solutions with $\Ratio \to 0$, we have $\Diss \to 1$
and conclude from \eqref{Dens:252} $f \to f^0 = \tfrac{\vau_E^\Ref +\Delta\vau}{\vau_S^\Ref}$.
We note that all curves $f(\sqrt{\Ratio})$ start at $\Ratio = 0$ with horizontal slope.
Such a horizontal slope is not consistent with a linear law of $f(\sqrt{\Ratio})$ according to \cite{RedlichMeyer:1964}.
However, this deviation from the extrapolated linear data fit
happens within a region where experimental accuracy is not sufficient to falsify any model hypothes,
cf.\ Fig.~\ref{fig:dens-ratio}.
When incomplete dissociation sets in, \eqref{Dens-SM:204} shows that the slope of the curves $f(\Ratio)$ is proportional to $\Delta\vau$ and a fit to experimental data in principle is possible.
For the
\begin{subequations}
\begin{align}
	&\text{chosen parameters} &
	\Delta g &= -0.075\Uev~, \qquad \nu_S = \kappa_A +\kappa_C = 12&
\\*
	&\text{we obtain from data fit} &
	\Delta\vau  &\approx -0.25\, \vau_S^\Ref \qquad
	\vau_E^\Ref +\Delta\vau  \approx 0.95\, \vau_S^\Ref ~. &
\end{align}
\end{subequations}
We note that this solution does not require to specify the ion individual parameters of the simple mixture model
like $\vau_\alpha^\Ref$, $\kappa_\alpha$ for $\alpha= A,C$, but only combinations thereof.

\paragraph*{The full model at $T=T^\Ref$.}
Compared to the simple mixture, the full model contains the susceptibility $\chi$ as additional input data.
At $T=T^\Ref$, we assume for $\chi$ the linear relation according to \eqref{Deb:342},
where the values of the parameters $\chi_S^\Ref$ and $d=d_A+d_C$ are as given in Table~\ref{tab:L05}.
In order to determine the constants $\chi_\alpha^\Ref$,
we need to specify $d_\alpha$, $\vau_\alpha^\Ref$ and $\kappa_\alpha$ for $\alpha= A,C$.
We choose $\kappa_A = \kappa_C =6$ and set $d_A = d_{\ce{Cl^-}} = 5$, $d_C = d_{\ce{Na^+}} = 7$
as suggested by \cite{Marcus:2013}.
The resulting values for $R_\alpha$,  $\vau_\alpha^\Ref$ and $\chi_\alpha^\Ref$ for $\alpha\in\{A,C\}$
are listed in Table~\ref{tab:chosen} below.
The conditions \eqref{Dens:257} to determine $f$ and $\Diss$ read for the full model at $T=T^\Ref$:
\begin{subequations}
\label{Dens:411}
\begin{align}
\label{Dens:411a}
	-p^\Deb &=
		K\cdot	\bigg( \dfrac{1 +f^\SM\,\Ratio}{1 +f\ \Ratio} -1 \bigg) ~,
\\
	\frac{\Delta g}{\kB T^\Ref} &=
		\sum_{\alpha\in\cI} \nu_\alpha	\ln(y_\alpha)
		+\frac{\Delta\vau}{\kB T^\Ref}\,K\,\ln\left( 1 -\frac{p^\Deb}{K} \right)
		+\frac{1}{\kB T^\Ref} 				\sum_{\alpha\in\cI} \nu_\alpha	\mu_{\alpha}^\Deb
		~.
\label{Dens:411b}
\end{align}
\end{subequations}

\begin{figure}
	\centering
	\includegraphics[width=.45\textwidth]{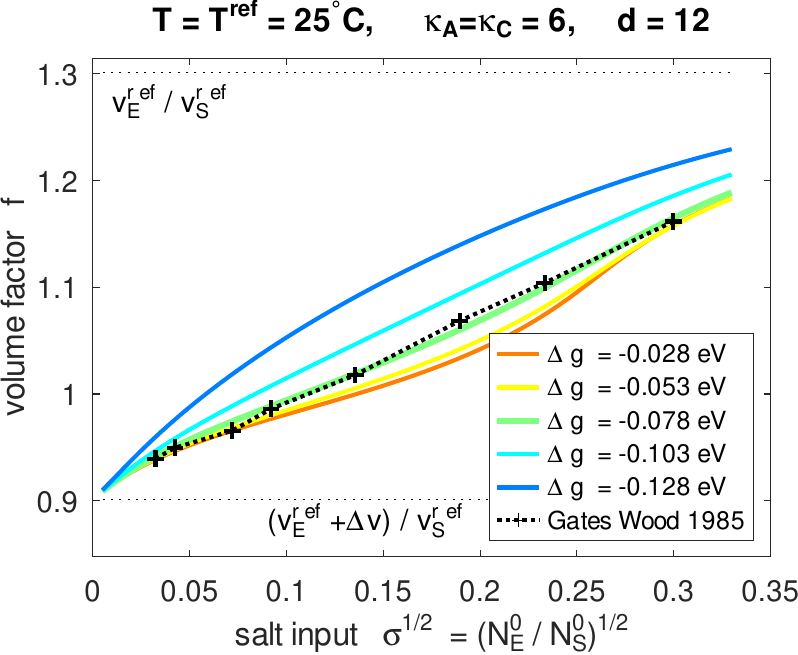}
	\hfill
	\includegraphics[width=.45\textwidth]{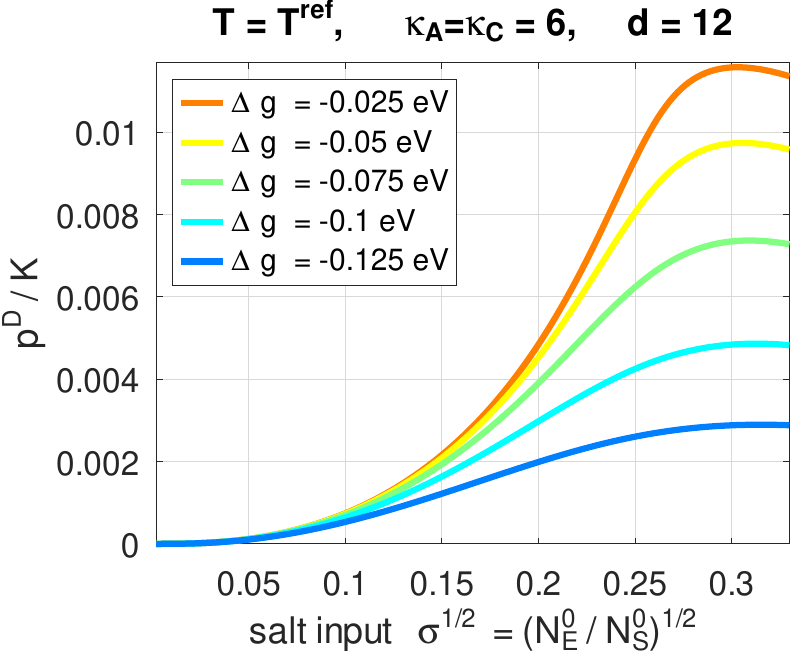}
	\caption{\label{fig:Deb-TRef}
		Full model at $T=T^\Ref$:
		Computed volume factor $f$ for different values of $\Delta{}g$ (left).
		The pressure contribution $p^\Deb$ is limited to smaller values for more negative $\Delta{}g$ (right).
		}
\end{figure}
Computed curves of the volume factor $f(\sqrt{\Ratio})$ are displayed in Fig.~\ref{fig:Deb-TRef}
for different values of the Gibbs energy $\Delta g$.
Since the curves $f(\sqrt{\Ratio})$ start with positive slope,
the limit value of the volume factor $f^0 = \lim_{\Ratio\to0} f = (\vau_E^\Ref +\Delta\vau)/\vau_S^\Ref$
is lower than in the simple mixture model.
The additional pressure contribution $p^\Deb$ remains rather small compared to the water bulk modulus $K$
and the upper bound of $p^\Deb$ decreases for smaller values of $\Delta{}g$. %
\begin{figure}
	\centering
	\includegraphics[width=.45\textwidth]{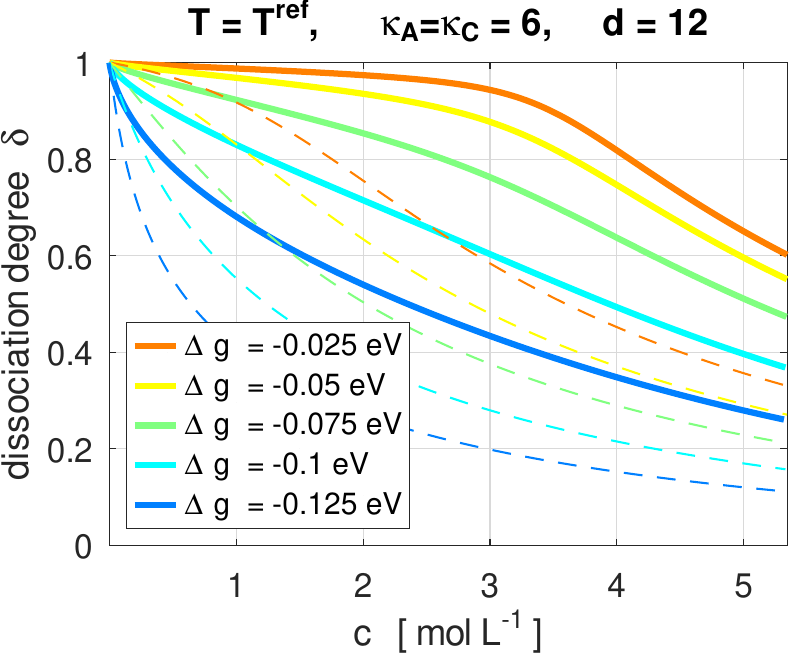}
	\hfill
	\includegraphics[width=.45\textwidth]{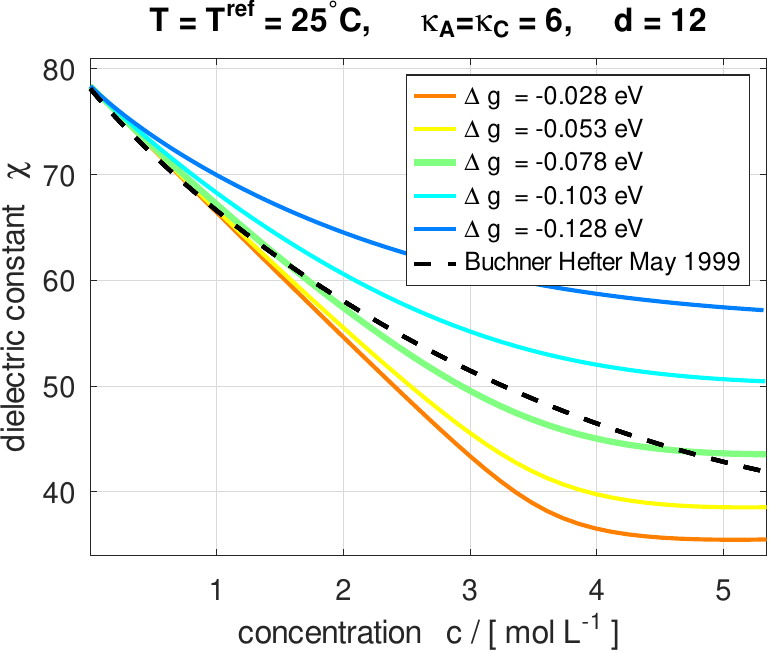}
	\caption{\label{fig:Deb-TRef-diss-chi}
		Full model at $T=T^\Ref$.
	Left:
		Computed dissociation degree $\Diss$ for different values of $\Delta g$.
		For comparison, the dashed lines are related to the simple mixture model.
	Right: resulting susceptibility $\chi$ shows initial linear decrement and
		later non-linear behavior due to incomplete dissociation.
	}
\end{figure}
The curves for the dissociation degree are shown in Fig.~\ref{fig:Deb-TRef-diss-chi}.
For the parameter value $\Delta g = -0.078\Uev$,
there is a non negligibly amount of the undissociated species $\ce{E}$ present in the electrolyte,
although for total salt concentration up to $c = 1\Umol\Ul^{-1}$
the dissociation degree remains larger than $0.9$.
Compared to the simple mixture model equipped with the same parameters,
the dissociation degree is considerably larger in full model.
Due to the assumed linear dependence of the susceptibility on the number densities,
there is an impact of the dissociation degree on the computed susceptibility that can be compared with experimental data.
For larger values $\Delta g > -0.07\Uev$, the curves $\chi(c)$ follow the linear dielectric decrement
over a too wide concentration range up to $c\approx 3\Umol\Ul^{-1}$
and then rapidly turn into a flat plateau.
For smaller values $\Delta g < -0.08 \Uev$, the curves $\chi(c)$ are too flat and $\chi$ does not decrease sufficiently.
Moreover, comparison of the computed curves $\chi(c)$ with experimental data from \cite{BuHeMa99}
justifies the initial choice $\kappa_A+\kappa_C = 12$,
because for values larger than $12$, the deviation of $\chi(c)$
from the linear decrement sets in more early and the susceptibility does not decrease sufficiently.
To the contrary, for $\kappa_A=\kappa_C = 5$, the linear decrement holds for a too wide
range of salt concentrations.
We note that the thus far unspecified
remaining parameter $\chi_E^\Ref$ only has a minor impact on the overall solution of \eqref{Dens:411}
that mainly can be observed in $\chi(c)$ for large concentrations $c> 4 \Umol\Ul^{-1}$.
For the computations, we made the rather arbitrary choice $\chi_E^\Ref = (\chi_A^\Ref +\chi_C^\Ref)/2$.

\paragraph{Temperatures $T\neq T^\Ref$.}
In order to compute the volume factor at temperatures $T\neq T^\Ref$,
the thermal expansion coefficients $\Thermal{\alpha}(T)$ need to be specified.
We choose $\Thermal{S}(T)$ according to the data for water in \cite{Apelblat:1999}.
Since we are not aware of any thermal expansion data for solvated ions,
we apply linearization at $T^\Ref$ with the coefficients $\beta_A^\Ref$, $\beta_C^\Ref$ as in Table~\ref{tab:chosen}.
Due to its minor impact on the computed results, we set the coefficient for the undissociated species $\ce{E}$
to coincide at $T=T^\Ref$ with the slope of $\Thermal{S}(T^\Ref)$, i.e. $\beta_E^\Ref = 255.75\cdot10^{-6}$.
For the evaluation of the functions $F(R_\alpha,\lambda)$,
the radii $R_\alpha$ at the temperature $T$ have to be determined in accordance with the thermal expansion.
Moreover, the potentials $\mu_\alpha^\Deb$, as well as the pressure contribution $p^\Deb$
contain derivatives $\tfrac{\partial}{\partial n_\alpha} \chi$,
which are a-priori unknown for $T\neq T^\Ref$ and have to be determined such that \eqref{Deb:127} is satisfied.
Therefore, we verify
\begin{align}
\notag%
	\tfrac{\partial}{\partial n_\alpha} &
		\left(
		\left( \frac{\chi}{D} \right)^2 \
		\sum_{\gamma\in\cI} z_\gamma^2 \frac{n_\gamma}{n^\Ref}		F(R_\gamma,\lambda)
		\right)
\\\notag
 &=
	\tfrac{\partial}{\partial n_\alpha} \chi
		\sum_{\gamma\in\cI} z_\gamma^2 \frac{n_\gamma}{n^\Ref}
		\left(
		\frac{\chi}{D^3} F(R_\gamma,\lambda) \,
	-
		\frac{\chi^2}{D^3} \tfrac{\lambda}{2} F'(R_\gamma,\lambda)
		\right)
\notag\\*&\rule{4ex}{0ex}
	+	\frac{\chi^2}{D^2} \frac{z_\alpha^2 }{\tau D \lambda^2}
		\sum_{\gamma\in\cI} z_\gamma^2 \frac{n_\gamma}{n^\Ref}		\tfrac{\lambda}{2} F'(R_\gamma,\lambda)
	+	\frac{\chi^2}{D^2}\,
							\frac{z_\alpha^2}{n^\Ref}			F(R_\alpha,\lambda) ~.
\label{Dens:721}
\end{align}
Then, we take the derivative of \eqref{Deb:3139} with respect to $n_\alpha$,
where on the right hand side we evaluate \eqref{Dens:721} for $T=T^\Ref$,
to finally determine $\tfrac{\partial}{\partial n_\alpha} \chi$ for $T\neq T^\Ref$.

\begin{figure}
	\centering
	\includegraphics[width=.45\textwidth]{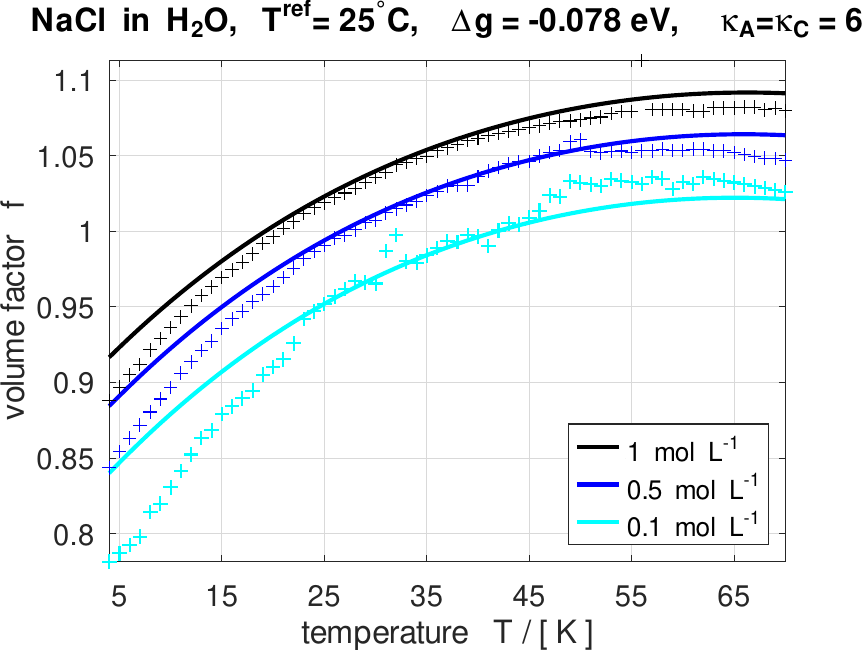}
	\hfill
	\includegraphics[width=.45\textwidth]{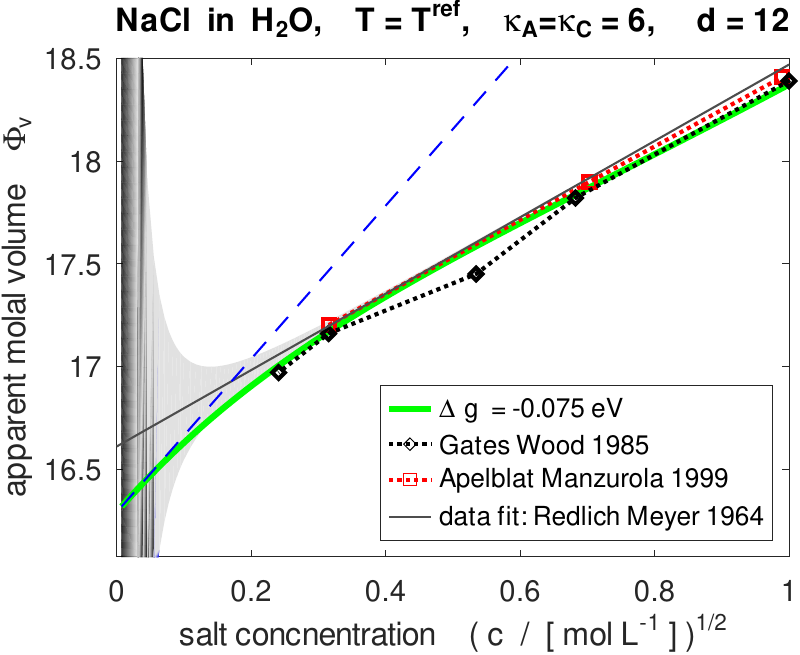}
	\caption{\label{fig:Deb-T}
	Left: Computed volume factor $f(T)$ for solutions of different molality (solid lines) and
			experimental data of \cite{Apelblat:1999} (crosses).
	Right:
		The limiting slope according to \eqref{Dens:834}
	}
\end{figure}
The computed volume factor $f(T)$ as a function of temperature
is shown in Fig.~\ref{fig:Deb-T} for different solutions of fixed salt molality.
Due to the linearization of the thermal expansion coefficients,
the accuracy of the computed curves is limited to a vicinity around the reference temperature
with a width in the order of $10 \UK$.
More results, showing the concentration and temperature dependence of the apparent molar volume,
are presented below on Sect.~\ref{sect:dens-vatious}.

\subsection{Limiting slopes for strong dilution}
The results of Debye-H\"uckel theory are most often expressed in the form of limiting laws for $\Ratio\to 0$.
An empirical limit law for the molal volume at infinite dilution was proposed by Masson \cite{Masson:1929}, viz.\
\begin{align}\label{Dens:812}
	\Phi_V	&=	\Phi_V^0 + S_V\, %
							\sqrt{c},
\end{align}
with individual slopes $S_V$ for each electrolyte.
A formally identic equation was derived by Redlich and Rosenfeld
from an extension of \cite{DH23} with temperature dependent susceptibility \cite{RedlichRosenfeld:1931}.
There, the slope is of the form $S_V = k\,\sqrt{w}^3$,
where $w$ is a factor depending only on the valency of the ions
and $k$ uniquely determined by the  Debye-H\"uckel theory.
Based on later experimental data, the parameters were identified as $\Phi_V^0 \approx 16.61$ and $S_V \approx 1.86$, cf.~\cite{RedlichMeyer:1964}.
The discrepancy between these theoretical values and the empirical slopes related to the Masson equation
caused a controversy that was resolved in such a way, cf.\  \cite{RedlichMeyer:1964,Millero:1971}
that the theoretical equation by \cite{RedlichRosenfeld:1931} is considered valid,
although possibly only for ion concentrations that are too low to allow accurate measurements,
whereas the empirical equation is also valid but
in an electrolyte concentration range where ion specific effects already caused
deviations from the Debye-H\"uckel type limiting law.

To determine the limit slope of the volume factor $f$ in the strong dilution limit,
we start from %
\eqref{Dens:257a} that we rearrange as
\begin{align}\label{Dens:822}
	f	&=	\frac{\frac{p^\Deb}{\Ratio} +K\,f^\SM }{K -p^\Deb} ~.
\end{align}
Then, we check $\lim_{\Ratio\to 0}n_\alpha = 0$ for $\alpha\in\{A,C,E\}$,
$\lim_{\Ratio\to 0}\Diss = 1$,
$\lim_{\Ratio\to 0}\chi = \chi_S^\Ref$,
$\lim_{\Ratio\to 0}p^\Deb = 0$
and $\lim_{\Ratio\to 0} \partial_{\sqrt{\Ratio}}\, p^\Deb = 0$,
to infer
\begin{align}\label{Dens:824}
	\lim_{\Ratio\to 0}
	\partial_{\sqrt{\Ratio} }\, f
	&=	\lim_{\Ratio\to 0}\left(
		\frac{ \partial_{\sqrt{\Ratio}}\left( \frac{p^\Deb}{\Ratio} \right)  }{K -p^\Deb}
	+	\frac{ \frac{p^\Deb}{K}  +K\,f^\SM }{ (K -p^\Deb)^2 }  \partial_{\sqrt{\Ratio}}\, p^\Deb
	\right)
	=
	\lim_{\Ratio\to 0}
	\frac{ \partial_{\sqrt{\Ratio}}\left( \frac{p^\Deb}{\Ratio} \right)  }{K} ~.
\end{align}
We abbreviate $F'_\gamma =\tfrac{\partial}{\partial\lambda}F(R_\gamma,\lambda)$.
With $\lim_{\Ratio\to 0}\lambda = 0$, we deduce
$\lim_{\Ratio\to 0}F(R_\gamma,\lambda) = \lim_{\Ratio\to 0}\left( \lambda F'_\gamma \right) = 0$.
Dividing the pressure contribution $p^\Deb$ from \eqref{Deb:138} by $\Ratio$
and applying product rule shows that
the remaining non-vanishing terms in \eqref{Dens:824} in the limit $\Ratio\to 0$ are
\begin{align}\label{Dens:826}
	\lim_{\Ratio\to 0} \partial_{\sqrt{\Ratio}} f
	&=	-\frac{a}{2K}
	\frac{\chi_S^\Ref}{(1+\chi_S^\Ref)^2}
	\sum_{\beta\in\{A,C\}} z_\gamma^2
	\lim_{\Ratio\to 0} \left(
		\frac{n_\gamma}{n^\Ref\Ratio} \cdot \partial_{\sqrt{\Ratio}} \lambda
		\right)
\quad \times \\*&\notag\rule{11ex}{0ex}
	\lim_{\Ratio\to 0} \partial_{\lambda}
			\Bigg[
				\frac{\lambda}{2}F'(R_\gamma,\lambda)	\frac{1+\chi -\cL_\chi}{1+\chi}
			-					F(R_\gamma,\lambda)	\frac{\chi-1}{1+\chi} \frac{\cL_\chi}{\chi}
			\Bigg] ~.
\end{align}
Applying again product rule in \eqref{Dens:826} and going to the limit $\Ratio\to 0$,
the only non-vanishing terms in the square brackets are
\begin{align}\label{Dens:836}
	\lim_{\Ratio\to 0} \partial_{\lambda}
		\Bigg[
			\frac{\lambda}{2}F'_\gamma	\cdot(1+\chi -\cL_\chi)
			-				F(R_\gamma,\lambda)	(\chi-1) \frac{\cL_\chi}{\chi}
		\Bigg]
	&=	\Bigg[
			\frac{1}{2}		\lim_{\Ratio\to 0} F'_\gamma
		-	(\chi_S^\Ref-1)	\lim_{\Ratio\to 0}  F'_\gamma
			\Bigg]&
\end{align}
We check that in the limit $\Ratio\to 0$ we have the identities
\begin{subequations}
\begin{align}\label{Dens:832}
	 \lim_{\Ratio\to 0} \frac{n_\gamma}{\Ratio}
	 &=	\lim_{\Ratio\to 0} \frac{\nu_\gamma\, \Diss\,N_S^0 }{V_S^\Ref\, (1 +f \Ratio)}
\quad
	=	\frac{\nu_\gamma}{\vau_S^\Ref} ~, &
\\
	 \lim_{\Ratio\to 0} \partial_{\sqrt{\Ratio}}\,\lambda
	 &=	 \frac{1}{\sqrt{(1+\chi_S^\Ref)\tau}} \sqrt{ \sum_{\beta\in\{A,C\}} \frac{\nu_\gamma\,z_\gamma^2}{\vau_S^\Ref n^\Ref} } ~, &
\end{align}
\end{subequations}
to conclude that the limit slope for the volume factor is
\begin{align}\label{Dens:844}
	\lim_{\Ratio\to 0} \partial_{\sqrt{\Ratio}} {f}
&=	\frac{a}{4K}\frac{\chi_S^\Ref}{(1+\chi_S^\Ref)^3}
		\frac{(2\chi_S^\Ref	-3)\,F'_0}{\sqrt{(1+\chi_S^\Ref)\tau}}
		\sqrt{ \sum_{\beta\in\{A,C\}} \frac{\nu_\gamma\,z_\gamma^2}{\vau_S^\Ref n^\Ref} }^3 ~,
\end{align}
where $F'_0$ denotes the unique limit $\lim_{\lambda\to 0}F'_\gamma$ for all $\gamma\in\cI$.
This limit slope has the same structure $\lim_{\Ratio\to 0} \partial_{\sqrt{\Ratio}} {f} = S_f =  k_f \sqrt{w_f}^3$
as the above mentioned slope derived by \cite{RedlichRosenfeld:1931}.
The numeric value of the limit slope is then obtained from
\begin{align}\label{Dens:834}
	\lim_{\Ratio\to 0} F'(R_\gamma,\lambda)
	&=\lim_{\Ratio\to 0}	\left(	\frac{3 +4\lambda R_\alpha/\LMicro}{(1+\lambda R_\alpha/\LMicro)^2}
	-2 \lambda R_\alpha/\LMicro	\frac{3 +2\lambda R_\alpha/\LMicro}{(1+\lambda R_\alpha/\LMicro)^3} \right)
	\quad
	=	3 ~.
\end{align}
For the apparent molar volume, we thus get a limit slope $S_V = 3.7306$,
approximately twice of the value by \cite{RedlichRosenfeld:1931}.
The computed apparent molar volume in the dilute regime,
together with experimental data and the limit law according to \cite{RedlichMeyer:1964},
are displayed in Fig.~\ref{fig:Deb-T}.
We observe, that the limit law only holds for rather low concentrations up to ca.\ $0.05 \Umol\Ul^{-1}$
and the deviation with respect to the standard Debye-H\"uckel limit according to \cite{RedlichRosenfeld:1931,RedlichMeyer:1964}
is within the shaded region indicating an uncertainty in the underlying mass density of the solution
less than $2.5\cdot10^{-6} \Ukg\Ul^{-1}$.
Since the here computed molar volume has a steeper slope for $c\to 0$,
the apparent volume at infinite dilution
$\Phi_V^0 = \lim_{c\to0}\Phi_V = 16.2904$ is smaller than $16.61$ according to \cite{RedlichMeyer:1964}.

\subsection{Application to different salts and various temperatures} \label{sect:dens-vatious}
So far, we computed the volume factor $f$, or the equivalent apparent molar volume $\Phi_V$,
using one set of parameters.
However, going from one salt to another related salt, sharing either the same anion or the same cation,
the model parameters can not be chosen completely independently.
None of the model parameters for solvated ions,
i.e.  $\vau_\alpha^\Ref$, $\chi_\alpha^\Ref$ for $\alpha\in\{A,C\}$,
is directly available from experimental data.
In addition also parameters $\kappa_\alpha$ and the Gibbs energy $\Delta g$ of the dissociation reaction are needed.
Table~\ref{tab:observed} lists parameter values that can be deduced from experimental data.
Here, the the values for the dielectric decrement $d$ are based on \cite{Marcus:2013}.
The values for the molar volume at infinite dilution $\Phi_V^0$
and for the volume change $\Delta\vau$ are based on
\cite{PDH:1984} for the fluoride salts,
\cite{GatesWood:1985} for the chloride salts and
\cite{HHLH:2016,HH:2021} for the perchlorate salts.

\begin{table}
\centering
	\caption{\label{tab:observed}
	Salt specific parameters for aqueous solutions deduced from experimental data at the reference temperature $T^\Ref=25^\circ{}C$.
	}
\begin{tabular}{l|lllllllll}
					&	$\ce{LiF}$	&	$\ce{NaF}$	&	$\ce{KF}$	&$\ce{LiCl}$&	$\ce{NaCl}$	&	$\ce{KCl}$	&	$\ce{LiClO_4}$	&	$\ce{NaClO_4}$ &	$\ce{KClO_4}$	 \\
\hline\hline%
$\phantom{\Longrightarrow} f^0$
					&	-0.149		&	-0.168		&	0.397		&0.921		&	0.902		&	1.47		&	 2.37			&	2.35		&	2.92	\rule{0pt}{3ex}\\
$\Longrightarrow \Phi_V^0$
					&	-2.684		&	-3.04		&	7.18		&16.65		&	16.29		&	26.51		&	42.89			&	42.53		&	52.75	\rule{0pt}{3ex}\\
\hline
$\phantom{\Longrightarrow} \Delta v / \vau_S^\Ref$
					&	-2.05		&	-0.5		&	-0.6		&-0.2		&	-0.4		&		-0.5	&	-0.1			& 	-0.35		&	-0.2	\rule{0pt}{3ex}\\
\hline
$\phantom{\Longrightarrow} d$
					&		13		&		12		&		11		&	13		&		12		&		11		&		15			&	14			&	13		\rule{0pt}{3ex}\\
\hline
$\Delta{}g / [ \Uev ]$
					&	-0.1		&	-0.105		&	-0.085		&-0.7		&	-0.78		&	-0.8		&	-0.1			&	-0.085		&	-0.14	\rule{0pt}{3ex}\\
\end{tabular}
\end{table}

The definition \eqref{Dens:251} together with $f^0\,{\vau_S^\Ref} = {\vau_E^\Ref +\Delta\vau}$ yield the relation
\begin{align}
\label{Dens:603}
	\vau_A^\Ref + \vau_C^\Ref  &= (f^0 +\kappa_A +\kappa_C)\vau_S^\Ref  &
\end{align}
that determines the difference in $f^0$ when exchanging one ion with another.
An analogous relation for the bare, unsolvated ions has been used
to define \emph{conventional partial molar volumes} relative to the $\ce{H^+}$ ion as a reference,
cf.~\cite{Millero:1971,Marcus:2011}. %
Due to singularity of experimental uncertainty,
values $\Phi_V^0$ necessarily have to rely on extrapolation of data
and thus depend on the assumed theoretical model.
Since the model proposed here differs in the limiting behavior from the standard Debye-Hückel theory,
the obtained values of $\Phi_V^0$  also differ from those found in the literature.
However, instead of relative partial apparent molar volumes,
our model requires as parameters absolute specific volumes $\vau_\alpha^\Ref$ of solvated ions
and the related assigned radii $R_\alpha$.
Therefore, a suitable choice for $\vau_\alpha^\Ref$ and $\kappa_\alpha$ needs to be done
in order to let \eqref{Dens:603} approximate the $f^0$ data of Tab.\ref{tab:observed}.
We keep the above choice $\kappa_\alpha = 6$
and the chosen values of $\vau_\alpha^\Ref$ are listed in Tab.~\ref{tab:chosen},
together with the related radii $R_\alpha$ of the solvated ions and $\tilde{r}_\alpha$
of the unsolvated center ions according to
\begin{align}
\label{Dens:605}
	R_\alpha		&= \sqrt[3]{\frac{3}{4\pi}\cdot \vau_\alpha^\Ref\Thermal{\alpha} } ~, &
	\tilde{r}_\alpha&= \sqrt[3]{\frac{3}{4\pi}\cdot(\vau_\alpha^\Ref\Thermal{\alpha} - (\kappa_\alpha \vau_S^\Ref\Thermal{S})} ~. &
\end{align}
The limiting molar volume at infinite dilution can be small, or even negative,
e.g.\ in the case of  $\ce{LiF}$ or $\ce{NaF}$ dissolved in water.
Thus, the specific volume $\vau_\alpha^\Ref$ can be smaller than the sum of $\kappa_\alpha$ solvent molecules.
Hence, while $\vau_\alpha^\Ref$ always has to be positive,
the radius $\tilde{r}_\alpha$ in general can also be negative, cf.\ $\tilde{r}_{\ce{F}}$ in Tab.~\ref{tab:chosen}.

\begin{table}
\centering
	\caption{\label{tab:chosen}
	Ion specific parameters at $T^\Ref=25^\circ{}C$ based on the choice $\kappa_\alpha=6$:
		specific volumes $\vau_\alpha^\Ref$ and the implied
		radii $R_\alpha$ of the solvated ion and $\tilde{r}_\alpha$ associated with the unsolvated ion in solution,
		additive dielectric decrement $d_\alpha$ %
		and $\beta_\alpha^\Ref$ for the linearization  of the thermal expansion at $T=T^\Ref$.
	}
\begin{tabular}{l|lll|lll}
					&	$\ce{Li^+}$	&	$\ce{Na^+}$	&$\ce{K^+}$	&$\ce{F^-}$&$\ce{Cl^-}$&$\ce{ClO4^-}$	 \\
\hline\hline%
$\phantom{\Longrightarrow} \vau_\alpha^\Ref\, 1000\Avo\ /\ [\Ul\Umol^{-1}]$
					&	0.12276		&	0.1224		&	0.13262	&	0.09138	&	0.11071		& 0.13695	\rule{0pt}{3ex}\\
$\Longrightarrow R_\alpha / \LMicro$
					&	7.5189		&	7.5116		&	7.7151	&	6.8144	&	7.2645	& 	7.7983		\rule{0pt}{3ex}\\
$\Longrightarrow \tilde{r}_\alpha\ /\ [\Upm]$
					&	178.5		&	177			&	212.5	&	-189	&	97		&	224.5		\rule{0pt}{3ex}\\
\hline
$\phantom{\Longrightarrow} d_\alpha$
					&		8		&	7			&	6		&	5		&	5		&	7			\rule{0pt}{3ex}\\
$\Longrightarrow \chi_\alpha^\Ref$
					&	13.21		&	21.19		&	33.14	&	23.66	&	33.22	&	27.27		\rule{0pt}{3ex}\\
\hline
$\phantom{\Longrightarrow} \beta_\alpha^\Ref\cdot10^{6}$
					&	306.9		&	754.5		&	690.5	&	306.9	&	319.7	&	1074.2		\rule{0pt}{3ex}\\
\hline%
\end{tabular}
\end{table}

The parameters for the linearized thermal expansion coefficients are chosen based on
\cite{Apelblat:1999,HH:2021} for $\ce{Na^+}$, $\ce{K^+}$, $\ce{Cl^-}$ and $\ce{ClO4^-}$,
\cite{Apelblat:2001,HHLH:2016} for $\ce{Li}^+$ and
\cite{MilleroDrostHansen:1968} for $\ce{F^-}$,
whereas the coefficient for the undissociated salt in the following is in all cases
set to the arbitrary value $\beta_E^\Ref = 255.75\cdot10^{-6}$ corresponding to the thermal expansion of water at $T=25^\circ{}C$.

\begin{figure}
	\centering
	\includegraphics[width=.45\textwidth]{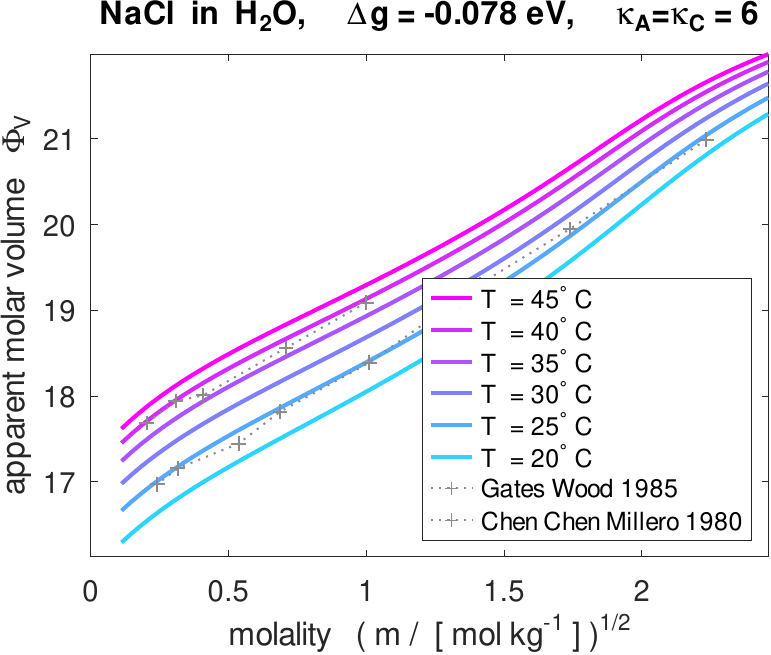}
	\hfill
	\includegraphics[width=.45\textwidth]{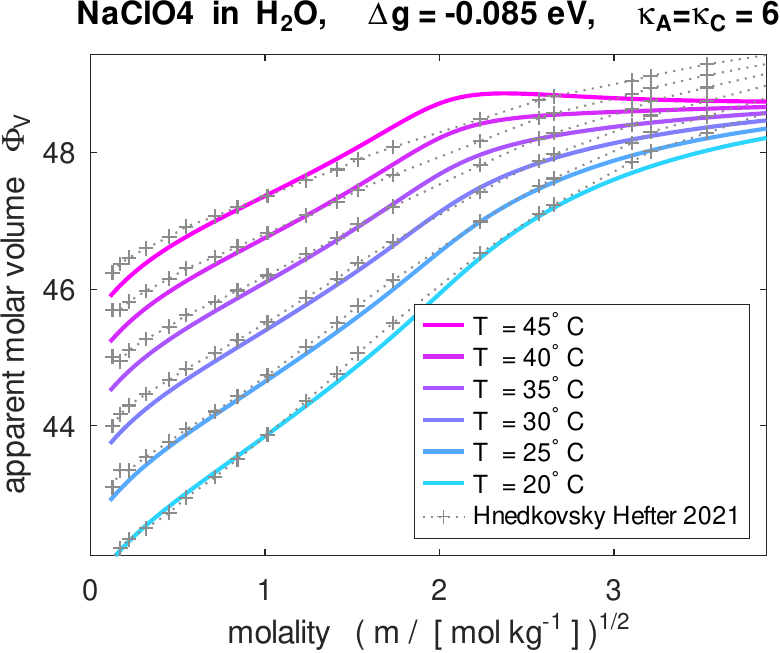}
\\[2ex]
	\includegraphics[width=.45\textwidth]{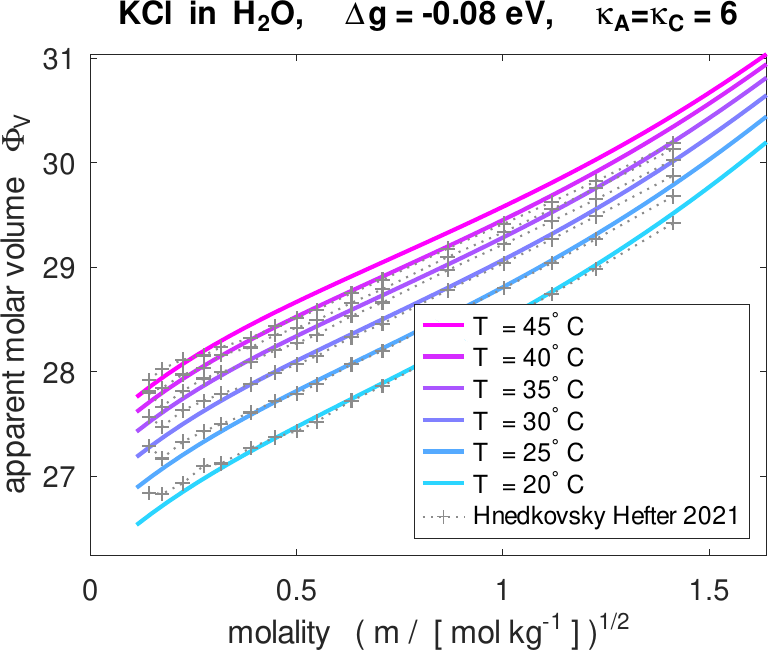}
	\hfill
	\includegraphics[width=.45\textwidth]{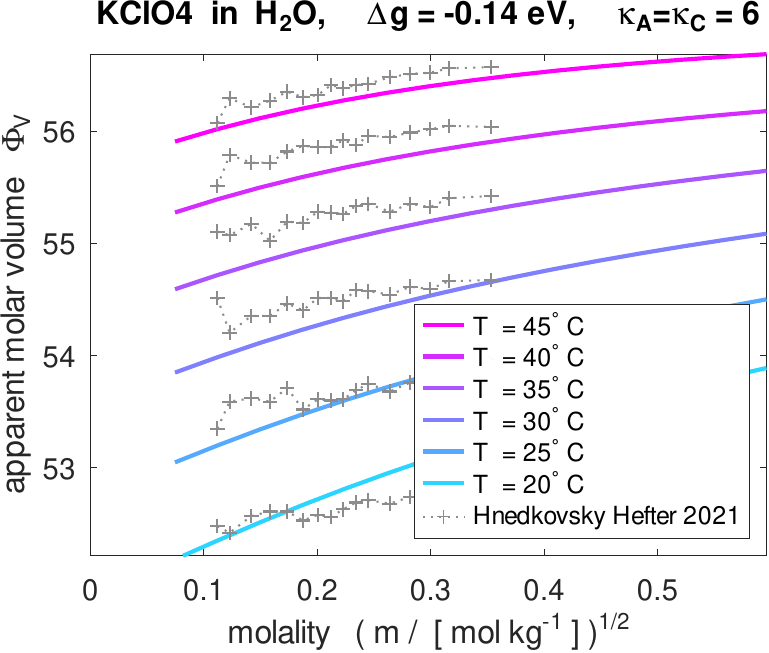}
	\caption{\label{fig:phi-related}
		Apparent molar volume for the related salts $\ce{NaCl}$, $\ce{KCl}$, $\ce{NaClO4}$ and $\ce{KClO4}$
		computed from a single set of parameters as given in Tab.~\ref{tab:chosen}
		and experimental data for comparison.
	}
\end{figure}
The apparent molar volume as function of the salt concentration and for different temperatures was computed
for the set of related salts $\ce{NaCl}$, $\ce{KCl}$, $\ce{NaClO4}$ and $\ce{KClO4}$.
The parameters specific to the ions were set as in Table~\ref{tab:chosen}.
In addition there are parameters specific to each salt.
These are the specific volume $\vau_E^\Ref$ that follows from $f^0$ and $\Delta\vau$ listed in Table~\ref{tab:chosen}
and susceptibility coefficient $\chi_E^\Ref$ and the Gibbs energy energy $\Delta{}g$ of the dissociation reaction.
The value $\Delta g_{\ce{NaCl}} = -0.78\Uev$ was checked against concentration dependent susceptibility data
of \cite{BuHeMa99}
and $\Delta g_{\ce{KCl}} = -0.1\Uev$ based on susceptibility data of \cite{ChenHefterBuchner:2003}.
Moreover, we set $\Delta g = -0.1\Uev$ for the perchlorate salts.
The coefficients $\chi_E^\Ref$ were arbitrarily set to $\chi_E^\Ref = (\chi_A^\Ref +\chi_C^\Ref)/2$.
Computed curves and good agreement with experimental data of \cite{AllredWoolley:1981,HHLH:2016,HH:2021}
are shown in Fig.~\ref{fig:phi-related}.

\begin{figure}
	\centering
	\includegraphics[height=.35\textwidth]{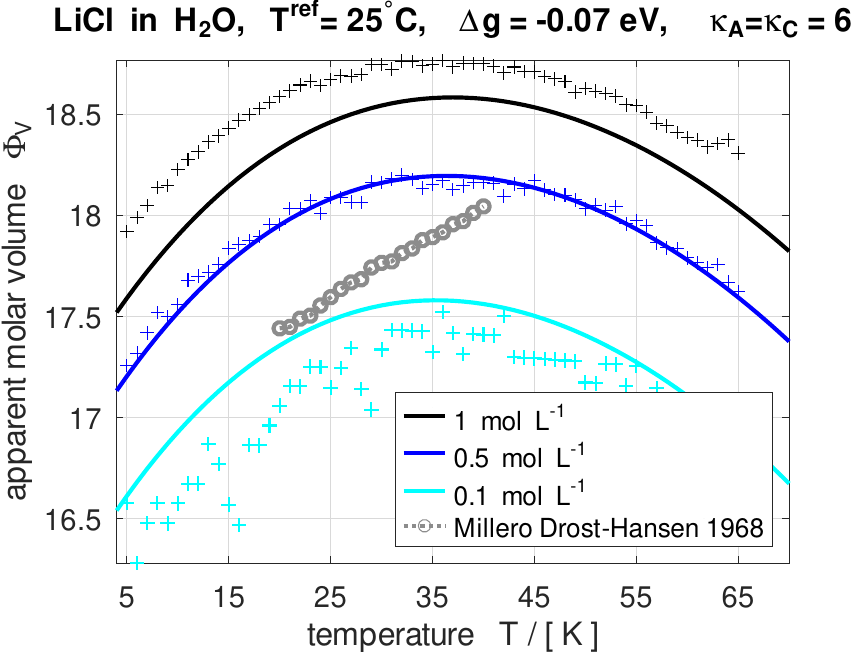}
	\hfill
	\includegraphics[height=.35\textwidth]{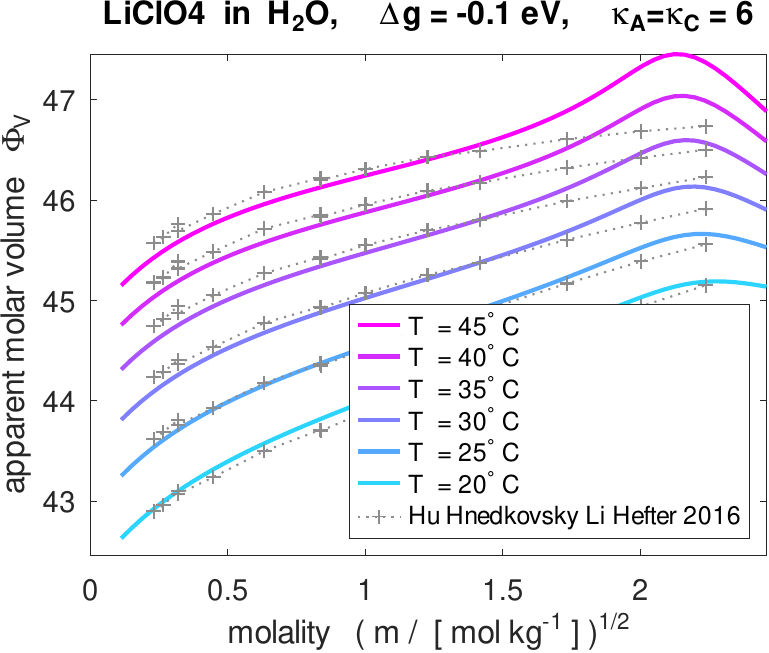}
	\caption{\label{fig:phi-Li}
		Left: 	Computed apparent molal volume of $\ce{LiCl}$ solutions of fixed concentration (solid lines)
				show a local maximum at $T\approx40^\circ{}C$ like experimentally observed in \cite{Apelblat:2001} (marker).
		Right: Computed apparent molal volume of aqueous $\ce{LiClO4}$ solutions and data of \cite{HHLH:2016} for comparison.
	}
\end{figure}

More related salts can be considered sharing the $\ce{Li^+}$ cation.
As stated in \cite{Apelblat:2001}, their data confirms earlier observations \cite{Ellis:1966,HarnedOwen1958}
that aqueous $\ce{LiCl}$ solutions show a local maximum of the apparent molar volume near $T\approx40^\circ{}C$,
much earlier than for solutions of $\ce{NaCl}$ or $\ce{KCl}$, where the maximum occurs around $T\approx60^\circ{}C$.
Remarkably, this maximum is well captured by the computed curves for $\ce{LiCl}$ in Fig.~\ref{fig:phi-Li},
whereas because of the large difference to the reference temperature for linearization
the computed apparent molar volume does not reproduce the local maximum for $\ce{NaCl}$, cf. Fig.~\ref{fig:Deb-T},
or similarly also for $\ce{KCl}$.
Here, the Gibbs energy $\Delta g_{\ce{LiCl}} = -0.095\Uev$ was adjusted according to \cite{GatesWood:1985} at $25^\circ{}C$.
As pointed out in \cite{Apelblat:2001}, their measured values for small molality $0.1 \Umol\Ul^{-1}$
are low compared to e.g.\ \cite{GatesWood:1985} at $25^\circ{}C$ or
\cite{MilleroDrostHansen:1968} for a slightly higher molality.

The apparent molar volume of $\ce{LiClO_4}$ in aqueous solutions as a function of temperature and molality
can be computed with the parameters fixed so far and $\Delta g_{\ce{LiClO_4}} = -0.1\Uev$
to reach good agreement with the experimental data of \cite{HHLH:2016}.

\PreprintClearpage
\section{Application to phase boundaries of binary electrolytes}	\label{sect:appl-phase}

\begin{figure}
	\centering
	\includegraphics[width=.6\textwidth]{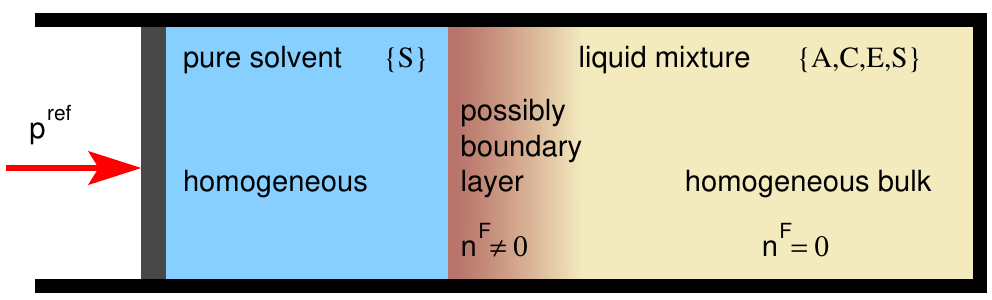}
	\caption{\label{fig:interf-pure-mix}
		Interface between pure solvent (solid or gas) and liquid mixture
		with possibly a charged boundary layer in the mixture.
	}
\end{figure}
We consider a planar interface,
where on the one side there is only the pure solvent $S$ present and
on the other side of the surface, there is a liquid electrolytic mixture
consisting of the species $\{A,C,E,S\}$.
For the pure solvent, we denote the Gibbs energy by $g_S^\Pure(T,p)$.
In the other subdomain, the electrochemical potentials are constant in equilibrium.
Since we assume the solvent $S$ electrically neutral,
the chemical potential $\mu_S$ is constant within the whole subdomain of the mixture,
even in the presence of a charged boundary layer at the surface.

Both phases are in equilibrium, if
\begin{align} \label{PB:03}
	g_S^\Pure	\overset{!}{=}	\mu_S ~.
\end{align}
In the mixture, we introduce
\begin{align}
	g_\alpha(T,p)
		&=	g_\alpha^\Ref
		+	m_\alpha	c_\alpha^\Ref \cdot \left( T -T^\Ref -T\ln\Big( \tfrac{T}{T^\Ref}\Big) \right	)
		-	m_\alpha	s_\alpha^\Ref \cdot		 (T-T^\Ref)
\\*\notag&
		+\vau_\alpha^\Ref \, \Thermal{\alpha}(T)\, (p -p^\Ref)
		+\vau_\alpha^\Ref \, (\Thermal{\alpha}(T)	 -1)\, p^\Ref
	~. &
\end{align}
Then, we apply the thermal equation of state from \eqref{Deb:320} to the chemical potential of the solvent
and eploy linearization of the logarithm of the pressure for large bulk modulus $K$,
such that
\begin{align}
	\mu_S
	&=	g_S(T,p^\Ref)								+\vau_S^\Ref\Thermal{S}(T)\cdot ( p -p^\Ref )
	+\kB T \ln\left( y_S \right)	+\mu_S^\Deb		-\vau_S^\Ref\Thermal{S}(T)\cdot  p^\Deb
	~.
\label{PB:04a}
\end{align}
In the particular case that there is only the solvent present on both of the interface,
but possibly in different phases,
there holds
\begin{align} \label{PB:04}
	\text{\uline{$y_S=1$: }}\quad
	g_S^\Pure(T,p)	\overset{!}{=}	g_S(T,p)~.
\end{align}
When the pressure in the subdomain of the pure phase is controlled,
then solution of \eqref{PB:04} defines either the \emph{melting temperature} $T^S(p)$
or the \emph{boiling temperature} $T^V(p)$ of the solvent, respectively,
depending on whether the pure phase is solid or vapour state.
On the other hand, we may alternatively control the temperature in the pure phase.
Then, solution of \eqref{PB:04} defines the \emph{vapour pressure} $p^V(T)$.

In the oversimplified ideal gas analogy of the early electrolyte theory,
the phase boundary was associated with a semipermeable membrane
and the phase equilibrium \eqref{PB:03} was used to introduce the notion of an 'osmotic pressure'.
Such an 'osmotic pressure' cannot be reconciled with
the very precisely defined concept of pressure within the applied electro-thermodynamic framework employed here.
And since real semipermeable membranes cannot provide reproducible results in practice,
they are also not suitable for testing a theory.
However, there is the commonly used quantity called '\emph{osmotic coefficient}', defined as
\begin{align} \label{PB:05}
	\phi
	&=-\frac{ \mu_S -g_S(T,p^\Ref)	}{\kB T \ (\nu_A +\nu_C)\,  m_S \Avo \cdot m}
	~,
\end{align}
which can serve as some kind of measure to compare
the results obtained from the here proposed model to results found in the literature.
Measurements of vapour pressure, freezing point depression and boiling point elevation
can be used to determine the 'osmotic coefficient'.

\subsection{Reduction of vapour pressure}
We assume that the pure solvent side is in vapour phase
and that it can be treated as an ideal gas
such that the Gibbs energy of the pure solvent is
\begin{align} \label{PB:505}
	g_S^\Pure(T,p)
	&=	g_S(T,p^V(T)) + \kB T \ln\left( \frac{p}{p^V(T)} \right)
	~.
\end{align}
Then, the equilibrium condition \eqref{PB:03} reads
\begin{align}
	\label{PB:511}
		\ln\left( \frac{p}{p^V(T)} \right)
	&= %
		\frac{\vau_S^\Ref\Thermal{S}(T)}{\kB T} (p^\Ref -p^V(T))
		+\frac{\mu_S -g_S(T,p^\Ref)}{\kB T} ~.
\end{align}
On the vapour side, the ideal gas law defines the specific volume $\vau_S^V$  by %
$p^V(T)\cdot\vau_S^V = \kB T$,
where $p^V(T)$ be the vapour pressure of the solvent determined by \eqref{PB:04} at temperature $T$.
Thus, %
\begin{align} \label{PB:513}
		\ln\left( \frac{p}{p^V(T)} \right)
	&= %
		\left( \frac{p^\Ref}{p^V(T)} -1 \right) \frac{\vau_S^\Ref\Thermal{S}(T)}{\vau_S^V}
		-\phi \cdot (\nu_A +\nu_C)\ m_S\Avo \cdot m ~.
\end{align}
For water at temperatures near $25^\circ{}C$,  %
the first term on the right hand side is small because
$\vau_S^\Ref \approx 2.3\cdot10^{-5}\vau_S^V$
and can be neglected.
We thus conclude that the vapour pressure of the electrolytic mixture is
\begin{align} \label{PB:514}
	\frac{p}{p^V(T)}
	&= \exp\bigg( -\phi \cdot (\nu_A +\nu_C)\  m_S\Avo \cdot m	\bigg) ~.
\end{align}
\begin{figure}
	\centering
	\includegraphics[width=.42\textwidth]{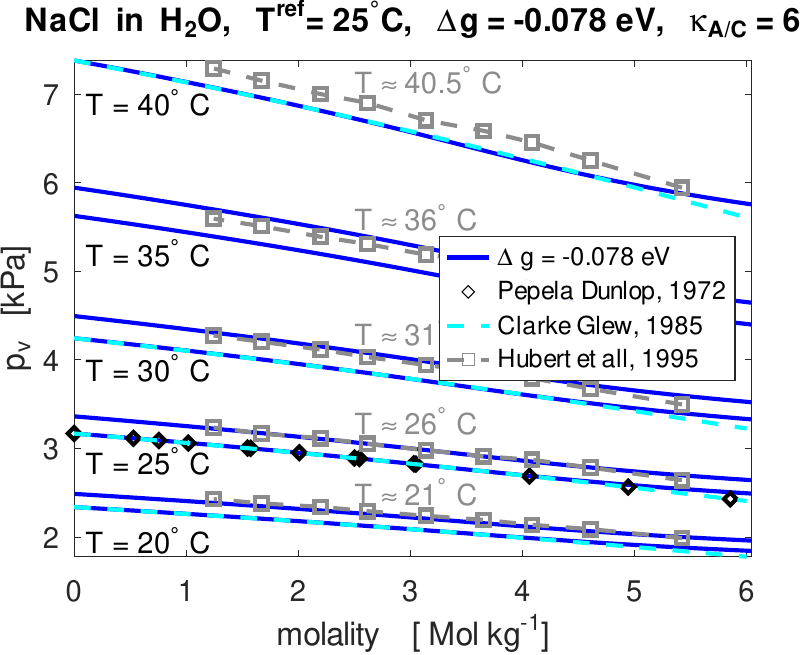}
	\hfill
	\includegraphics[width=.42\textwidth]{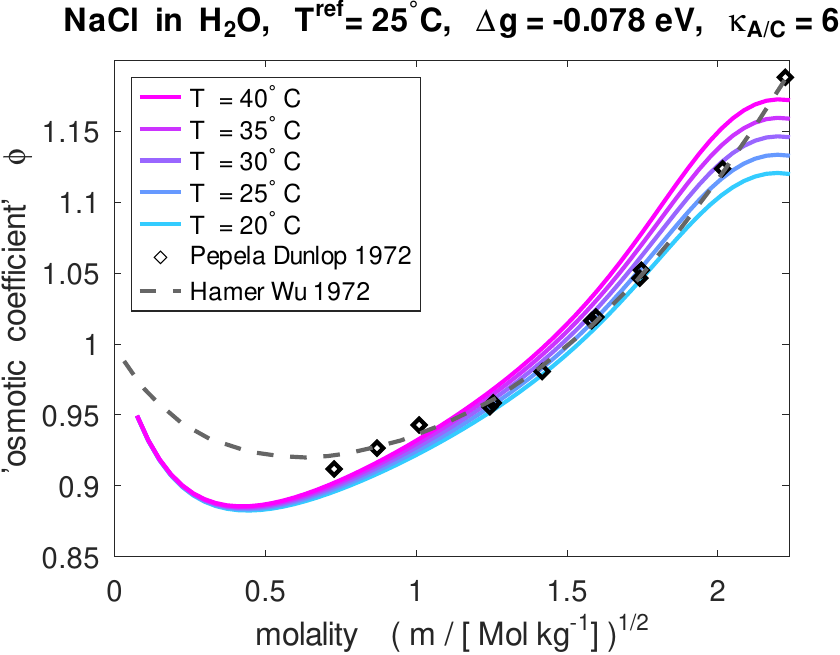}
	\caption{\label{fig:interf-vp}
	Left: Computed concentration dependent vapour pressure for different temperatures $T$
			follows the literature data \cite{PepelaDunlop:1972,ColinGlew:1985,HGBS:1995}
			closely up to $5\Umol\Ul^{-1}$.
	Right:
		'Osmotic coefficient' based on the complete model
		shows weak temperature dependence for higher salt concentrations.
		For comparison 'osmotic coefficient'
		derived from vapour pressure data \cite{PepelaDunlop:1972} at $T=25^\circ{}C$
		and reference data from \cite{HamerWu:1972}.
	}
\end{figure}
To compute the vapour pressure of an aqueous $\ce{NaCl}$ solution,
wet set $T^\Ref=25^\circ{}C$ and $p^\Ref$ equal to atmospheric pressure and
use the same set of model parameters that have been employed before
in Sect.~\ref{sect:appl}, cf.\ Tables~\ref{tab:L05}-\ref{tab:chosen}.
For the vapour pressure of pure water, we used the values given in \cite[Tab.~6-8]{CRC05}.
The results are shown in Fig.~\ref{fig:interf-vp}
and compared with experimental data \cite{PepelaDunlop:1972,HGBS:1995}
and tabulated values based on a regression model \cite{ColinGlew:1985}.
We observe that
the computed vapour pressure follows the experimental results very closely
over the whole tested temperature and concentration range,
although the model parameters have been chosen on the basis of different independent experiments,
i.e.\ apparent molar volume and concentration dependent susceptibility.
The 'osmotic coefficients' based on our model show at higher salt concentrations some temperature dependence
that vanishes for strongly diluted solutions.
The 'osmotic coefficients' derived from vapour pressure measurements \cite{PepelaDunlop:1972} at $T=25^\circ{}C$
agree well with the computed curves.
For low concentrations, where we have no data from vapour pressure measurements,
the reference data \cite{HamerWu:1972} based on a regression model incorporating the Debye-H\"uckel theory
shows some deviations to the results of our model.
In comparison, our model yields a steeper slope in the limit of infinite dilution
and has an earlier and deeper local minimum.

\subsection{Freezing point depression at $p=p^\Ref$} %
We consider the pure solvent in solid phase
and fix the pressure in this subdomain to the reference pressure $p^\Ref$
such that $T^S$ is given by solution of \eqref{PB:04} with $p = p^\Ref$.
For the freezing point of water at atmospheric pressure, we have $T^S = 273.15\UK$.
Applying \eqref{PB:04a} with $p=p^\Ref$ to the equilibrium condition \eqref{PB:03} yields
\begin{align} \label{PB:11}
	g_S^\Pure( T,p^\Ref)	&= g_S( T,p^\Ref)
		+\kB T \, \ln\left( y_S \right)
		+\mu_S^\Deb -\vau_S^\Ref\Thermal{S}(T) \cdot p^\Deb
	~.
\end{align}
We linearize $g_S$ and $g_S^\Pure$ around the state $(T^S,p^\Ref)$.
Therefore, we introduce
\begin{align} \label{PB:13}
\Delta	&:= T^S -T ~, &
s		&:= -\partial_T\, g_S(		T^S,p^\Ref) ~, &
s^\Pure	&:= -\partial_T\, g_S^\Pure(	T^S,p^\Ref)~. &
\end{align}
With \eqref{PB:04} and linearization of the thermal expansion $\Thermal{\alpha} \approx 1 -\beta_\alpha^\Ref\cdot\Delta$,
we infer
\begin{subequations}
\begin{align} \label{PB:15}
	g_S^\Pure(T,p^\Ref)	-g_S(T,p^\Ref)
	&= (s^\Pure -s)\,\Delta \ +\cO(\Delta^2) ~,
\end{align}
\end{subequations}
such that neglecting the higher order terms we get from \eqref{PB:11}
\begin{align} \label{PB:21}
	(s^\Pure -s)\, \Delta
	&=	\kB T\,\ln(y_S)		+\mu_S^\Deb
	-\vau_S^\Ref\Thermal{S}(T) p^\Deb	%
	~.
\end{align}
With the enthalpy of fusion $-q = \tfrac{T^S}{m_S} ( s^\Pure -s )$, %
\eqref{PB:05} and \eqref{Dens:233b} the lowering of the freezing point $\Delta$ is determined by
\begin{align}
	-		\tfrac{m_S}{\kB T}\,q 						\frac{\Delta}{T^S}
	&=	\ln(y_S)
	+\frac{\mu_S^\Deb -\vau_S^\Ref\Thermal{S}(T)  p^\Deb %
			}{\kB T}
\notag\\*&
	=	-\phi \cdot  1000 \Avo\, \vau_S^\Ref\Thermal{S}(T) (1 +f \Ratio) (\nu_A +\nu_C)\cdot  c
	~.
	\label{PB:23}
\end{align}

\paragraph*{Simple mixture and classical limit.}
In the simple mixture model, the contributions from the electric interaction disappear in \eqref{PB:23}.
For comparison with the full model, we introduce $\Delta_{\SM}$ defined as
\begin{align} \label{PB:33}
	-		\tfrac{m_S}{\kB T}\,q 				\,	\frac{\Delta_{\SM}}{T^S}
	&=	\ln(y_S) ~.
\end{align}
In the limit of strong dilution the simple mixture approaches the standard ideal electrolyte theory,
i.e.\ we can assume complete dissociation and $n_\alpha \ll n_S$ for $\alpha\in\{A,C\}$,
such that $\vau_S^\Ref\Thermal{S}(T)\, n_S \approx 1$.
Then, we approximate
$\ln(y_S)	%
			\approx -\sum_{\alpha\neq S} \frac{n_\alpha}{n_S}$.
The ideal electrolyte limit of \eqref{PB:33} thus reads
\begin{align} \label{PB:35}
	 {\Delta_\Ideal}{}
	&=	T^S \tfrac{\kB T}{m_S\,q}\, \vau_S^\Ref\Thermal{S}(T) \, \sum_{\alpha\in\{A,C\}} {n_\alpha}
\end{align}
When we express the ionic number densities in terms of the concentration in $\Umol\Ul^{-1}$
as $n_A +n_C = (\nu_A +\nu_C)\cdot \,1000\Avo \cdot c$,
we can conclude the relations
\begin{align} \label{PB:36}
	\frac{\Delta}{\Delta_\Ideal} &= (1 +f\Ratio)\cdot \phi~, &
	\Delta_\Ideal
	&\approx 1.8623  \frac{\UK\Ul}{\Umol} \cdot 2c ~, &
\end{align}
where we used $q \approx$ $333.55 \UJ \Ug^{-1}$ for water at atmospheric pressure $p^\Ref$
and approximated $T \approx T^S$.

\begin{figure}
	\centering
	\includegraphics[width=.42\textwidth]{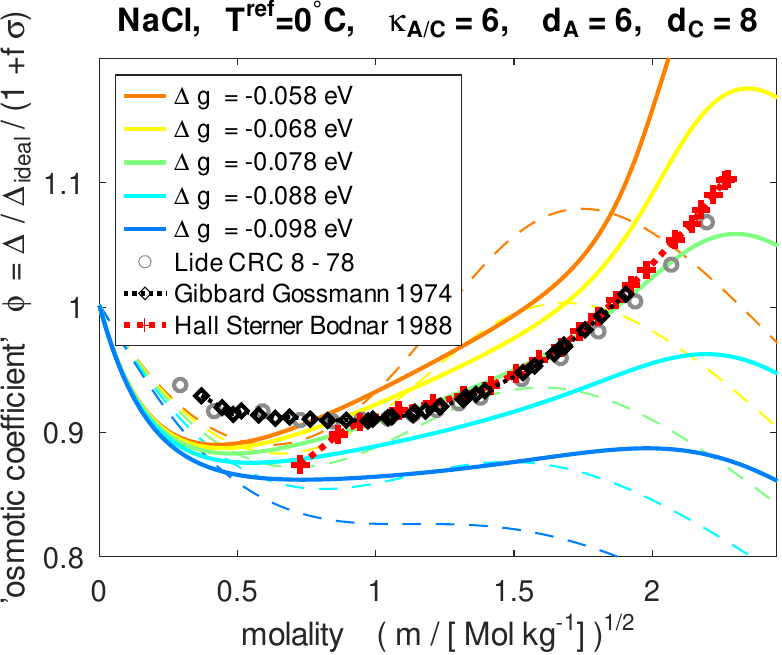}
	\caption{\label{fig:interf-osm}
	Left:
		Computed 'osmotic coefficient' derived from freezing point depression for different values of $\Delta g$
		and comparison with data from \cite{CRC05,GibbardGross:1974,HallSternerBodnar:1988}.
		The dashed lines are related to the Debye-H\"uckel energy of \cite{DH23}.
	}
\end{figure}
\paragraph*{Comparison to experiments.}
When computing the freezing point depression $\Delta$ and 'osmotic coefficient' $\phi$,
we have to apply model parameters different than in Sect.~\ref{sect:appl},
in order to account for the reference temperature $T^\Ref =T^S =0^\circ{}C$.
We set $\vau_S^\Ref = (55.5\cdot n^\Ref)^{-1}$
and $d_A=6$ and $d_C=8$ to achieve $d=d_A+d_C=14$ as motivated by data of \cite{BuHeMa99} for $5^\circ{}C$.
Using the density data of \cite{ChenChenMillero:1980},
and with $\Delta g = -0.78\Uev$,
we adjust the size parameters of the ionic species to $\tilde{r}_A = 82 \Upm$ and $\tilde{r}_C = 163 \Upm$
and set $\Delta\vau = -0.9$. %
For simplicity, we evaluated \eqref{PB:23} at $T^S$ instead of $T$.
Fig.~\ref{fig:interf-osm} shows that the computed 'osmotic coefficient' for different values of the Gibbs energy $\Delta g$.
We observe that the curve agrees best with the experimental data for $m \geq 1 \Umol\Ul^{-1}$
for the same $\Delta g = -0.78\Uev$ that also matched the computed apparent molar volume
to the mass density data of \cite{ChenChenMillero:1980}.
For molality less than  $m \leq 1 \Umol\Ul^{-1}$, there is a visible deviation
between the experimental data and the computed curves.
This appears to be caused by the rather steep slope in the beginning of the curves
and the local minimum that is approached early.
The dashed lines in Fig.~\ref{fig:interf-osm} show for comparison also
the computed 'osmotic coefficient' from the simple mixture with the original Debye-H\"uckel energy added.
From the experimental data, no conclusions about the correct limit slope can be drawn and
for $m \approx 0.2 \Umol\Ul^{-1}$ the distance of both sets of computed curves
to the experimental data is comparable in size.
Remarkably, there is a gap between the different experimental data,
which opens up at $m \approx 0.8 \Umol\Ul^{-1}$ and encloses the computed curves.

\paragraph*{Limiting law.}
To analyze the freezing point depression in the limit $c\to0$,
we first check on the left hand side of \eqref{PB:23} that
\begin{align} \label{PB:16}
	\mu_S^\Deb 	-\vau_S^\Ref p^\Deb
	&=	+\frac{a}{2}\frac{\chi}{D^2}
			\bigg(
				\vau_S^\Ref\Thermal{S}(T) \frac{\lambda}{2}d\cF
				-\left(	\frac{\chi-1}{\chi} \cF		+\frac{\lambda}{2}d\cF \right)
				\frac{\vau_S^\Ref\Thermal{S}(T)\cL_\chi -\frac{\partial}{\partial n_S} \chi}{D} %
			\bigg)~.
\end{align}
For a strongly diluted solution we may assume $1 +f\Ratio \to 1$, as well as
$\vau_S^\Ref\Thermal{S}(T) \cL_\chi = \vau_S^\Ref\Thermal{S}(T) \sum_{\gamma\in\cI} n_\gamma \frac{\partial}{\partial n_\gamma} \chi
\to \tfrac{\partial}{\partial n_S} \chi$. %
We thus approximate \eqref{PB:23} as
\begin{align} \label{PB:213}
	-		\tfrac{m_S}{\kB T^S}\,q 													\, \frac{\Delta}{T}
	&=	\ln(y_S) +\frac{a\,\vau_S^\Ref\Thermal{S}(T)}{4\kB T} \frac{\chi}{D^2} \, \lambda\, d\cF ~.
\end{align}
Subtracting  from this the simple mixture relation \eqref{PB:33}
and applying the strong dilution limit \eqref{PB:35},
the difference of the full model to the standard ideal theory can be expressed as
\begin{align}
	\frac{\Delta_{\SM} -\Delta}{\Delta_\Ideal}
	&=	\frac{1}{\Delta_\Ideal}\frac{T^S}{m_S\,q}\ \frac{a\,\vau_S^\Ref\Thermal{S}(T)}{4} \frac{\chi}{D^2} \lambda\, d\cF
\notag\\
	&=\frac{a}{4\kB T}\, \frac{\LMicro}{n^\Ref} \frac{\chi}{D^2} \	\Lambda \
			\frac{\sum_{\alpha\in\{A,C\}}  z_\alpha^2 {n_\alpha}{} \cdot  F'(R_\alpha,\lambda)}{\sum_{\alpha\in\{A,C\}} {n_\alpha}{} }
\notag\\
	&=	\frac{e_0^2}{4\pi (1+\chi)\eps_0 \kB T} \ \frac{1}{8} \frac{\chi}{D}  \	\Lambda \
			\frac{\sum_{\alpha\in\{A,C\}}  z_\alpha^2 {n_\alpha}{} \cdot F'(R_\alpha,\lambda)}{\sum_{\alpha\in\{A,C\}} {n_\alpha}{} }
	~.
\label{PB:215}
\end{align}
For strong dilution, $\Delta_\SM$ approaches $\Delta_\Ideal$,
$F'(R_\alpha,\lambda)$ converges for all $R_\alpha$ to the same value $F'_0=3$  %
and the neutral salt dissociates completely into $\nu_A$ anions and $\nu_C$ cations.
We thus can write  $n_\alpha = \nu_\alpha\, n^\Ref \cdot c$.
Then, the definition of $\Lambda$ according to \eqref{Deb:06} implies
\begin{align} \label{PB:217}
	\frac{\Delta_\Ideal -\Delta}{\Delta_\Ideal}
	&=	\frac{F'_0}{8}  \frac{e_0^2}{4\pi (1+\chi)\eps_0 \kB T} \ \frac{\chi}{D}  \
			\sqrt{\textstyle  \frac{e_0^2}{     (1+\chi)\eps_0 \kB T} 	\sum_{\alpha\in\{A,C\}}  z_\alpha^2 \nu_\alpha \cdot c \ } \	\cdot
			\frac{\sum_{\alpha\in\{A,C\}}  z_\alpha^2 \nu_\alpha}{\sum_{\alpha\in\{A,C\}} \nu_\alpha } ~.
\end{align}
We conclude, that in the strong dilution limit the same three theorems formulated in \cite{DH23}
are also valid here,
i.e.\  $(\Delta_\Ideal -\Delta)/\Delta_\Ideal$ is:
\begin{enumerate}
\item	 proportional to $\sqrt{c}$,
\item	 proportional to $( \sum_{\alpha\in\{A,C\}} \nu_\alpha z_\alpha^2 / \sum_{\alpha\in\{A,C\}} \nu_\alpha )^{3/2} $,
\item	 proportional to $(1 +\chi)^{-3/2}$.
\end{enumerate}
Applying $\nu_A=\nu_C=1$ and $F'_0=3$ to \eqref{PB:217}, we get for $c\to0$ the approximation
\begin{align}
\label{PB:219}
	\Delta/\Delta_\Ideal & \approx 1 -S_\Delta\cdot \sqrt{2 c} &
	&\text{ with the slope }  \ S_\Delta = 0.5823 ~. &
\end{align}
This limit slope is more steep compared to the slope $S_\Delta^{DH23} = 0.2588$
that is obtained for the Debye-H\"uckel energy according to \cite{DH23} with $(F^{DH23})' \to 4/3$ for $c\to 0$,
cf.\ the supporting material.

\PreprintClearpage
\section{Discussion and conclusions}\label{sect:discussion}
\paragraph*{Failure of pure steric models.}
As shown in Sect.~\ref{sect:appl}, the simple mixture,
which serves as a representative of the class of sterically modified Poisson-Boltzmann models,
cannot reproduce the apparent molar volume in aqueous solutions.
Admittedly, this type of models were not developed with this goal in mind,
but rather to limit charge accumulation in highly concentrated boundary layers.
However, the main feature of these models is the introduction of a size parameter for the ions,
and the most obvious way to identify this parameter should be the mass density of the solutions.
Although fitting the model is in principle possible in the regime
where sufficiently accurate experimental data can be found, two objections remain:
i.) The simple mixture predicts a horizontal slope of $\Phi_V(\sqrt{c})$ at infinite dilution.
ii.) The direct relationship of the apparent molar volume to the degree of dissociation necessarily requires
that a considerable amount of undissociated ion pairs are already present in fairly dilute solutions of strong electrolytes,
which seems unrealistic.

We thus find that the correct representation of the molar volume of electrolytes
must take into account the electrical interaction of the ions, which the simple mixture cannot adequately do,
since it treats all constituents as uncharged.
On the other hand, there is no reason to exclude the simple mixture from the model,
as it incorporates the mechanical and thermal properties of the electrolyte.

\paragraph*{Simple mixture in coupled model.}
The coupling of the simple mixture model with another free energy contribution proposed here
requires careful modifications of the axiomatic derivation procedure described in \cite{BoDrDr23}.
Since in the coupling also the other contribution must in general also be expected to contribute to the total pressure,
as is the case with the electric interaction energy derived above,
the simple mixture model can not be built on an assumed law for the specific heat capacity at constant pressure.
Instead, the derivation must be based on the specific heat capacity at constant volume.
Another essential ingredient of the simple mixture model is the thermal equation of state,
which relates the pressure contribution to the volume.
It contains the bulk modulus, which in the case of water typically is large compared to the ambient pressure,
leading to the notion of incompressibility.
Although the incompressibility can not be exploited
if the balancing of the different pressure contributions in the coupling is the focus of interest,
the coupled system should also fulfill the requirement that the molar volume function
$\nInv^\SM(T,\bm{y})$ in \eqref{SM331} depends linearly on the composition.
%
% \todo{smallness of pressure contribution SiLiKo:2022 (nach eqn (15))}

\paragraph*{Ion interaction energy contribution.}
For the electric ion-ion interaction, we derived a free energy contribution that shares some simplifying assumptions with the derivation of Debye and H\"uckel.
In particular, the local perturbation of the free charge in the electrolyte caused by the ion is approximated on the microscopic level by a linearized Boltzmann distribution around the state in the far field.
The local temperature $T$ and number densities $n_\alpha$ from the macroscopic scale are used to define the far field $n_\alpha^\infty = n_\alpha$  and, consistent with the linearization, the spatially homogeneous temperature and susceptibility $\chi(T,(n_\alpha^\infty)_\alpha)$ on the microscopic scale.
The linearization used in the derivation does not appear to be problematic due to the small amount of charge of the single selected ion.
Likewise, on the microscopic scale, the concentration dependence of $\chi$ will have only minor impact due to the low charge accumulation in the boundary layer around the ion,
whereas the dependence on the concentration is generally taken into account on the macroscopic scale.

The model uses solvated ion radii, which depend on temperature.
This does not exclude the possibility that differently charged ions approach each other to form ion pairs.
However, it makes it possible to take into account a reduction in volume during dissociation, as in $\ce{LiF}$ and $\ce{NaF}$, without having to introduce e.g.\ holes or other additional species into the mixture model.
We also use the value of the susceptibility in the solvation shell, which is assumed locally on the macroscopic scale, since no other value is obvious to us.
The value $\chi_S^\Ref$ of the pure solvent is not suitable in the solvation shell because the solvent molecules there differ from the free ones in that they are bound and no longer freely orientable.

The free energy density due to the electric interaction is then determined directly based on the general thermodynamic relations of Sect.~\ref{sect:gen-thermodyn}.
In contrast to the standard Debye-H\"uckel energy, neither an integration of the thermodynamic relations with respect to temperature is required, nor is the thought experiment of the charging process required as an aid, but instead the general thermodynamic consistency conditions lead to a new restriction.
The temperature dependence of the susceptibility can not be freely chosen, but is predetermined by the thermodynamic conditions!
We are not aware of a similar compatibility condition in the existing literature.

Although the energy derived here differs from standard Debye-H\"uckel energy,
it shares some structural properties.
In particular, the resulting limit laws for infinite dilution coincide in qualitative way
but yield different limit slopes.
However, the experimental uncertainty in mass density measurements does not allow a decision what is the correct limit slope.

\paragraph*{Non-constant susceptibility.}
Experiments with %
dilute solutions suggest a linear dependence of the susceptibility on the number densities at the reference temperature of the experiment.
The thermodynamic consistency condition then prescribes in general a nonlinear dependence for temperatures different from the reference temperature.
As can be seen in Sect.~\ref{sect:const-discuss}, the deviation from linearity is small enough to maintain consistency with a linear dependence as model input even in alternative model approaches with a deviating reference temperature.
In the limit of infinite dilution, the temperature dependence of the susceptibility is independent of the dissolved salt and is determined exclusively by the thermal expansion of the solvent.

\paragraph*{Incomplete dissociation.}
The numerical experiments show that the electric ion interaction energy provides the coupled model
with an alternative mechanism for non-constant apparent molar volume in addition to incomplete dissociation.
We include incomplete dissociation in general into the coupled model,
even for strong electrolytes such as aqueous $\ce{NaCl}$ solutions,
and observe that when using identical parameters, the dissociation degree is considerably higher in the coupled model
compared to the pure simple mixture.
For a solution of $0.5\Umol\Ul^{-1} \ce{NaCl}$ the coupled model yields a fraction of undissociated ion pairs less than $0.05$.

In principle, one might suspect that the inclusion of incomplete dissociation in the model is an attempt to increase the degrees of freedom to facilitate the fit of the results to experimental data.
However, we note that the model targets dilute solutions of strong electrolytes as well as concentrated solutions of weak electrolytes and no clearly defined point is apparent where incomplete dissociation would set in.
Moreover, the additional free model parameters introduced by incomplete dissociation are themselves also tied to independent experiments such as the dielectric decrement.

More generally, in the material modeling we rely almost entirely on linear mixing laws that require only a minimal number of parameters.
Linear laws are applied for
the molar volume function $\nInv^\SM(	T,					\bm{y})$,
specific heat				$c_v(		T, 		\nInv^\SM,	\bm{y})$,
specific internal energy $\bar{u}(		T^\Ref,	\nInv^\SM,	\bm{y})$ and
susceptibility				$\chi(		T^\Ref,			(n_\alpha)_\alpha)$.
Only the specific entropy $\bar{s}(		T^\Ref,	\nInv^\SM,	\bm{y})$
contains an additional logarithmic contribution of Boltzmann type, but without adjustable parameters.
Instead of modelling nonlinear material behaviour by adding nonlinear terms
with additional parameters to the mentioned material laws,
which in the case of $\nInv^\SM$ would even be incompatible with an incompressible limit,
the nonlinearity in our model is realized by the dissociation reaction.

\paragraph*{Numerical results.}
In the above numerical studies, the coupled model is successfully applied to reproduce the apparent molar volume
over a wider range of salt concentration and temperature.
From experiments with related salts that share either an anion or a cation,
a consistent set of ion specific parameters can be deduced.
Notably, for aqueous $\ce{LiCl}$ solutions of fixed salt concentration and varying temperature,
the model predicts the local maximum of the apparent molar volume at $T\approx40^\circ{}C$,
much earlier than for solutions of $\ce{NaCl}$ or $\ce{KCl}$.
The solvated ion radii deduced  here from mass density measurements
are considerably smaller than those derived from steric Poisson-Boltzmann models
applied to double layer capacitance experiments, cf.\ e.g.\ \cite{LGD16,LM22}.
Independent experiments related to phase boundaries, %
such as vapour pressure reduction or freezing point depression
confirm the previously identified model parameters.
In terms of the 'osmotic coefficient' as a comparison measure,
the numerical results show good agreement with available direct experimental data.

We remark that here no particular emphasis was placed on parameter optimization.
An effort to tune model parameter for an optimal fit of experimental results
should be based on more high precision experimental data.

\paragraph*{Outlook.}
The model proposed here should next be applied to other the classical tests of electrolyte theory,
such as the heat of dilution and the concentration dependent specific heat capacity $c_p$ at fixed external pressure.
However, such tests also require further input parameter related to the thermal properties of the electrolyte, i.e.\
the specific heat capacities $c_\alpha^\Ref$ which
define $c_v(T,\nInv^\SM,\bm{y})$ (at fixed volume) but are unknown for the solvated ions.
In addition, the model should be applied to electromotive force of Galvanic cells,
which are used to evaluate activity coefficients as characteristic measures for electrolytes.
This requires careful modeling of the involved electrode surface reactions in the way of \cite{DGM16,DGM18}.

In order to limit the scope of this paper,
the applications and numerical studies here have been restricted to
solutions of binary electrolytes of monovalent ions.
For example, the autoprotolysis of water, possibly different solvation shells of ions depending on the salt input,
or different types of ion pairs have been neglected here.
The model can and should also be applied to solutions with multivalent ions whose description with the classical Debye-Hückel theory has weaknesses. %
However, the mere application of the above equations with only increased charge numbers may not be sufficient
and careful consideration of the different dissociation steps
and properties of the solvation shell of multivalent ions may be required.

The striking symmetry of the measured differential double layer capacitance
of a non-adsorbing electrolyte such as $\ce{KPF_6}$ on single crystal surfaces, cf.\ e.g.~\cite{Valette81},
suggests that in a steric Poisson-Boltzmann model $\vau_A^\Ref = \vau_C^\Ref$
and one could even be led to conjecture
that the specific volume of an ion depends only on the solvation number $\kappa_\alpha$,
which in turn is determined solely by the valence of the ion.
Taking the solvated ion radii deduced here from the apparent molar volume at infinite dilution
as input parameters of the steric model,
it would be interesting to analyze,
whether or not  the dependence of the susceptibility on the strong electric field inside the double layer, cf.\ e.g.\ \cite{BaKiStAj09,LM22}, instead of the elastic volume exclusion effects,
can explain the experimentally observed symmetry of the double layer capacitance.

\PreprintClearpage
\bibliographystyle{alpha}
% \bibliography{bibliography_debye}

\begin{thebibliography}{KMMT18}

\bibitem[AM99]{Apelblat:1999}
A.~Apelblat and E.~Manzurola.
\newblock Volumetric properties of water, and solutions of sodium chloride and
  potassium chloride at temperatures from ${T}= 277.15 {K}$ to ${T}= 343.15
  {K}$ at molalities of ($0.1$, $0.5$, and $1.0$) $mol\cdot{}kg^{-1}$.
\newblock {\em J. Chem. Thermodyn.}, 31(7):869--893, 1999.

\bibitem[AM01]{Apelblat:2001}
A.~Apelblat and E.~Manzurola.
\newblock Volumetric properties of aqueous solutions of lithium chloride at
  temperatures from $278.15 {K}$ to $338.15 {K}$ and molalities ($0.1$, $0.5$,
  and $1.0$)$mol\cdot{}kg^{-1}$.
\newblock {\em J. Chem. Thermodyn.}, 33(9):1133--1155, 2001.

\bibitem[Arr87]{Arrhenius:1887}
S.~Arrhenius.
\newblock {\"Uber} die {Dissociation} der in {Wasser} {gel\"osten} {Stoffe}.
\newblock {\em Z. Physik. Chem.}, 1:631--648, 1887.

\bibitem[AW81]{AllredWoolley:1981}
G.~C. Allred and E.~M. Woolley.
\newblock Heat capacities of aqueous {$HCl$}, {$NaOH$}, and {$NaCl$} at
  $283.15$, $298.15$ and $313.15$ {K}: {$\Delta{}C^\circ_p$} for ionization of
  water.
\newblock {\em J. Chem. Thermodyn.}, 13(2):147--154, 1981.

\bibitem[BAO97]{BoAnOr97}
I.~Borukhov, D.~Andelman, and H.~Orland.
\newblock Steric effects in electrolytes: A modified {Poisson}--{Boltzmann}
  equation.
\newblock {\em Phys. Rev. Lett.}, 79:435--438, 1997.

\bibitem[BDD23]{BoDrDr23}
D.~Bothe, W.~Dreyer, and P.-E. Druet.
\newblock Multicomponent incompressible fluids--{An} asymptotic study.
\newblock {\em J. Appl. Math. Mech.}, 103(7):e202100174, 2023.

\bibitem[BHM99]{BuHeMa99}
R.~Buchner, G.~T. Hefter, and P.~M. May.
\newblock Dielectric relaxation of aqueous {NaCl} solutions.
\newblock {\em J. Phys. Chem. A}, 103(1):1--9, 1999.

\bibitem[Bje26]{Bjerrum:1926}
N.~Bjerrum.
\newblock Untersuchungen {\"uber} {Ionenassoziation}.
\newblock {\em K. Dan. Vidensk. Selsk.}, 7(9), 1926.

\bibitem[BKSA09]{BaKiStAj09}
M.Z. Bazant, M.S. Kilic, B.D. Storey, and A.~Ajdari.
\newblock Towards an understanding of induced-charge electrokinetics at large
  applied voltages in concentrated solutions.
\newblock {\em Advances in Colloid and Interface Science}, 152(1-2):48 -- 88,
  2009.

\bibitem[BYAP11]{BeYaAnPo11}
D.~Ben-Yaakov, D.~Andelman, and R.~Podgornik.
\newblock Dielectric decrement as a source of ion-specific effects.
\newblock {\em J. Chem. Phys.}, 134(7):074705, 2011.

\bibitem[CBL{\etalchar{+}}07]{Chu:2007}
V.~B. Chu, Y.~Bai, J.~Lipfert, D.~Herschlag, and S.~Doniach.
\newblock Evaluation of ion binding to dna duplexes using a size-modified
  poisson-boltzmann theory.
\newblock {\em Biophys. J.}, 93(9):3202 -- 3209, 2007.

\bibitem[CCM80]{ChenChenMillero:1980}
C.-T.~A. Chen, J.~H. Chen, and F.~J. Millero.
\newblock Densities of {$NaCl$}, {$MaCl_2$}, {$Na_2SO_4$}, and {$MgSO_4$}
  aqueous solutions at $1 atm$ from $0$ to $50^\circ{}{C}$ and from $0.001$ to
  $1.5 m$.
\newblock {\em J. Chem. Eng. Data}, 25(4):307--310, 1980.

\bibitem[CG85]{ColinGlew:1985}
E.~C.~W. Colin and D.~N. Glew.
\newblock Evaluation of the thermodynamic functions for aqueous sodium chloride
  from equilibrium and calorimetric measurements below $154^\circ{}${C}.
\newblock {\em J. Phys. Chem. Ref. Data}, 14(2):489--610, 1985.

\bibitem[CHB03]{ChenHefterBuchner:2003}
T.~Chen, G.~Hefter, and R.~Buchner.
\newblock Dielectric spectroscopy of aqueous solutions of {$KCl$} and {$CsCl$}.
\newblock {\em J. Phys. Chem. A}, 107(20):4025--4031, 2003.

\bibitem[Deb24]{Debye:1924}
P.~Debye.
\newblock Osmotische {Zustandsgleichung} und {Aktivit\"at} {verd\"unnter}
  starker {Elektrolyte}.
\newblock {\em Physik. Z.}, 25:97--107, 1924.

\bibitem[dGM84]{dGM84}
S.~R. de~Groot and P.~Mazur.
\newblock {\em Non-equilibrium Thermodynamics}.
\newblock Dover Publications, New York, 1984.

\bibitem[DGM13]{DGM13}
W.~Dreyer, C.~Guhlke, and R.~M\"uller.
\newblock Overcoming the shortcomings of the {N}ernst--{P}lanck model.
\newblock {\em Phys. Chem. Chem. Phys.}, 15:7075--7086, 2013.

\bibitem[DGM16]{DGM16}
W.~Dreyer, C.~Guhlke, and R.~M\"uller.
\newblock A new perspective on the electron transfer: recovering the
  {Butler}-{Volmer} equation in non-equilibrium thermodynamics.
\newblock {\em Phys. Chem. Chem. Phys.}, 18:24966--24983, 2016.

\bibitem[DGM18]{DGM18}
W.~Dreyer, C.~Guhlke, and R.~{M\"uller}.
\newblock Bulk-surface electro-thermodynamics and applications to
  electrochemistry.
\newblock {\em Entropy}, 20(12):939, 2018.

\bibitem[DGM19]{DGM19}
W.~Dreyer, C.~Guhlke, and R.~{M\"uller}.
\newblock The impact of solvation and dissociation on the transport parameters
  of liquid electrolytes: continuum modeling and numerical study.
\newblock {\em Eur. Phys. J.-Spec. Top.}, 227(18):2515--2538, 2019.

\bibitem[DH23a]{DH23}
P.~Debye and E.~{H\"uckel}.
\newblock Zur {Theorie} der {Elektrolyte} {I}. {G}efrierpunktserniedrigung und
  verwandte {E}rscheinungen.
\newblock {\em Physik. Z.}, 24(9):185--206, 1923.

\bibitem[DH23b]{DH23II}
P.~Debye and E.~{H\"uckel}.
\newblock Zur {Theorie} der {Elektrolyte} {II}. {D}as {G}renzgesetz für die
  elektrische {L}eitf{\"a}higkeit.
\newblock {\em Physik. Z.}, 24:305--325, 1923.

\bibitem[Ell66]{Ellis:1966}
A.~J. Ellis.
\newblock Partial molal volumes of alkali chlorides in aqueous solution to
  $200^\circ$.
\newblock {\em J. Chem. Soc. A}, pages 1579--1584, 1966.

\bibitem[FCBM16]{FiChBuMo:2016}
B.~Figliuzzi, W.H.R. Chan, C.R. Buie, and J.L. Moran.
\newblock Nonlinear electrophoresis in the presence of dielectric decrement.
\newblock {\em Phys. Rev. E}, 94:023115, Aug 2016.

\bibitem[GG74]{GibbardGross:1974}
H.~F. Gibbard and A.~F. Gossmann.
\newblock Freezing points of electrolyte mixtures. {I}. mixtures of sodium
  chloride and magnesium chloride in water.
\newblock {\em J. Solution Chem.}, 3(5):385--393, 1974.

\bibitem[GH25]{GrHa:1925}
P.~{Gro\ss} and O.~Halpern.
\newblock {\"Uber} die {temperaturabh\"angigen} {Parameter} in der {Statistik}
  und die {Debyesche} {Elektrolyttheorie}.
\newblock {\em Physik. Z.}, 26:403--407, 1925.

\bibitem[GS18]{GuSt:2018}
A.~Gupta and H.A. Stone.
\newblock Electrical double layers: Effects of asymmetry in electrolyte valence
  on steric effects, dielectric decrement, and ion–ion correlations.
\newblock {\em Langmuir}, 34(40):11971--11985, 2018.

\bibitem[GW85]{GatesWood:1985}
J.~A. Gates and R.~H. Wood.
\newblock Densities of aqueous solutions of sodium chloride, magnesium
  chloride, potassium chloride, sodium bromide, lithium chloride, and calcium
  chloride from $0.05$ to $5.0$ $mol\cdot kg^{-1}$ and $0.1013$ to $40 {MPa}$
  at $298.15 {K}$.
\newblock {\em J. Chem. Eng. Data}, 30(1):44--49, 1985.

\bibitem[Hey96]{Heyrovska:1996}
R.~Heyrovsk{\'{a}}.
\newblock Physical electrochemistry of strong electrolytes based on partial
  dissociation and hydration: Quantitative interpretation of the thermodynamic
  properties of {NaCl}(aq) from {\textquotedblleft}zero to
  saturation{\textquotedblright}.
\newblock {\em J. Electrochem. Soc.}, 143(6):1789--1793, jun 1996.

\bibitem[HGBS95]{HGBS:1995}
N.~Hubert, Y.~Gabes, J.-B. Bourdet, and L.~Schuffenecker.
\newblock Vapor pressure measurements with a nonisothermal static method
  between 293.15 and 363.15 {K} for electrolyte solutions. application to the
  {$H_2O + NaCl$} system.
\newblock {\em J. Chem. Engin. Data}, 40(4):891--894, 1995.

\bibitem[HH21]{HH:2021}
L.~Hnedkovsky and G.~Hefter.
\newblock Densities and apparent molar volumes of aqueous solutions of
  {$NaClO4$}, {$KClO4$}, and {$KCl$} at temperatures from 293 to 343 {K}.
\newblock {\em J. Chem. Eng. Data}, 66(9):3645--3658, 2021.

\bibitem[HHLH16]{HHLH:2016}
B.~Hu, L.~Hnedkovsky, W.~Li, and G.~Hefter.
\newblock Densities and molar volumes of aqueous solutions of {$LiClO_4$} at
  temperatures from 293 {K} to 343 {K}.
\newblock {\em J. Chem. Eng. Data}, 61(4):1388--1394, 2016.

\bibitem[HO59]{HarnedOwen1958}
H.S. Harned and B.B. Owen.
\newblock {\em The Physical Chemistry of Electrolytic Solutions}.
\newblock Reinhold, New York, third edition, 1959.

\bibitem[HRC48]{HaRiCo:1948}
J.~B. Hasted, D.~M. Ritson, and C.~H. Collie.
\newblock Dielectric properties of aqueous ionic solutions. parts {I} and {II}.
\newblock {\em J. Chem. Phys.}, 16(1):1--21, 1948.

\bibitem[HSB88]{HallSternerBodnar:1988}
D.~L. Hall, S.~M. Sterner, and R.~J. Bodnar.
\newblock Freezing point depression of {$NaCl$}-{$KCl$}-{$H_2O$} solutions.
\newblock {\em Econ. Geol.}, 83(1):197--202, 1988.

\bibitem[H{\"u}c25]{Hueckel:1925}
E.~H{\"u}ckel.
\newblock Zur {Theorie} konzentrierterer {w\"asseriger} {L\"osungen} starker
  {Elektrolyte}.
\newblock {\em Physik. Z.}, 26:93--147, 1925.

\bibitem[HvRL12]{HavRLu:2012}
M.M. Hatlo, R.~van Roij, and L.~Lue.
\newblock The electric double layer at high surface potentials: {The} influence
  of excess ion polarizability.
\newblock {\em EPL}, 97(2):28010, Jan 2012.

\bibitem[HW72]{HamerWu:1972}
W.~J. Hamer and Y.‐C. Wu.
\newblock Osmotic coefficients and mean activity coefficients of
  uni‐univalent electrolytes in water at 25$^\circ${C}.
\newblock {\em J. Phys. Chem. Ref. Data}, 1(4):1047--1100, 1972.

\bibitem[KBA07]{KiBaAj07I}
M.~S. Kilic, M.~Z. Bazant, and A.~Ajdari.
\newblock Steric effects in the dynamics of electrolytes at large applied
  voltages. {I}. {Double}-layer charging.
\newblock {\em Phys. Rev. E}, 75:021502, 2007.

\bibitem[KII96]{KrIgIg:1996}
V.~Kralj-{Igli\v{c}} and A.~{Igli\v{c}}.
\newblock A simple statistical mechanical approach to the free energy of the
  electric double layer including the excluded volume effect.
\newblock {\em J. Phys. II}, 6(4):477--491, 1996.

\bibitem[KMMT18]{KoMaMoTh:2018}
G.~M. Kontogeorgis, B.~Maribo-Mogensen, and K.~Thomsen.
\newblock The {Debye}-{H\"uckel} theory and its importance in modeling
  electrolyte solutions.
\newblock {\em Fluid Ph. Equilib.}, 462:130--152, 2018.

\bibitem[LE14]{LiEi14}
J.-L. Liu and B.~Eisenberg.
\newblock {Poisson}-{Nernst}-{Planck}-{Fermi} theory for modeling biological
  ion channels.
\newblock {\em J. Chem. Phys.}, 141(22):22D532, 2014.

\bibitem[LE15]{LiuEisenberg:2015}
J.-L. Liu and B.~Eisenberg.
\newblock Poisson–{Fermi} model of single ion activities in aqueous
  solutions.
\newblock {\em Chem. Phys. Lett.}, 637:1--6, 2015.

\bibitem[LE18]{LiuEisenberg:2018}
J.-L. Liu and B.~Eisenberg.
\newblock Poisson-{Fermi} modeling of ion activities in aqueous single and
  mixed electrolyte solutions at variable temperature.
\newblock {\em J. Chem. Phys.}, 148(5):054501, 02 2018.

\bibitem[LGD16]{LGD16}
M.~Landstorfer, C.~Guhlke, and W.~Dreyer.
\newblock Theory and structure of the metal-electrolyte interface incorporating
  adsorption and solvation effects.
\newblock {\em Electrochim. Acta}, 201:187--219, 2016.

\bibitem[Li09]{Li:2009}
B.~Li.
\newblock Continuum electrostatics for ionic solutions with non-uniform ionic
  sizes.
\newblock {\em Nonlinearity}, 22(4):811--833, feb 2009.

\bibitem[Lid05]{CRC05}
D.P. Lide, editor.
\newblock {\em CRC Handbook of Chemistry and Physics}.
\newblock CRC PRESS, 2005.

\bibitem[LM29]{LangeMeixner:1929}
E.~Lange and J.~Meixner.
\newblock Zur {Individualit\"at} der integralen {Verd\"unnungs}{w\"armen}
  starker {Elektrolyte}.
\newblock {\em Physik. Z.}, 30:670--678, 1929.

\bibitem[LM22]{LM22}
M.~Landstorfer and R.~{M\"uller}.
\newblock Thermodynamic models for a concentration and electric field dependent
  susceptibility in liquid electrolytes.
\newblock {\em Electrochim. Acta}, 428:140368, 2022.

\bibitem[Mar11]{Marcus:2011}
Y.~Marcus.
\newblock Electrostriction in electrolyte solutions.
\newblock {\em Chem. Rev.}, 111(4):2761--2783, 2011.

\bibitem[Mar13]{Marcus:2013}
Y.~Marcus.
\newblock Evaluation of the static permittivity of aqueous electrolytes.
\newblock {\em J. Solution Chem.}, 42(12):2354--2363, 2013.

\bibitem[Mas29]{Masson:1929}
D.~O. Masson.
\newblock Solute molecular volumes in relation to solvation and ionization.
\newblock {\em Philos. Mag.}, 8(49):218--235, 1929.

\bibitem[MDH68]{MilleroDrostHansen:1968}
F.~J. Millero and W.~Drost-Hansen.
\newblock Apparent molal volumes of aqueous monovalent salt solutions at
  various temperatures.
\newblock {\em J. Chem. Eng. Data}, 13(3):330--333, 1968.

\bibitem[MH06]{MarcusHefter:2006}
Y.~Marcus and G.~Hefter.
\newblock Ion pairing.
\newblock {\em Chem. Rev.}, 106(11):4585--4621, 2006.

\bibitem[Mil71]{Millero:1971}
F.~J. Millero.
\newblock Molal volumes of electrolytes.
\newblock {\em Chem. Rev.}, 71(2):147--176, 1971.

\bibitem[ML23]{ML23}
R.~M{\"u}ller and M.~Landstorfer.
\newblock Galilean bulk-surface electrothermodynamics and applications to
  electrochemistry.
\newblock {\em Entropy}, 25(3):416, 2023.

\bibitem[MR17]{MayRowland:2017}
P.~M. May and D.~Rowland.
\newblock Thermodynamic modeling of aqueous electrolyte systems: Current
  status.
\newblock {\em J. Chem. Engin. Data}, 62(9):2481--2495, 2017.

\bibitem[M{\"u}l85]{Mu85}
I.~M{\"u}ller.
\newblock {\em Thermodynamics}.
\newblock Pitman Publishing, London, 1985.

\bibitem[NA15]{NaAn15}
Y.~Nakayama and D.~Andelman.
\newblock Differential capacitance of the electric double layer: {The}
  interplay between ion finite size and dielectric decrement.
\newblock {\em J. Chem. Phys.}, 142(4):044706, 2015.

\bibitem[Ons27]{OnsagerII:1927}
L.~Onsager.
\newblock Zur {Theorie} der {Elektrolyte} {II}.
\newblock {\em Physik. Z.}, 28(8):277--298, 1927.

\bibitem[PD72]{PepelaDunlop:1972}
C.~N. Pepela and P.~J. Dunlop.
\newblock A re-examination of the vapour pressures of aqueous sodium chloride
  solutions at 25 $^\circ${C}.
\newblock {\em J. Chem. Thermodyn.}, 4(2):255--258, 1972.

\bibitem[PDH84]{PDH:1984}
T.~G. Pedersen, C.~Dethlefsen, and A.~Hvidt.
\newblock Volumetric properties of aqueous solutions of alkali halides.
\newblock {\em Carlsberg Res. Commun.}, 49(3):445--455, 1984.

\bibitem[PG08]{Perrys}
R.~H. Perry and D.~W. Green, editors.
\newblock {\em Perry's chemical engineers' handbook}.
\newblock McGraw-Hill, New York, 8 edition, 2008.

\bibitem[Red46]{Redlich:1946}
O.~Redlich.
\newblock The dissociation of strong electrolytes.
\newblock {\em Chem. Rev.}, 39(2):333--356, 1946.

\bibitem[RM64]{RedlichMeyer:1964}
O.~Redlich and D.~M. Meyer.
\newblock The molal volumes of electrolytes.
\newblock {\em Chem. Rev.}, 64(3):221--227, 1964.

\bibitem[RR31]{RedlichRosenfeld:1931}
O.~Redlich and P.~Rosenfeld.
\newblock Das partielle molare {Volumen} von {gel\"osten} {Elektrolyten}. {I.}
\newblock {\em Z. Phys. Chem.}, 155A(1):65--74, 1931.

\bibitem[RS02]{RoSt2002}
R.A. Robinson and R.H. Stokes.
\newblock {\em Electrolyte Solutions}.
\newblock Dover Publications, New York, second revised edition, 2002.

\bibitem[SL15]{ShLy:2015}
I.~Y. Shilov and A.~K. Lyashchenko.
\newblock The role of concentration dependent static permittivity of
  electrolyte solutions in the {Debye}–{H\"uckel} theory.
\newblock {\em J. Phys. Chem. B}, 119(31):10087--10095, 2015.

\bibitem[SLK23]{SiLiKo:2023}
G.~M. Silva, X.~Liang, and G.~M. Kontogeorgis.
\newblock How to account for the concentration dependency of relative
  permittivity in the {Debye}–{H\"uckel} and {Born} equations.
\newblock {\em Fluid Ph. Equilib.}, 566:113671, 2023.

\bibitem[Val81]{Valette81}
G.~Valette.
\newblock Double layer on silver single-crystal electrodes in contact with
  electrolytes having anions which present a slight specific adsorption: Part
  {I}. the (110) face.
\newblock {\em J. Electroanal. Chem.}, 122:285--297, 1981.

\bibitem[VB14]{ValiskoBoda:2014}
M.~{Valisk\'o} and D.~Boda.
\newblock The effect of concentration- and temperature-dependent dielectric
  constant on the activity coefficient of {NaCl} electrolyte solutions.
\newblock {\em J. Chem. Phys.}, 140(23):234508, 2014.

\bibitem[VB23]{ValiskoBoda:2023}
M.~{Valisk\'o} and D.~Boda.
\newblock Resurrection of {H\"uckel}’s idea: {Decoupling} ion–ion and
  ion–water terms in activity coefficients via the state-dependent dielectric
  constant.
\newblock {\em Fluid Ph. Equilib.}, 572:113826, 2023.

\bibitem[VVB10]{VinczeValiskoBoda:2010}
J.~Vincze, M.~{Valisk\'o}, and D.~Boda.
\newblock The nonmonotonic concentration dependence of the mean activity
  coefficient of electrolytes is a result of a balance between solvation and
  ion-ion correlations.
\newblock {\em J. Chem Phys}, 133(15):154507, 10 2010.

\bibitem[Zav01]{Zavitsas:2001}
A.~A. Zavitsas.
\newblock Properties of water solutions of electrolytes and nonelectrolytes.
\newblock {\em J. Phys. Chem. B}, 105(32):7805--7817, 2001.

\bibitem[ZH18]{ZhangHuang:2018}
Y.~Zhang and J.~Huang.
\newblock Treatment of ion-size asymmetry in lattice-gas models for electrical
  double layer.
\newblock {\em J. Phys. Chem. C}, 122(50):28652--28664, 2018.

\end{thebibliography}
\newcommand{\etalchar}[1]{$^{#1}$}

\end{document}